\documentclass[12pt, final]{article}
\usepackage{amssymb}
\usepackage{amsmath}
\usepackage{mathtools}
\usepackage[margin=1in]{geometry}
\usepackage{titling}
\usepackage{titlesec}
\usepackage[usenames, dvipsnames]{xcolor}
\usepackage{setspace}
\usepackage{graphicx}
\usepackage{siunitx}
\usepackage{array}
\usepackage{booktabs}
\usepackage[font={normal,sc}]{caption}
\usepackage[capposition=top]{floatrow}
\usepackage[font={small}]{subcaption}
\usepackage[bottom]{footmisc}
\definecolor{mycolor}{RGB}{140, 26, 17}
\usepackage[most]{tcolorbox}
\usepackage{tikz}

\usepackage{natbib}
\usepackage[colorlinks=true,
			urlcolor=mycolor,
			citecolor=mycolor,
			linkcolor=mycolor]{hyperref}

\ExplSyntaxOn
\cs_if_exist:NF \expandableinput
  {
    \cs_new:Npn \expandableinput #1
      { \use:c { @@input } { \file_full_name:n {#1} } }
  }
\ExplSyntaxOff

\newcommand{\lmat}{\left( \begin{matrix} }
\newcommand{\rmat}{\end{matrix} \right)}

\newcommand{\nin}{\noindent}
\newcommand{\veps}{\varepsilon}

\makeatletter \renewcommand\d[1]{\ensuremath{\;\mathrm{d}#1\@ifnextchar\d{\!}{}}}
\makeatother

\newcommand{\ee}{\mathbb{E}}	 	 	             		  

\newcommand*{\QED}{\hfill\ensuremath{\square}}

\titleformat{\title}{\normalfont\Large\scshape}{\thetitle}{1em}{}									\titleformat{\section}{\centering\normalfont\fontsize{14}{15}\scshape}{\thesection}{1em}{}				\titleformat{\subsection}{\normalfont\fontsize{13}{15}\scshape}{\thesubsection}{1em}{}		\titleformat{\subsubsection}{\normalfont\fontsize{12}{15}\itshape}{\thesubsubsection}{1em}{}

\newtcbtheorem[auto counter]{definition}{Definition}{after title = {\smallskip},
	colback = white,
	colbacktitle = white,
	coltitle = black,
	fonttitle = \normalsize\scshape,
	colframe = white,
	boxrule = 0pt,
	titlerule = 1pt,
	arc = 1pt,
	boxsep = 1pt,
	title = {\strut#1},
	before skip = 20pt plus 4pt,
	after skip = 20pt plus 4pt,
	breakable
}{def}

\newtcbtheorem[auto counter]{assumption}{Assumption}{after title = {\smallskip},
	colback = white,
	colbacktitle = white,
	coltitle = black,
	fonttitle = \normalsize\scshape,
	colframe = white,
	boxrule = 0pt,
	titlerule = 1pt,
	arc = 1pt,
	boxsep = 1pt,
	title = {\strut#1},
	before skip = 20pt plus 4pt,
	after skip = 20pt plus 4pt,
	breakable
}{assump}

\newcounter{proposition}
\newtcbtheorem[auto counter]{proposition}{Proposition}{before title={\stepcounter{proposition}},
	after title={\smallskip},
	colback = gray!10,
	colbacktitle = gray!30,
	coltitle = black,
	fonttitle = \normalsize\scshape,
	colframe = white,
	boxrule = 0pt,
	titlerule = 0pt,
	arc = 5pt,
	boxsep = 5pt,
	title = {\strut#1},
	before skip = 20pt plus 4pt,
	after skip = 20pt plus 4pt,
	breakable
}{prop}

\newcounter{lemma}
\newtcbtheorem[auto counter]{lemma}{Lemma}{before title={\stepcounter{lemma}},
	after title={\smallskip},
	colback = gray!10,
	colbacktitle = gray!30,
	coltitle = black,
	fonttitle = \normalsize\scshape,
	colframe = white,
	boxrule = 0pt,
	titlerule = 0pt,
	arc = 5pt,
	boxsep = 5pt,
	title = {\strut#1},
	before skip = 20pt plus 4pt,
	after skip = 20pt plus 4pt,
	breakable
}{lemma}

\newtcbtheorem[auto counter, number within = proposition]{corollary}{Corollary}{after title = {\smallskip},
	colback = white,
	colbacktitle = white,
	coltitle = black,
	fonttitle = \normalsize\scshape,
	colframe = white,
	boxrule = 0pt,
	titlerule = 1pt,
	arc = 1pt,
	boxsep = 1pt,
	title = {\strut#1},
	before skip = 20pt plus 4pt,
	after skip = 20pt plus 4pt,
	breakable
}{coro}

\newcommand{\defref}[1]{\hyperref[#1]{Definition~\ref*{#1}}}
\newcommand{\assumptionref}[1]{\hyperref[#1]{Assumption~\ref*{#1}}}
\newcommand{\propref}[1]{\hyperref[#1]{Proposition~\ref*{#1}}}
\newcommand{\lemmaref}[1]{\hyperref[#1]{Lemma~\ref*{#1}}}

\begin{document}

\title{\bf From Long to Short: \\ How Interest Rates Shape Life Insurance Markets\thanks{We are grateful to Beatriz Garcia for excellent research assistance. We thank Jennie Bai, Alexandru Barbu (discussant), Enrico Biffis, Emilio Bisetti (discussant), Patrick Bolton, Markus Brunnermeier, Gilles Chemla, Leonardo D’Amico, Andrew Ellul, Johan Hombert, Nicolas Hommel, Rajkamal Iyer, Alessandro Previtero, Tarun Ramadorai, Ishita Sen (discussant), Jakob Ahm Sørensen (discussant), Motohiro Yogo, and seminar and conference participants at Imperial College London, Indiana University, Bayes Business School, CICF, Wabash River Conference, HKUST-GZ, HEC Paris, AFA, MFA, University of Porto, USC, Princeton University, Columbia Workshop in New Empirical Finance, NBER Insurance Working Group Meeting, and BIS-CEPR-Gerzensee-SFI Conference on Financial Intermediation for constructive feedback.}}
\author{{Ziang Li}\thanks{Imperial College London. Email: \href{mailto:ziang.li@imperial.ac.uk}{ziang.li@imperial.ac.uk}} \and {Derek Wenning}\thanks{Kelley School of Business, Indiana University. Email: \href{mailto:dtwennin@iu.edu}{dtwennin@iu.edu}}}
\date{August 5, 2026}

\thispagestyle{plain}
\maketitle

\medskip

\begin{center}
	{\scshape Abstract}
\end{center}

{
\nin This paper explores how financial institutions pass interest rate risk through to product markets using the life insurance industry as a setting. We show theoretically that it is optimal for insurers to distort product issuance across maturities to offset duration gaps. We examine insurers exogenously exposed to interest rate risk through their variable annuity liabilities after the 2008 financial crisis. Consistent with our mechanism, exposed insurers developed negative duration gaps, increased markups on long-duration products, and rebalanced product issuance toward shorter-duration products to hedge. This response reduced long-duration life insurance coverage by 12.1\% of GDP between 2005 and 2023.

}

\pagebreak

\section{Introduction}

\nin Financial institutions play two central roles in the financial system: they sell financial products, which become their liabilities, and they manage investments funded by the proceeds. This duality gives them several ways to manage balance-sheet risk. For example, when regulatory or market frictions limit asset-side adjustment, institutions may instead hedge through product markets. Such rebalancing can be socially costly, however, if doing so makes certain products less desirable and reduces financial participation. To understand this tradeoff, this paper studies how U.S. life insurers rebalance their product portfolios in response to heightened interest rate risk.

\begin{figure}[h!]
    \begin{subfigure}{0.49\textwidth}
        \includegraphics[width = \textwidth]{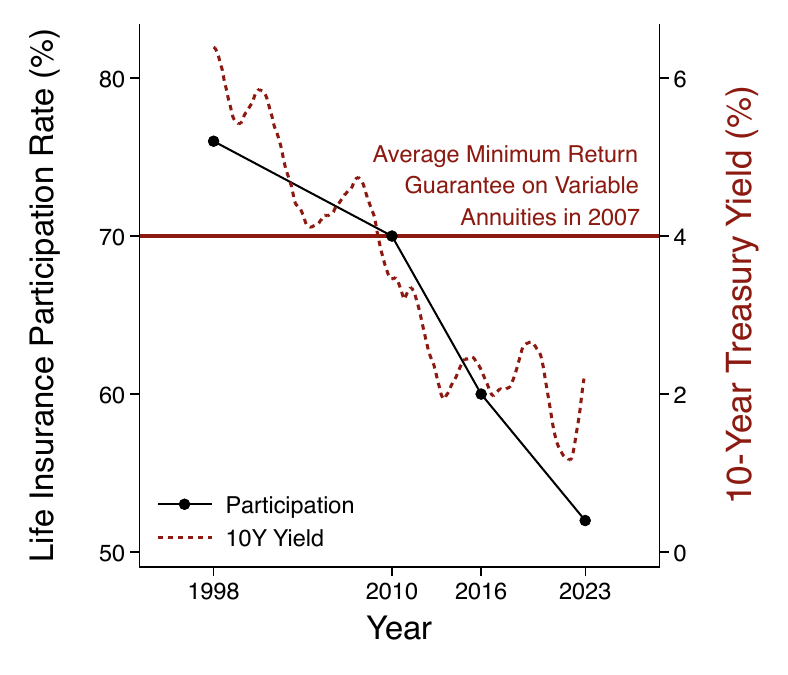}
        \caption{Participation}
    \end{subfigure}
    \begin{subfigure}{0.49\textwidth}
        \includegraphics[width = \textwidth]{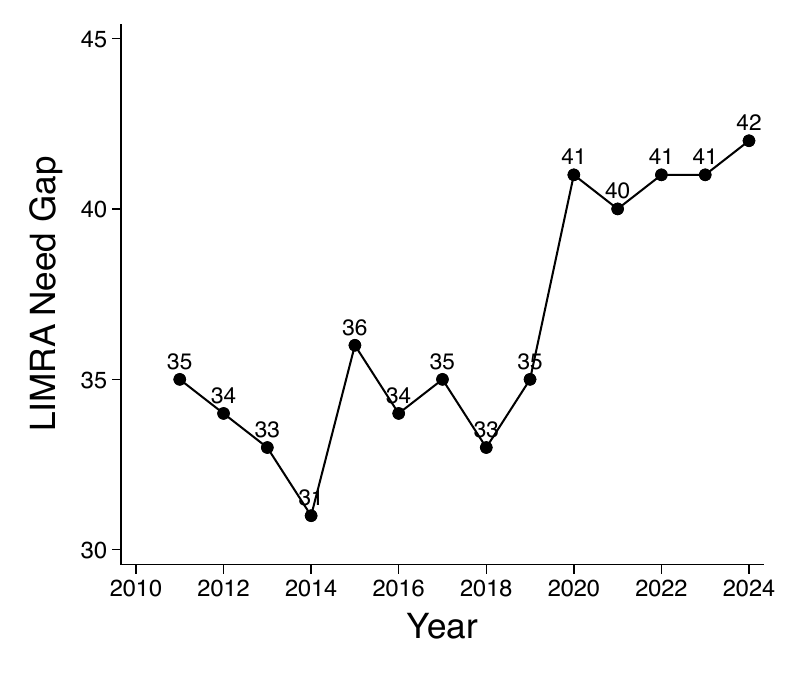}
        \caption{Need Gaps}
    \end{subfigure}
    \caption{Life Insurance Participation and Need Gaps}
    \label{fig:motivation}
    \vspace{-0.3cm}
    \floatfoot{Note: Panel (a) plots life insurance participation rates (left axis) and 2-year rolling average 10-year Treasury yields (right axis) over time. Data on life insurance participation come from \cite{guardian2023}, which is itself derived from LIMRA Barometer reports. Monthly 10-year Treasury yields are taken from FRED and cover 1998 to 2022. The average minimum return guarantee is taken from \cite{koijen2022fragility}. Panel (b) plots the life insurance need gap reported in the 2024 LIMRA Barometer report \citep{limra2024barometer}, defined as the share of US adults aged 18–75 who either do not own life insurance but say they need coverage or own life insurance but say they need additional coverage.}
\end{figure}

Life insurance serves as an essential risk-sharing tool for households. As of 2023, U.S. life insurers provide coverage to the household sector that exceeds 150\% of GDP. However, life insurance participation in the U.S. has steadily declined for the past three decades, and at an accelerating pace (Figure \ref{fig:motivation}). According to a report by the Guardian Life Insurance Company \citep{guardian2023}, life insurance participation declined from 76\% in 1998 to 70\% in 2010, a rate of 0.5 percentage points per year. Participation continued to decline to 60\% just six years later --- an accelerated rate of 1.67 percentage points per year --- and today sits at 52\%. The sharp drop in participation has important consequences: among households that experience the loss of an income-earner, 84\% that did not have life insurance report living paycheck-to-paycheck as opposed to the 36\% that did \citep{guardian2023}. This is reflected in survey-reported life insurance need gaps, which widened from 31\% to 42\% over the decade leading up to 2024 \citep{limra2024barometer}.

It is therefore reasonable to suspect that the sharp decline in participation was driven by forces beyond household demand. In particular, the post-crisis recovery was accompanied by historically low interest rates, as shown in Figure \ref{fig:motivation}. Life insurers --- financial institutions with particularly long-lived liabilities --- are generally sensitive to the revaluation effects of interest rate changes. Modern life insurance and annuity products are especially exposed due to their minimum return guarantees, embedded options whose valuation grows dramatically when interest rates are low. In particular, \cite{koijen2022fragility} highlight that the average minimum return guarantee of variable annuities issued in 2007 sat at 4\%, approximately 2 percentage points higher than ensuing Treasury yields just a few years later. As a result, the reserve value of the embedded options grew substantially, leaving life insurers exposed to elevated interest rate risk.

This paper explores a new channel through which life insurers may hedge interest rate risk: \textit{liability rebalancing}.  As we discuss in Section \ref{sec:institutional setting}, life insurance product markets are segmented by maturity, and therefore, degrees of interest rate risk. Ordinary life insurance products (term or whole life) provide long-term coverage, while life insurance accessed through employers (group life) typically provides coverage for a single year. Given limits to duration matching through asset rebalancing \citep{ozdagli2019interest,sen2023regulatory,alfaro2024lash}, insurers may naturally transition from ordinary life to group life issuance to reduce their interest rate risk and thereby transfer that exposure to employers and households. However, since group life insurance is only accessible through (large) employers, there could be negative consequences for participation at the market level. Moreover, since group life policies typically provide lower levels of coverage than ordinary life policies \citep{guardian2023}, life insurance coverage as a whole may shrink.

We explore these insights formally in Section \ref{sec:model}. We provide a model of insurance product markets in which risk-averse insurance companies are exposed to interest rate risk. Insurers care about their operating profits as well as the volatility of their capital returns. Further, they add duration to their capital through new liability issuance, which could amplify the risk of their capital returns when interest rates are uncertain. As a result, insurers trade off current-period profits with future interest rate risk when issuing new policies.

We first show formally that insurers hedge interest rate risk through product markets, consistent with our concept of liability rebalancing. In particular, when interest rate uncertainty rises, insurers issue fewer long-duration policies but increase their issuance of short-duration policies. This effect is especially pronounced for insurers with more negative duration gaps and larger capital convexity: because their capital returns respond more to declines in interest rates, they rebalance toward short-duration policies with greater intensity.

We then cast the model in general equilibrium to study how liability rebalancing affects product markets. In contrast to the partial equilibrium setting, we show that less exposed (but not unexposed) insurers may increase their long-duration product issuance due to the decline in competition. In this sense, less exposed insurers try to fill the gap left by more exposed insurers. However, due to decreasing returns to scale, substitution across insurers is insufficient to stabilize the market, and total issuance of long-duration policies declines.

With these predictions in hand, we next turn to our empirical analysis in Section \ref{sec:empirics}. Our data are taken from life insurers' annual statutory filings. For each insurer, we have access to both new issuance and insurance in force for their term life, whole life, and group life businesses from 2005 to 2023. We also collect data on monthly term life prices from Compulife, a quotation software used by life insurance agents. We use information on the account value of insurers' variable annuities in the pre-crisis period to classify them into exposed and non-exposed groups. The exposed insurers are relatively large in terms of assets and capital, but they are 2-3 times more leveraged. Beyond their exposure to variable annuity guarantees, they also hold a higher share of interest-sensitive life insurance reserves. 

We next use Actuarial Guideline 43, a 2009 regulatory reform, to further classify exposed insurers as having or not having risk-sensitive (RS) liabilities. Among variable annuity guarantees with comparable economic risks, the regulation treated some as sensitive to interest rate risk, while treating the rest as if they carried no interest rate risk \citep{sen2023regulatory}. As a result, the new regulatory regime induced unexpected cross-sectional variation in duration hedging motives among exposed insurers by increasing the need to hedge for insurers with RS liabilities, thereby sharpening our identification.

We begin by revisiting duration gap estimates for our insurer classifications. We first replicate the finding in the literature that at the industry level, duration gaps became negative after the financial crisis (e.g., \citealp{hartley2016measuring,koijen2022fragility,sen2023regulatory}). We take one step further and utilize liability duration estimates from \cite{huber2022} to examine the differences in duration gaps between exposed and non-exposed insurers following the financial crisis. Consistent with our narrative, we find that the duration gaps of exposed insurers became relatively more negative during the low-interest-rate period, and the effects are concentrated among exposed insurers with RS liabilities. Our results are robust to controlling for year and insurer fixed effects.

According to our theory, product prices should reflect widening duration gaps. We use our data on term life insurance prices to test this hypothesis. We first show that relative maturity markups --- the difference between long- vs. short-maturity policy prices sold by exposed relative to non-exposed insurers --- correlates negatively with 10-year Treasury yields at the monthly level.\footnote{We show this for 20-year relative to 10-year term life products.} This finding is consistent with our theory: during low interest rate periods, exposed insurers face relatively wider (negative) duration gaps, which increases the spread between the prices of their long- and short-duration products. 

We then show that the result holds more generally in a regression format, where we can control for a variety of granular fixed effects. We obtain the same results when we further compare exposed insurers with and without RS variable annuities. Similar to low rates, interest rate uncertainty also amplifies insurers' exposure to interest rate risk. Accordingly, we find that relative maturity markups also rise when interest rate uncertainty is heightened. The results indicate that insurers pass through interest rate risk to their products on the maturity dimension, which we interpret as indirect evidence of liability rebalancing.

We then turn to a more direct test of liability rebalancing using our data on the issuance of insurance coverage. Since we cannot observe issuance at the individual product level, we instead explore how insurers rebalance their issuance between generally long-duration products (term and whole life insurance) and short-duration group life insurance. We demonstrate this in both the raw data and through a regression specification with insurer and year fixed effects, showing that exposed insurers rebalance their liabilities toward group life insurance relative to non-exposed insurers throughout the low-interest-rate period. As exposed insurers shift the mix of their product offerings towards group life products, they also issue smaller quantities of long-term life products than non-exposed insurers. We interpret these results as more direct evidence of liability rebalancing. We again verify that, among the exposed insurers, the rebalancing pattern is stronger and, at times, entirely driven by insurers exposed to RS liabilities. 

We further show that the rebalancing did not occur through an increase in group life issuance, but rather through a sharp contraction in ordinary life issuance, consistent with our results on pricing. Beyond pricing, our theory also predicts that insurers may rebalance their liabilities by distorting commissions paid to their agents. We show that this is the case, and that the reduction in commission rates is concentrated among renewal commissions rather than first-year commissions. These results suggest a lapsation channel: by removing the incentives for agents to nudge existing policyholders to renew, insurers may be implicitly encouraging lapsation. Consistent with this mechanism, we find that lapse rates increase for exposed insurers relative to non-exposed insurers, as well as for insurers exposed to RS liabilities relative to exposed insurers that are not.

We then aggregate policy issuance across insurers to reinterpret product market trends as consequences of insurers' risk management strategies. We document that the yearly aggregate issuance of ordinary life insurance as a percentage of GDP declined by 48\% between 2005 and 2023, with two-thirds of the decline stemming from exposed insurers.\footnote{This does not imply that nominal issuance of non-exposed insurers declined; in fact, total non-exposed insurer issuance increased by 34\% by the end of the sample period. The discrepancy is due to relative changes in real GDP. See Section \ref{sec:empirics:agg} for more details.} Group life insurance issuance also declined as a percentage of GDP, but to a much lesser extent. The results suggest that both supply-side and demand-side effects contributed to the reduction in life insurance issuance, but that the supply-side effects, primarily driven by exposed insurers, exacerbated the demand-side effects in ordinary life markets. 

Despite the decline in issuance rates, the size of the insurance market could remain stable if new issuance exceeded claims and lapsation. We show that this was not the case: between 2005 and 2023, ordinary life insurance in force as a percentage of GDP declined from 138.9\% to 107.1\%, over a third of which we estimate was due to the liability rebalancing response of RS-exposed insurers. In contrast, group life insurance in force only declined from 56.6\% to 53\% of GDP, the majority of which came after the COVID-19 crisis. Combining the two implies an 18\% market-wide contraction of approximately 35.4\% of GDP.

Our results suggest that the interest rate risk exposure of variable annuity issuers had severe consequences for life insurance markets, highlighting a growing need for regulation to address insurers' risk exposures and mitigate resulting cross-product spillovers. We stress that these insights apply not only to insurance markets, but to any financial product market in which sellers carry risk exposures on their balance sheet.

\subsection*{Related Literature}

Our work relates most closely to the literature on how regulation and the financial health of life insurers spill over into their product markets.\footnote{In the context of P\&C insurance, 
\cite{gron1994capacity}, \cite{froot2001market}, and \cite{zanjani2002pricing} show that insurers' capital constraints can affect their insurance supply and demand. \cite{damast2025homeowners} study how monetary policy affects homeowner insurance products through P\&C insurers' interest rate risk exposure.} \cite{koijen2015cost} document that the wedge between actuarial and statutory reserve valuation methods affects pricing behavior. \cite{koijen2022fragility} show that insurers increased fees on variable annuities and offered less generous guarantees post-GFC. \cite{ge2022financial} further shows that for insurance groups with both P\&C and life divisions, P\&C losses worsen their financial health, spilling over to their life insurance division and leading to nuanced pricing behavior. \cite{barbu2024a} find that the introduction of risk-based capital accounting affects the supply and demand of life insurance. \cite{knox2024insurers} show that insurance prices reflect the gains and losses stemming from insurers' asset performance. \cite{verani2024s} show that interest rate risk management plays a key role in annuity pricing, and the cost of interest rate risk hedging has pushed up annuity premia post-2008. \cite{heinrich2026liability} show that balance-sheet structure affects how monetary tightening transmits to the retail-institutional funding wedge of insurers. \cite{giambona2025hedging} find that reducing hedging costs increases hedging, especially for insurers that are likely to face costly external finance and closer to default, allowing such insurers to sell more policies. Our contribution highlights the long-term effects on product markets, particularly on the composition and availability of insurance products, when interest rate risk cannot be perfectly hedged.  

Complementing our findings on interest rate risk and the life insurance market, insurers also appear to manage their market risk exposure using different types of annuity products. \cite{barbu2023ex} shows that insurers reduce their exposure to variable annuities by having customers exchange them for less generous products. \cite{barbu2024b} document that insurers can reduce exposure to downside market risk by selling index-linked annuities that resemble long-dated short puts. \cite{ellis2025impact} further find that the roll-out of enterprise risk management mandates prompted insurers to switch to such index-linked annuities. Relative to these studies, we provide evidence of indirect liability rebalancing across less-connected product markets and distortions along the maturity dimension.

We build on the recent literature highlighting the risk management challenges facing the life insurance industry. \cite{ellul2022insurers} show both theoretically and empirically that insurers only partially hedge their variable annuity guarantee exposures by rebalancing their bond portfolios, but in doing so, exacerbate systemic risk. \cite{ozdagli2019interest} show that transaction costs make it difficult for insurers to fully hedge their duration gap with corporate bonds. \cite{sen2023regulatory} conducts a detailed study of variable annuity hedging and finds that differences in accounting methods used for assets and liabilities can further lead to imperfect hedging. Building on the UK LDI crisis, \cite{alfaro2024lash} argue that liquidity risk increases the cost of hedging with derivatives. We offer a new channel in this paper --- liability rebalancing --- through which insurers reduce their exposure to variable annuities and interest rate risk by reducing issuance of long-duration products and increasing issuance of shorter-duration products, such as group life insurance. 

In particular, our paper adds to earlier work on the interest rate risk of U.S. life insurers. Since \cite{berends2013sensitivity}, a large literature has established that the duration gap of US life insurers switched from positive to negative after the 2007-2008 Financial Crisis (e.g., \citealp{hartley2016measuring}; \citealp{ozdagli2019interest}; \citealp{koijen2021evolution}, \citeyear{koijen2022fragility}; \citealp{huber2022}; \citealp{sen2023regulatory}; \citealp{Drexler2024InterestRateRisk}; \citealp{kirti2024}; \citealp{li2024}).\footnote{\cite{domanski2017hunt}, \cite{kirti2024}, and \cite{li2024} further show that interest rate risk has important asset pricing implications through insurers' asset demand.} The majority of existing work identifies insurers' duration gap by estimating the sensitivity of U.S. life insurers' stock returns with respect to Treasury yields, while \cite{huber2022} and \cite{sen2023regulatory} provide more direct evidence from insurers' balance sheets. We defer a more detailed discussion of insurers' duration gaps to Section \ref{sec:inst setting:dmm and irr}. Importantly, we show that insurers'  interest rate risk spills over to product markets and changes the duration mix of their product offerings.

Our paper also belongs to the literature that aims to understand household participation in insurance and annuity products. While existing work has mostly focused on the demand-side factors driving market dynamics (\citealp{inkmann2011deep, koijen2016health, hartley2017explains, gropper2025wealth, briggs2023risky}), we offer a new supply-side perspective for the recent period. In particular, we emphasize that life insurers' risk-management motives can constrain their product supply, exacerbating demand forces behind low insurance participation.

Last, our paper connects to the literature on the industrial organization of insurance markets, particularly life insurance. \cite{koijen2015cost,koijen2022fragility} estimate demand for life insurance and variable annuities, respectively, and use their framework to understand how regulation affects product markets. \cite{tang2022} uses a structural model to evaluate the effects of regulatory competition across US state insurance regulators and the establishment of captive reinsurance. \cite{wenning2024} estimates a model of life insurance agent distribution across a rich geography to explore the consequences of national price-setting behavior. While we have not done so in this paper, our model is amenable to estimation and can be used to carry out counterfactual analyses in future work.

\section{Institutional Setting}\label{sec:institutional setting}

We begin with a broad description of insurance product markets. We then discuss the interaction between insurance reserves and interest rates. We end with a discussion of the regulatory and economic motives for life insurance companies to hedge interest rate risk and highlight why product markets are a feasible outlet for hedging.

\subsection{An Overview of Life Insurance Products}

Life insurance markets have evolved considerably since their inception. The earliest forms of life insurance were short-duration policies with minor payouts. Prominent insurers by today's standards often began with such policies \citep{knight1920}: for example, at its inception in 1875, Prudential Financial, which today has over \$1.4 trillion in assets under management, primarily sold industrial life insurance --- small policies with maturities of about a week that targeted laborers in poor urban neighborhoods \citep{carr1975}. 

Since then, life insurance products have evolved considerably. The closest category to the traditional industrial life policy is what is known as group life insurance.\footnote{Industrial life insurance still exists today but the market is minuscule: as of 2023, it only accounts for 0.016\% of gross life insurance coverage in force.} Insurers write group contracts with firms rather than individuals, and the firm itself issues insurance certificates to their workers. These certificates function primarily as yearly-renewable policies with premium rates that are renegotiated at renewal. Employer-sponsored group policies are especially small, typically covering only one to two years of an employee's salary \citep{guardian2023}, and are less accessible, since not all employers offer group life insurance as a benefit. Group life coverage totaled about 55\% of GDP in 2005.

Ordinary life insurance departs from group life insurance in both the coverage and the time dimension. On the coverage dimension, policyholders are free to choose their desired level of coverage rather than being fixed at one year's wage.\footnote{In our data that we discuss in Section \ref{sec:empirics}, the average ordinary life insurance policy covers \$144,281 in 2023, while the average group life policy covers \$67,185.} On the time dimension, products can be split into two broad categories: term life policies and whole or permanent policies. Term life policies pay out a pre-specified benefit upon the death of the insured, conditional on the death happening during a set number of years.\footnote{Some policies allow for yearly renewals with adjusted premiums, but policyholders are not permitted to renew for the full term. Other provisions may allow the policies to be converted to permanent contracts.} For example, a 10-year term life policy pays out if the insured dies between the time of issuance and 10 years. Whole life policies, on the other hand, do not expire unless premiums are not paid.\footnote{These policies technically expire at a very old age, such as 100 or 121. Since most individuals do not live this long or lapse well before this, the restriction is typically not binding.} 

Whole life policies are notable due to their embedded savings components. These policies typically have lower coverage but redirect a fraction of the premium revenue toward a savings vehicle that accrues interest. This is referred to as the cash value of the policy. Traditionally, the cash value is invested in fixed income assets whose investment returns are fairly stable. New innovations in whole life policies have emerged over time, such as variable, indexed, and universal life products, that invest the cash value in a variety of non-fixed-income assets and may come with additional embedded options, such as minimum return guarantees.

Life insurers also issue annuities, products that insure longevity as opposed to mortality. Standard annuity products are paid for upfront and provide a fixed stream of payments until the death of the insured. The payments can either start alongside the initial payment (immediate annuities) or after a set number of years (deferred annuities). Similar to whole life insurance, insurers have innovated on annuity products by allowing the payments to fluctuate with an underlying mutual fund. These are known as variable annuities. A key similarity with variable life insurance is that the returns often come with a minimum return guarantee. For example, if the return guarantee is 4\% per year and the mutual fund only returns 2\% in a given year, the insurance company must pay the remaining 2\% out of pocket.

\subsection{Insurance Reserves and Interest Rate Risk}\label{sec:inst setting:res and irr}

Insurance companies must hold reserves to ensure available payment for policyholders. The value of the reserves for traditional policies directly accounts for mortality risk conditional on the age, gender, and health status of the policyholder. Since many policies have a time component, the value of a given policy's reserves may change over time due to a higher loading on a higher mortality risk or due to changes in the discounted value of future payouts (\citealp{koijen2015cost}; \citealp{huber2022}). As such, these policies, especially whole life and long maturity term life, carry implicit interest rate risk. Group policies, which are often yearly renewable, typically have a low reserve requirement and are not sensitive to interest rate risk due to their short maturities.

Insurers hold non-traditional policy reserves in their separate accounts rather than their general accounts \citep{koijen2022fragility}. This is due to the fluctuating nature of the savings components. However, when these policies are bundled with minimum return guarantees, the value of the separate account does not cover the residual returns between the underlying mutual fund and the minimum return guarantee when the guarantee is in the money. Insurers therefore hold reserves in their general accounts to account for these options.

Variable annuity and life insurance reserves are therefore convex. When interest rates and stock market returns are high, the likelihood that the minimum return guarantee will be exercised is low. Reserve positions are therefore small since insurers are less likely to have to cover the gap in returns. However, when rates and stock returns are low and declining, the guarantees are more likely to be exercised, and the reserve valuations increase substantially. For example, as discussed in \cite{huber2022}, MetLife's ``5 Year Ratchet \& ROP-d, GMIB w/ 10y, 7 to 8'' variable annuity had a reserve value that increased 4-fold between 2009 and 2011. In general, \cite{sen2023regulatory} estimates that the duration of minimum return guarantees is between 9 and 17 years. The high duration and convexity of minimum return guarantees have also raised concerns among insurance practitioners. For example, a report by AM Best \citep{panko2012} also argues that large blocks of legacy annuities with minimum return guarantees created severe pressures on insurers' balance sheets post-2008.

\subsection{Duration Matching Motives in the Life Insurance Industry}\label{sec:inst setting:dmm and irr}

Insurers that specialize in ordinary life insurance hold reserves with long maturities, often spanning more than 30 years. Minimum return guarantees on their variable liabilities add both duration and convexity to their total reserve positions. Given the sensitivity of their reserves to interest rates, a natural interest rate risk management strategy is to hold assets that match the duration of their reserves.

However, duration matching is not always a successful or even feasible strategy. Market incompleteness may prevent insurers from perfectly matching the duration between their assets and liabilities. Corporate bonds, which account for the majority of insurers' asset portfolios \citep{koijen2023understanding}, have an average duration of only around 7-8 years. While Treasury bonds can have a longer duration, their maturities are also capped at 30 years, and insurers in general dislike Treasuries for their relatively low returns. 

Beyond market incompleteness, insurers also face a variety of other frictions that push against duration-matching motives. First, insurance regulations might inadvertently distort insurers' hedging motives. \cite{sen2023regulatory} argues that the mismatch in the accounting methods used for assets and liabilities discourages insurers from using interest rate derivatives to hedge variable annuities. Second, \cite{ozdagli2019interest} find that illiquidity and transaction costs in the corporate bond market are potentially important factors preventing insurers from closing their duration gaps, as doing so requires insurers to turn over large fractions of their bond holdings, which could be prohibitively expensive.\footnote{Furthermore, \cite{domanski2017hunt} and \cite{greenwood2018impact} suggest that, due to their large scale, the reach-for-duration by insurers could lead to a substantial increase in the total demand for long-term assets, which could further push down long-term interest rates, resulting in a vicious cycle.} This is consistent with the evidence in \cite{huber2022}, which shows that the asset duration of individual life insurers did increase somewhat after the financial crisis, but not substantially. Third, \cite{alfaro2024lash} argue that liquidity risk increases the cost of hedging with derivatives and leads insurers to hedge interest rate risk imperfectly.

Consequently, life insurers' duration gaps became negative after the financial crisis. Several existing studies (e.g., \citealp{berends2013sensitivity}; \citealp{hartley2016measuring}; \citealp{ozdagli2019interest}; \citealp{koijen2021evolution}, \citeyear{koijen2022fragility}; \citealp{Drexler2024InterestRateRisk}; \citealp{kirti2024}; \citealp{li2024}) arrived at this conclusion by examining how insurers' stock returns co-move with interest rates. After carefully studying insurers' balance sheets, \cite{sen2023regulatory} finds direct evidence that many insurers failed to hedge a significant proportion of their variable annuity liabilities. By calculating the insurance companies' asset and liability durations directly, \cite{huber2022} finds that the aggregate gap switched from positive to negative after 2010.\footnote{Note that despite the duration estimation of minimum return guarantees by \cite{koijen2022fragility} and \cite{sen2023regulatory}, \cite{huber2022} sets the duration of the minimum return guarantees for variable annuities and life insurance policies to zero. Incorporating these liabilities would likely lead to an even stronger decline in duration gaps.} Additionally, \cite{li2024} shows that after the financial crisis, the market leverage of life insurance companies co-moved negatively with long-term Treasury yields. The negative impacts of low interest rates on the life insurance sector have also been voiced frequently in practitioner publications (e.g., \citealp{panko2012}; \citealp{ambest2015yielding}; \citealp{wsj2023})

Given the limits to duration matching through asset rebalancing, we explore an alternative channel: \textit{liability rebalancing}. Insurers can reduce the duration of their liabilities by allowing their legacy reserves to expire and shifting new issuance toward shorter-duration policies. In the following section, we present a model of insurance product markets in the presence of interest rate risk to explore how liability rebalancing can be used as a risk management strategy.

\subsection{Regulation-induced Variations in Duration Hedging Motives}\label{sec:inst setting: regulation and hedging}

In 2009, a new regulatory framework, Actuarial Guideline 43, created cross-sectional variations in duration hedging motives by changing the regulatory payoff to hedging for insurers, depending on the pre-existing mix of variable-annuity guarantees they already had on their books. Before 2009, the regulatory regime largely treated variable annuity guarantees as if they had no interest rate risk. \cite{sen2023regulatory} finds that while different types of VA guarantees have broadly similar underlying economic exposures to interest rate risk, the 2009 reform made regulatory liabilities risk-sensitive in different ways: GMAB/GMWB guarantees became sensitive to interest rates, while GMIB/GMDB guarantees remained insensitive. This distinction arises from the regulatory valuation framework (Standard or Stochastic) that binds for each guarantee type, which depends on product-specific and contract-design features rather than on the products’ economic risk. As in \cite{sen2023regulatory}, we label GMAB/GMWB guarantees risk-sensitive or simply RS.

Section 6.2 of \cite{sen2023regulatory} discussed potential interpretations of this regulatory reform. The findings in this paper support the hypothesis that the reform strengthened hedging motives by formally recognizing interest rate risk in the risk-based capital framework: insurers care about their regulatory risk (in addition to economic risk), and the new regulation increased the regulatory risk exposure for insurers exposed to risk-sensitive liabilities, thereby necessitating hedging for them. Because the products are economically similar but treated differently by the regulation, this setup approximates a quasi-random shock to hedging incentives across otherwise comparable VA-issuing insurers. 

Life insurers with risk-sensitive variable annuities could potentially hedge across several dimensions. For example, \cite{sen2023regulatory} shows that they increased their use of interest rate derivatives, \cite{barbu2023ex} shows that they exchanged risk-sensitive variable annuities into less generous products, and \cite{barbu2024b} show that they partially offset risk exposures by issuing index-linked annuities. However, insurers would still carry residual interest rate risk on their balance sheets if derivative hedging is costly --- because of frictions such as liquidity risk \citep{alfaro2024lash}, model uncertainty \citep{coleman2007robustly, koijen2022fragility}, and basis risk \citep{li2026improving} --- or if household demand for product exchanges and index-linked annuities is limited. Consistent with partial hedging with other instruments, we find in Section \ref{sec:empirics:duration} that the (regulatory) duration gaps of insurers exposed to risk-sensitive annuities became significantly more negative post-2009. Accordingly, we expect those insurers to hedge more aggressively in the product market. In Sections \ref{sec:empirics:pricing} and \ref{sec:empirics:quantity}, we formally test whether insurers with risk-sensitive annuity exposure generated more product-market distortions after the reform.

\section{A Model of Product Markets and Interest Rate Risk}\label{sec:model}

We first present a simple model of duration matching to organize the empirical exploration. We discuss the structure of the model in Section \ref{sec:model:setup}. We then explore how duration mismatch affects product pricing and liability rebalancing in Section \ref{sec:model:liability rebalancing}. We end with a discussion on the cross-market equilibrium outcomes in Section \ref{sec:model:equilibrium}.

\subsection{Setup}\label{sec:model:setup}

\nin Time is discrete, $t \in \mathbb{N}$. There are a large number of insurance companies, $j \in \mathcal{J}$, that sell a variety of insurance and annuity products, $i \in \mathcal{I}$, to a unit measure of households. Insurers have two functions. First, they sell insurance to households, strategically setting prices and the extent of their market penetration for each product. Second, they manage a portfolio of assets with exogenous insurer-specific returns. These two activities shape the behavior of insurers' capital.

We denote insurer $j$'s portfolio's return between periods $t-1$ and $t$ as $R_{jt}^A$, which the insurer takes as given.\footnote{In practice, insurers hold 60-70\% of their asset portfolios in corporate bonds and, therefore, have asset returns close to the average return of the bond market \citep{koijen2023understanding}. This assumption can in principle be relaxed to allow for reaching-for-duration by insurers (\citealp{ozdagli2019interest}).} Insurers can expand their balance sheets and increase their assets by selling new insurance policies. When selling new products, insurers can attract more demand by setting lower prices, $P_{ijt}$, or by hiring more agents to market their products, $T_{ijt}$. We assume demand for each product-insurer pair takes the form 
    \begin{equation}\label{eq: demand PE}
        Q_{ijt} \equiv \overline{Q}_{ijt}\kappa(T_{ijt})P_{ijt}^{-\veps_{it}},
    \end{equation} 
\nin where $\overline{Q}_{ijt}$ is an insurer-product-specific component that we elaborate on in Section \ref{sec:model:equilibrium}, $\kappa(T_{ijt})$ is an increasing and concave function of $T_{ijt}$ that varies between 0 and 1, and $\veps_{it}>1$ is the demand elasticity for policy $i$ at time $t$. We assume for simplicity that the total number of agents attracted to sell the insurer's products is linear in the commissions paid, $T_{ijt} = \eta_{it}^{-1}F_{ijt}$, for some constant $\eta_{it} > 0$. 

Let $A_{jt}$ denote insurer $j$'s assets at the \textit{beginning} of period $t$ --- i.e., the level of assets inherited from period $t-1$ and before issuing new products at time $t$. New policies issued during period $t$ contribute to the insurer $j$'s assets at the beginning of the next period, $t+1$. Hence, insurers $j$'s assets evolve according to the law of motion
    \begin{equation}\label{eq:asset law of motion}
        A_{jt+1} = R_{jt+1}^A 
        \left[ A_{jt} + \sum_{i\in\mathcal{I}} \left( P_{ijt}Q_{ijt} - F_{ijt} \right) \right].
    \end{equation}
\nin When issuing new policies, insurers add to their existing liabilities, $L_{jt}$, through the creation of reserves. We refer to $V_{it}$ as product $i$'s reserve value. The total reserves created through the issuance of policy $i$ at time $t$ is then $V_{it}Q_{ijt}$.\footnote{Statutory values for insurance policies are typically more conservative than their actuarial value, which can also affect pricing \citep{koijen2015cost}. For our purposes, this distinction is not necessary.} We denote the return on an insurer's stock of existing reserves as $R_{jt}^{L}$. We then refer to the return on a particular product's reserves as $R_{it}$, which we assume is constant across insurers.\footnote{This implies that $R_{jt}^{L}$ is determined through the composition of insurer $j$'s outstanding insurance policies.} Similarly, we use $L_{jt}$ to denote insurer $j$'s liabilities at the beginning of period $t$, before the issuance of new products. Insurer $j$'s liabilities therefore evolve according to

    \begin{equation}\label{eq:liability law of motion}
        L_{jt+1} = R_{jt+1}^{L} L_{jt} + \sum_{i \in \mathcal{I}} R_{it+1} V_{it}Q_{ijt}.
    \end{equation}
\nin Combining (\ref{eq:asset law of motion}) and (\ref{eq:liability law of motion}) therefore gives us the evolution of insurers' capital:
    \begin{align}
        K_{jt+1} & = A_{jt+1} - L_{jt+1} \nonumber \\
        & = \underbrace{R_{jt+1}^{A}A_{jt} - R_{jt+1}^{L}L_{jt}}_{\text{Legacy Capital}, \ \equiv \widetilde{K}_{jt+1}} + \underbrace{\sum_{i\in\mathcal{I}} \Big[ \left( R_{jt+1}^A P_{ijt}-R_{it+1}V_{it}\right)Q_{ijt} - R_{jt+1}^AF_{ijt} \Big]}_{\text{Return on New Policy Issuance in Period } t}  \label{eq:capital law of motion}
    \end{align}
\nin where insurer $j$'s legacy capital, $\widetilde{K}_{jt+1}$, is their level of capital at $t+1$ if they did not issue any new policies in period $t$. Insurer $j$'s capital evolution therefore depends on the financial returns from their legacy capital and the return on new policy issuance.

Asset and reserve returns have two components: a guaranteed component (e.g., coupon payments, bonds maturing, policy claims, and lapsation) and a revaluation component due to changes in market interest rates, $\Delta R_{t+1} = R_{t+1} - R_{t}$.\footnote{One could argue that claims and lapsation rates themselves are both inherently random (e.g., \citealp{gottlieb2021lapse}; \citealp{koijen2024aggregate}; \citealp{kubitza2023life}). Since our framework considers atomistic households, after aggregating, we treat the idiosyncratic components of such risks as diversified. We also let the returns be time-dependent, which allows for aggregate claim and lapsation risks.} We assume returns take the form 
\begin{align*}
    R_{jt+1}^{A}  = \overline{R}_{jt+1}^{A} -D_{jt}^{A}\Delta R_{t+1},  \quad
    R_{jt+1}^{L}  = \overline{R}_{jt+1}^{L} - D_{jt}^{L}\Delta R_{t+1},  \quad
    R_{it+1}  = \overline{R}_{it+1} - D_{it}\Delta R_{t+1},
\end{align*}

\nin where the guaranteed components of returns $\overline{R}_{jt+1}^{A}, \overline{R}_{jt+1}^{L}$ and $\overline{R}_{it+1}$ are assumed to be exogenous, reflecting the characteristics of the underlying securities. We refer to $D_{jt}^{A}$ as insurer $j$'s asset duration, $D_{jt}^{L}$ as insurer $j$'s liability duration, and $D_{it}$ as policy $i$'s duration, as they measure the sensitivities of the returns to the market interest rate.

Insurers have two objectives. First, they maximize their operating profits. Second, they minimize the volatility of their growth rate. Let $R_{jt+1}^K$ denote insurer $j$'s capital growth rate from period $t$ to $t+1$,
\begin{align*}
    R_{jt+1}^K = \frac{K_{jt+1}}{K_{jt}} = \frac{\widetilde{K}_{jt+1}}{K_{jt}} +  \sum_{i\in\mathcal{I}}\frac{\left( R_{jt+1}^A P_{ijt}-R_{it+1}V_{it}\right)Q_{ijt} - R_{jt+1}^AF_{ijt}}{K_{jt}}.
\end{align*}
We assume insurers are risk averse, and capture their risk management motives through an decreasing and concave function $\Lambda_j(R_{jt+1}^K - \ee_t[R_{jt+1}^K])$, where the expectation $\ee_t[\cdot]$ is taken over the distribution of policy rate innovations, $\Delta R_{t+1}$. Therefore, their objective function can be summarized as
\begin{align*}
    \max_{\{P_{ijt}, F_{ijt}\}} \ \underbrace{\sum_{i\in\mathcal{I}}\Big[ \left(P_{ijt} - V_{it}\right)Q_{ijt} - F_{ijt}\Big]}_{\text{Operating Profits}} + \underbrace{\ee_t\Bigg[\Lambda_j\Big(R_{jt+1}^K - \ee_t[R_{jt+1}^K]\Big)\Bigg]}_{\text{Risk Management}}.
\end{align*}
One can interpret $\Lambda_j(\cdot)$ as insurer $j$'s disutility for the volatility of its capital growth rate. For example, if $\Lambda_j(x) = - \gamma_j x^2$, the risk management component of the objective function becomes $-\gamma_j \text{Var}_t(R_{jt+1}^K)$, and we can interpret the objective function as a mean-variance preference that trades off the operating profits and the variance of capital growth.

In what follows, we will use a first-order approximation of $\Lambda_j(\cdot)$ around legacy returns, $R_{jt+1}^K \approx \widetilde R_{jt+1}^K$, where the legacy return $\widetilde R_{jt+1}^K$ is the return on capital from $t$ to $t+1$ absent any new policy issuance during period $t$:
$$\widetilde R_{jt+1}^K \equiv \frac{\widetilde{K}_{jt+1}}{K_{jt}} = \frac{R_{jt+1}^{A}A_{jt} - R_{jt+1}^{L} L_{jt}}{K_{jt}}.$$
Then, the approximation of $\Lambda_j(\cdot)$ can be written as follows:
    \begin{align}  \nonumber
        & \Lambda_j\Big(R_{jt+1}^K - \ee_t[R_{jt+1}^K]\Big) \approx \Lambda_j\Big(\widetilde R_{jt+1}^K - \ee_t[\widetilde R_{jt+1}^K]\Big) + \frac{\Lambda_j'\Big(\widetilde R_{jt+1}^K - \ee_t[\widetilde R_{jt+1}^K]\Big)}{K_{jt}}  \times \\
        & \qquad \qquad \sum_{i\in\mathcal{I}}\Bigg[ (R_{jt+1}^{A} - \overline R_{jt+1}^A)(P_{ijt}Q_{ijt} - F_{ijt}) - (R_{it+1} - \overline{R}_{it+1})V_{it} Q_{ijt}\Bigg]. \label{eq:rm approx 1} 
    \end{align}

\nin The first term is independent of the product issuance decisions made by the insurer, and therefore is taken as given. The second term captures the marginal value of risk management from the issuance of new products, and is the relevant piece of our model. For notational convenience, we denote $\lambda_{jt+1} \equiv \Lambda_j'\Big(\widetilde R_{jt+1}^K - \ee_t[\widetilde R_{jt+1}^K]\Big)/K_{jt}$, which depends on the realization of interest rates $\Delta R_{t+1}$ but not on products issued in period $t$. Formally, insurer $j$ solves
    \begin{multline}
        \max_{\{P_{ijt}, F_{ijt}\}} \ \overbrace{\sum_{i\in\mathcal{I}}\Big[(P_{ijt} - V_{it})Q_{ijt} - F_{ijt}\Big]}^{\text{Operating Profits}} \\ + \underbrace{\ee_t\Bigg[\lambda_{jt+1}\sum_{i\in\mathcal{I}}\Big( (R_{jt+1}^{A} - \overline R_{jt+1}^A)(P_{ijt}Q_{ijt} - F_{ijt}) - (R_{it+1} - \overline{R}_{it+1})V_{it} Q_{ijt}\Big)\Bigg]}_{\text{Expected Value of Risk Management}}.
    \end{multline}
\nin The insurer trades off its immediate profits with its expected return on its capital in the next period. The expectation is taken over the distribution of market rate innovations, $\Delta R_{t+1}$. The choice of product prices and agent distribution in the current period will therefore depend on the insurer's \textit{interest rate risk} and, in particular, the strength of the insurer's risk management motive, $\lambda_{jt+1}$.

\subsection{Duration Gaps and Liability Rebalancing}\label{sec:model:liability rebalancing}

Given the trade-off between profits and return risk, how should an insurer design its product portfolio? To study this question, we first need to understand the determinants of pricing and agent distribution and, therefore, their product issuance. We begin by characterizing the optimal decisions of a given insurer in the following lemma. 

\begin{lemma}{Optimal Issuance Decisions}{decisions}
    \textit{Insurer $j$'s optimal price for product $i$ and the optimal number of agents hired to sell product $i$ satisfy}
        \begin{equation*}
            \frac{P_{ijt}}{V_{it}} = \Bigg(\frac{\veps_{it}}{\veps_{it} - 1}\Bigg)\mathcal{M}_{ijt}, \qquad T_{ijt} = \max \Bigg\{(\kappa')^{-1}\Bigg(\frac{\eta_{it}}{\mathcal{E}_{it}\overline{Q}_{ijt}V_{it}^{1-\veps_{it}}\mathcal{M}_{ijt}^{1-\veps_{it}}}\Bigg), \ 0\Bigg\},
        \end{equation*}
    \nin \textit{where $\mathcal{E}_{it} \equiv \veps_{it}^{-\veps_{it}}(\veps_{it}-1)^{\veps_{it}-1}$ and the risk management markup,} $\mathcal{M}_{ijt}$, \textit{satisfies}
        \begin{equation*}
            \mathcal{M}_{ijt} = \frac{1 + \ee_t\Big[\lambda_{jt+1}\big(R_{it+1} - \overline R_{it+1}\big)\Big]}{1 + \ee_t\Big[\lambda_{jt+1}\big(R_{jt+1}^A - \overline R_{jt+1}^A\big)\Big]}.
        \end{equation*}

    \bigskip
    \textbf{Proof sketch:} See Appendix \ref{app_sec:proofs:decisions}.
\end{lemma}

\nin For a given product, both prices and agent distribution depend explicitly on the returns to that product's reserve value as well as to its interaction with the insurer's marginal value of risk management. Risk management markups, $\mathcal{M}_{ijt}$, are increasing in $\ee_t[\lambda_{jt+1}(R_{it+1} - \overline R_{it+1})]$. In other words, the insurer charges higher markups on liabilities that grow faster than expectation ($R_{it+1} > \overline R_{it+1}$) when the marginal benefit of risk management ($\lambda_{jt+1}$) is high. To examine this case, we consider an approximation of $\lambda_{jt+1}$ around $\Delta R_{t+1} = 0$:
    \begin{equation}\label{eq:rm approx 2}
        \lambda_{jt+1} \approx \underbrace{\frac{\Lambda_j'(0)}{{K}_{jt}}}_{\displaystyle \equiv \bar\lambda_{jt+1}} - \underbrace{\frac{\Lambda_j''(0)}{{K}_{jt}}}_{\displaystyle \equiv \bar\lambda_{jt+1}'<0}D_{jt}^{K}\Delta R_{t+1}.
    \end{equation}
\nin where $D_{jt}^{K} \equiv (D_{jt}^{A} A_{jt} - D_{jt}^{L}{L}_{jt})/{K}_{jt}$ is insurer $j$'s duration gap. Since the function capturing the risk management motive $\Lambda_j(\cdot)$ is concave, $\bar\lambda_{jt+1}' \equiv \Lambda_j''(0)/{K}_{jt}<0$. As highlighted in Section \ref{sec:inst setting:dmm and irr}, many life insurers faced a negative duration gap after the financial crisis. This fact is of first order when analyzing insurers' pricing and issuance patterns. To do so, we use the following lemma to understand how a product's duration affects its pricing.

    \begin{lemma}{Approximate Risk Management Markups}{rmm}
        \textit{Suppose that the market interest rate $R_{t}$ follows a martingale process with variance $\sigma_t^2$. Then under the approximation (\ref{eq:rm approx 2}), risk management markups $\mathcal{M}_{ijt}$ can be written as}
            \begin{equation}
                \mathcal{M}_{ijt} = \frac{1 + (\bar\lambda_{jt+1}'D_{jt}^{K}\sigma_{t+1}^2) D_{it}}{1 + (\bar\lambda_{jt+1}'D_{jt}^{K}\sigma_{t+1}^2) D_{jt}^A}.
            \end{equation}

        \bigskip
        \textbf{Proof sketch:} See Appendix \ref{app_sec:proofs:rmm}.
    \end{lemma}

\nin The lemma highlights an important result: if insurers face a negative duration gap, $D_{jt}^{K} < 0$, then long duration policies have higher markups, all else equal. Since insurers are risk-averse over capital returns, they put a higher weight on capital losses than they do capital gains. Therefore, when they have a negative duration gap, their value of capital losses due to interest rate declines outweighs their value of capital gains due to interest rate hikes. They therefore set a higher price on long-duration policies when this gap is larger to justify the higher potential losses. Higher prices further translate into reduced agent distribution and commissions as they lower the profitability of long-duration policies. Equipped with this insight, we present our first result.

\begin{proposition}{Interest Rate Uncertainty and Product Issuance}{iss risk}
    \textit{Consider two interest rate environments, 1 and 2. The interest rate uncertainty in the second environment is higher, $\sigma_{2,t+1}^2 > \sigma_{1,t+1}^2$. Then for any insurer $j$ such that $D_{jt}^{K} < 0$, }
    \begin{align*}
            Q_{ijt}^2 > Q_{ijt}^{1} & \qquad \textit{if } D_{it} < D_{jt}^{A} \\[1.5ex]
            Q_{ijt}^{2} < Q_{ijt}^{1} & \qquad \textit{if } D_{it} > D_{jt}^{A}
    \end{align*}
    \bigskip
    \textbf{Proof sketch:} See Appendix \ref{app_sec:proofs:iss risk}.
\end{proposition}

\nin \propref{prop:iss risk} says that if interest rate uncertainty increases, then relative to their asset duration, insurers with a negative duration gap decrease the issuance of long-duration products and increase the issuance of short-duration products. Since their duration gap is negative, their capital is already exposed to interest rate risk. Therefore, they optimally move away from long-duration products that exacerbate their duration gap in an attempt to hedge additional interest rate risk. 

We next explore how this result changes in the cross-section of insurers in different interest rate environments. In particular, we are interested in the role of capital \textit{convexity}. If some insurers have especially convex liabilities --- such as insurers that previously issued variable life insurance or annuities with generous minimum return guarantees \citep{koijen2022fragility,sen2023regulatory} --- then in a low rate environment, their duration gap should increase. This makes them especially susceptible to interest rate risk, even if interest rate uncertainty remains unchanged. 

Denote the convexity of an insurer's capital as $\gamma_{jt}^{K} = -\partial D_{jt+1}^{K}/\partial R_{t+1} < 0$. The following proposition considers how two insurers with different capital convexity respond to a decline in interest rates, holding fixed interest rate volatility.

\begin{proposition}{Capital Convexity and Product Issuance}{iss convex}
    \textit{Consider two interest rate environments, 1 and 2, that are identical except that interest rates are lower in the second environment, $R_{t}^2 < R_{t}^1$. In addition, consider two insurers, $j$ and $j'$, that are identical except that insurer $j'$ has more convex capital, $|\gamma_{j't}^K| > |\gamma_{jt}^{K}|$. Suppose the insurers have the same negative duration gap in environment 1 and both gaps remain negative in environment 2. Then,}
    \begin{align*}
            \displaystyle \frac{Q_{ij't}^{2}}{Q_{ij't}^{1}} > \frac{Q_{ijt}^{2}}{Q_{ijt}^{1}} > 1 & \qquad \textit{if } D_{it} < D_{jt}^{A} \\[1.5ex]
            \displaystyle \frac{Q_{ij't}^{2}}{Q_{ij't}^{1}} < \frac{Q_{ijt}^{2}}{Q_{ijt}^{1}} < 1 & \qquad \textit{if } D_{it} > D_{jt}^{A} 
    \end{align*}

    \bigskip
    \textbf{Proof sketch:} See Appendix \ref{app_sec:proofs:iss convex}.
\end{proposition}

\nin To understand the proposition, consider the product issuance distribution of two insurers, $j$ and $j'$. Suppose that initially, in a high-interest-rate environment, the two insurers have the same duration gaps and issue products with the same intensity; however, insurer $j'$ has higher capital convexity than insurer $j$, $|\gamma_{j't}^{K}|>|\gamma_{jt}^{K}|$, for example due to previously issuing variable annuities with generous guarantees. As they transition into an environment with lower rates, the duration gap of $j'$ becomes more negative than the duration gap of $j$. Both insurers respond to lower rates by shifting their issuance toward low-duration policies, but since insurer $j'$ is especially sensitive, their response is more pronounced.\footnote{See Appendix Figure \ref{fig:props 2-3} for a depiction of this explanation.}

It is important to note that the results of this section are partial equilibrium results. If a large insurer such as MetLife responds to a decline in rates by no longer selling long-duration policies, less exposed insurers may step in to fill the gap in demand despite also having some exposure to the decline in rates. We therefore turn to an analysis of product market equilibrium to study the market-level effects of interest rate risk and duration gaps.

\subsection{Duration Mismatch and the Size of Insurance Markets}\label{sec:model:equilibrium} 

\nin We begin by zooming in on household purchasing behavior. For simplicity, we assume that households may hold multiple life insurance policies and treat each product market in isolation.\footnote{According to data from the 2018 Health and Retirement Survey, of the 54\% of households that hold a life insurance policy, 38\% of households hold more than one policy. While there may be correlated preferences within a household across markets, this assumption adds complexity to the solution without materially altering the core mechanism.} We assume households have identical preferences within a product class, but that their preferences may differ across product classes. Household $h$'s indirect utility from purchasing product $i$ sold from insurer $j$ is
    \begin{equation*}
        u_{ijt}^{h} = \log \alpha_j + \log \kappa_{ijt}(T_{ijt}) - (\veps_{it} - 1)\log \Bigg(\frac{P_{ijt}}{V_{it}}\Bigg) + \nu_{ijt}^{h}
    \end{equation*}
\nin where $\alpha_j$ is an insurer-specific characteristic (``quality'') and $\nu_{ijt}^{h}$ is an idiosyncratic taste shock distributed according to an extreme value type I distribution with unit variance.\footnote{We include market penetration explicitly in indirect utility for simplicity. The interpretation is that if insurer $j$ has more agents, they are more accessible, which reduces the cost of search or travel for households.} Household $h$ spends a constant amount, $Y_{it}^{h}$, on coverage through product $i$. They therefore purchase $Q_{ijt}^{h} = Y_{it}^{h}/P_{ijt}$ units of coverage conditional on buying from insurer $j$. Households may also choose an outside option $0$ (e.g., cash) with preferences satisfying $u_{i0t}^{h} = \log \alpha_{it}^0$. We normalize the price of the outside option to 1. With these assumptions, insurer $j$ faces the following demand curve
    \begin{equation*}
        Q_{ijt}(P_{ijt}, T_{ijt}) = \alpha_j\kappa(T_{ijt})\frac{Y_{it}}{P_{ijt}}\Bigg(\frac{P_{ijt}/V_{it}}{\mathcal{P}_{it}}\Bigg)^{1-\veps_{it}},
    \end{equation*}
\nin We introduce a functional form for the market penetration function, $\kappa(T_{ijt}) = 1 - \exp(-T_{ijt})$, This functional form allows us to solve for $\mathcal{P}_{it}$ in closed form, which simplifies the analysis. Aggregate expenditures, $Y_{it}$, and the product market price index, $\mathcal{P}_{it}$, respectively satisfy
    \begin{equation*}
        Y_{it} \equiv \int_{0}^{1}Y_{it}^{h}dh, \qquad \mathcal{P}_{it}^{1-\veps_{it}} \equiv \alpha_{it}^0 + \sum_{j\in\mathcal{J}}\alpha_j\kappa(T_{ijt})\Bigg(\frac{P_{ijt}}{V_{it}}\Bigg)^{1-\veps_{it}}.
    \end{equation*}

\nin We begin by addressing the question asked at the end of Section \ref{sec:model:liability rebalancing}: in response to a decline in interest rates, how do insurers adjust within a product market when we account for cross-sectional differences in capital convexity? As highlighted by \cite{huber2022}, some insurers did not see a large decline in their duration gaps post-2008 and should therefore respond differently than insurers whose duration gaps widened. The following result highlights a condition that determines whether or not the competitive effects of reduced issuance by highly exposed firms outweigh the direct effects of additional exposure by other firms.

\begin{proposition}{Insurer Substitution}{sub}
    \textit{Consider two otherwise identical interest-rate environments. Suppose $D_{jt}^{K,2}\leq D_{jt}^{K,1}\leq0$ for all $j$ (strictly for some $j$), and assume no participation ties. Suppose also that all insurers active in environment 1 have the same initial market penetration. Let $\psi_{ijt}=\mathcal{M}_{ijt}^{2}/\mathcal{M}_{ijt}^{1}$ denote the risk-management markup ratio.} 

    \textit{\qquad If $D_{it}>D_{jt}^{A}$ for all $j$, then there exists a unique common threshold $\overline{\psi}_{it}$ such that, for every insurer $j$ active in environment 1,}
    \begin{align*}
            \displaystyle Q_{ijt}^{2} < Q_{ijt}^{1} & \qquad \textit{if } \psi_{ijt} > \overline{\psi}_{it} \\[1.5ex]
            \displaystyle Q_{ijt}^{2} > Q_{ijt}^{1} & \qquad \textit{if } \psi_{ijt} < \overline{\psi}_{it}
    \end{align*}
    \textit{\qquad If $D_{it}<D_{jt}^A$ for all $j$, the same cutoff characterization holds, but falling rates push more-convex insurers toward expanding rather than contracting supply.}

    \bigskip
    \textbf{Proof sketch:} See Appendix \ref{app_sec:proofs:sub}.
\end{proposition}

\nin The partial equilibrium setting of Section \ref{sec:model:liability rebalancing} suggested that even slightly more exposed insurers alter their behavior, and that insurers whose interest rate risk exposure does not change ($D_{jt}^{K} = 0$) do not adjust their issuance. Instead, in equilibrium, \propref{prop:sub} says that the retreat of the exposed insurers opens up demand for the unexposed insurers, leading them to increase their issuance. This occurs both due to an increase in the number of agents and, therefore, the share of households that they reach, as well as cross-insurer substitution by market participants. We will see in the following section that this pattern holds in the data.

Nevertheless, it is unclear whether unexposed insurers can fully pick up the slack left by the exposed insurers. For example, if lower-quality insurers are the ones with higher exposure, we might expect the higher-quality insurers to easily buy up the policies that they left on the table. However, this may not be sufficient if households' preferences are sufficiently dispersed or if the decreasing returns to scale implied by their market penetration is too strong. 

To study this trade-off, note that we can write the share of expenditures that accrue to the outside option as
    \begin{equation}\label{eq:outside option share}
        \frac{Q_{it}^{0}}{Y_{it}} = \alpha_{it}^0 \mathcal{P}_{it}^{\veps_{it}-1} = \frac{\alpha_{it}^0}{\displaystyle \alpha_{it}^0 + \sum_{j\in\mathcal{J}}\alpha_j\kappa_{ijt}(P_{ijt}/V_{it})^{1-\veps_{it}}}.
    \end{equation}

\nin Holding fixed the outside option value $\alpha_{it}^0$, a ubiquitous increase in prices at the market level points to an increase in the outside option share, and therefore, a decline in the expenditures spent on insurance. Since prices are increasing while expenditures are falling, this would immediately imply that total new coverage issued should decline as well. The following result confirms this finding conditional on insurers having the same initial exposure. 

\begin{proposition}{Product Market Issuance Dynamics}{iss market}
    \textit{Consider two interest rate environments, 1 and 2, that are equivalent except that all insurers' duration gaps are nonpositive and become weakly more negative: $D_{jt}^{K,2}\leq D_{jt}^{K,1}\leq0$ for all $j$, with a strict inequality for at least one insurer active in the product market. Additionally, assume that $\mathcal{M}_{ijt}^{1}$ is constant across insurers. Then the total issuance of product $i$ satisfies}
    \begin{align*}
            \displaystyle Q_{it}^{2} < Q_{it}^{1} & \qquad \textit{if } D_{it} > D_{jt}^{A} \textit{ for all }j \\[1.5ex]
            \displaystyle Q_{it}^{2} > Q_{it}^{1} & \qquad \textit{if } D_{it} < D_{jt}^{A} \textit{ for all }j
    \end{align*}

    \bigskip
    \textbf{Proof sketch:} See Appendix \ref{app_sec:proofs:iss market}.
\end{proposition}

\nin Therefore, according to \propref{prop:iss market}, a decline in rates that renders all insurers' duration gap more negative leads to a reduction in market issuance for long-duration policies but increases market issuance for short-duration policies. With these results in hand, we now turn to our empirical setting: life insurance markets during the post-GFC, low-interest-rate period.

\section{The State of Life Insurance After the Financial Crisis}\label{sec:empirics}

Equipped with the model predictions, we now turn to our empirical analysis. We begin by discussing our data sources and our method for identifying interest rate risk exposures across insurers. We then present results on liability rebalancing and issuance dynamics for exposed and non-exposed insurance groups. We end with an exploration of aggregate issuance dynamics and the evolution of life insurance markets over the last two decades.

\subsection{Data Construction}\label{sec:empirics:data}

\paragraph{Statutory Filings} Much of our data are sourced from life insurers' statutory filings, which we access through S\&P Global. Every insurer in the United States must prepare these filings annually for the National Association of Insurance Commissioners (NAIC), who then provides these data to institutions for research purposes. 

We pull from a variety of exhibits in the statutory filings. Our primary data is from the Exhibit of Life Insurance, which provides detailed information on coverage issued and in force. The exhibit separately identifies ordinary life (term and whole life policies) and group life lines of business. We complement these data with reserve positions, premiums, and commissions for each product category. Reserves are taken from the Aggregate Reserves for Life Contracts. The filings record the reserve positions (gross and net) for each product category at the end of the fiscal year. Premiums and commissions come from Exhibit 1.

Data on variable annuity issuance and holdings come from the General Interrogatories. These filings record the total related account value for each annuity product sold, the reserves held in the general account, and any reserves reinsured by the issuing insurer. Note that the account values and reserves only reflect minimum return guarantees since insurers hold the principal of the annuities in their separate accounts.

Finally, we use information on insurers' assets and liabilities, which further allows us to produce leverage ratios. For summary statistics, we use data from the Interest Sensitive Life Insurance Products Report. We also use stock prices of publicly traded insurers\footnote{We adopt the same list of insurers as in \cite{koijen2022fragility}.} from CRSP. Data on Treasury yields and annual GDP are taken from FRED. 

\paragraph{Insurance Prices} Our data on insurance prices are taken from Compulife, a software system used by insurance agents that generates quotes for various product categories and insurance companies. We collect monthly quotes for 10-, 15-, and 20-year term life products between January 2008 and December 2022. Quotes are for 40-year-old, non-smoking men in regular health. The data only contain quotes for a select sample of insurers. We discuss the representativeness later in this section, and \cite{wenning2024} provides a detailed breakdown of a broader subset of annual data.

\paragraph{Sample Construction} Our units of analysis are insurance group headquartered in the United States. We choose to use insurance groups over individual companies for two reasons. First, many insurance groups organize their subsidiaries according to their product specialization. For example, among the subsidiaries of the insurance group MetLife Inc., Brighthouse Financial was a large issuer of variable annuities and variable life insurance. Separating Brighthouse Financial from other subsidiaries, such as the flagship company Metropolitan Life Insurance Company, would paint an incomplete picture of MetLife as a whole. Second, insurers are publicly traded at the insurance group level. Since most public insurers also issued variable annuities, it is consistent with existing evidence on duration gaps and stock returns to use insurance groups (e.g., \citealp{hartley2016measuring}; \citealp{koijen2022fragility}; \citealp{li2024}).

Our theory predicts that insurers whose liabilities are more convex are more exposed to interest rate risk. Variable annuities are a particularly convex liability due to their minimum return guarantees as discussed in Section \ref{sec:inst setting:res and irr}. We therefore follow \cite{koijen2022fragility} split insurance groups by their variable annuity exposure, measured as
    \begin{equation*}
        \frac{\text{Total Related VA Account Value} + \text{Gross VA Reserves} - \text{Reinsurance Reserve Credits}}{\text{Total Liabilities}}
    \end{equation*}
\nin We label an insurer as ``exposed'' if its variable annuity exposure is in the top decile of insurers between 2005 and 2007. Note that only about 25\% of insurance groups in our sample issue variable annuities during this time period, so our cutoff corresponds to approximately the top 40\% of variable annuity issuers.

To further sharpen our identification, we utilize variation in variable annuities created by Actuarial Guideline 43. As we discussed in Section \ref{sec:inst setting: regulation and hedging}, \cite{sen2023regulatory} documents that the regulatory change, which took effect in 2009, rendered certain types of variable annuity guarantees sensitive to interest rates from a regulatory perspective, while keeping otherwise similar guarantees insensitive to interest rates. As a result, among insurers that issued variable annuities with comparable economic risk, those that had issued risk-sensitive variable annuities before the financial crisis had stronger incentives to hedge their interest rate risk post-crisis. We therefore split our sample of variable annuity issuers into two groups: exposed risk-sensitive (RS) issuers, which issued positive amounts of risk-sensitive variable (GMAB/GMWB) annuities between 2005 and 2007, and exposed non-risk-sensitive (non-RS) issuers, which did not.

For several robustness tests, we also use a continuous measure of risk-sensitive variable annuity exposure. We calculate the net reserves --- gross reserves minus reinsurance reserve credits --- for both risk-sensitive variable annuities and all variable annuities for each insurer, following \cite{sen2023regulatory}. We then calculate the share of a given insurer's net variable annuity reserves that are risk sensitive. For insurers that do not issue variable annuities, we set this measure to zero.

We exclude insurers that were not in an insurance group between 2005-2007 for most of the analysis. This is done to provide a clean comparison between exposed and non-exposed insurers prior to the crisis. We bring these insurers back into the sample when we explore aggregate product market trends for completeness. 

We also exclude captive reinsurers from our insurance group definitions. This is of little consequence when studying trends in product issuance since reinsurers typically do not issue new policies. Additionally, as we will see later in this section, adding them back to the sample when studying market-level trends does not change the results, as their life insurance holdings remain stable over time. This exclusion also prevents large jumps in the exposed insurers' insurance in force due to the split between MetLife and RGA.

\paragraph{Summary Statistics} We provide summary statistics for our primary sample in Table \ref{tab:sum stats}. We split the table on two dimensions. First, we report summary statistics for exposed and non-exposed insurance groups separately. Second, we report the statistics for 2005-2008 and 2009-2023 separately. We refer to the first time period as the pre-crisis period and the second time period as the post-crisis period. There are 27 (26) exposed insurers and 239 (198) non-exposed insurers in the pre-crisis (post-crisis) period. Exposed insurers are relatively more represented in the Compulife data, although there is still sufficient variation across the two groups in both time periods.

\begingroup
\setlength{\tabcolsep}{16pt}
\renewcommand{\arraystretch}{1.3}
\sisetup{input-symbols = {( )}}
\begin{table}[t!]\small 
    \begin{center}
        \vspace{-0.5em}

        \begin{tabular}{@{}l *{4}{S[table-format = -1.1, table-space-text-post=$^{***}$]}@{}c}
            \hline
             & \multicolumn{2}{c}{Exposed Insurers} & \multicolumn{2}{c}{Non-Exposed Insurers} \\  \cmidrule(l){2-3} \cmidrule(l){4-5}
             & {2005-2008} & {2009-2023} & {2005-2008} & {2009-2023} \\
             \hline\\[-2.5ex]
             Number of Groups&&&&\\
\cmidrule{1-1} \hfill Full Sample&{27}&{26}&{239}&{198}\\
\hfill Compulife Sample&{11}&{15}&{40}&{43}\\
\\[-1em] Assets&91.12&97.78&8.62&14.77\\
Surplus&4.91&5.21&0.69&1.27\\
Leverage Ratio&19.62&18.81&6.54&8.98\\
Leverage Ratio (Weighted)&20.16&21.30&17.97&16.19\\
\\[-1em] VA Liability Share&0.54&0.42&0.01&0.01\\
IS Reserve Share&0.65&0.63&0.24&0.25\\
\\[-1em] Issuance Market Share&&&&\\
\cmidrule{1-1} \hfill Ordinary&0.40&0.28&0.58&0.62\\
\hfill Group&0.45&0.42&0.54&0.51\\
\\[-1em] In Force Market Share&&&&\\
\cmidrule{1-1} \hfill Ordinary&0.37&0.28&0.39&0.40\\
\hfill Group&0.48&0.44&0.49&0.47\\
\hline

        \end{tabular}
        \caption{Summary Statistics}
        \label{tab:sum stats}
        \vspace{-3em}
        \floatfoot{Note: This table reports summary statistics for our primary sample. Assets and surplus are reported in billions of dollars. All variables except market shares, the weighted leverage ratio, and the number of groups are unweighted averages across insurers. The weighted leverage ratio is weighted by insurer assets within each period. Market shares are calculated across all years within each period.}
    \end{center}
\end{table}
\endgroup

Exposed insurers are systematically larger than non-exposed insurers. In particular, the average exposed insurer is 10.57 times as large as the average non-exposed insurer in the pre-crisis period and 6.62 times as large in the post-crisis period. This is consistent with variable annuity issuance being dominated by large insurers: since variable annuities are among the most complex products issued by life insurers, it is likely that only large insurance groups have adequate resources to manage them. Exposed insurers also have more capital (surplus), though only by an order of 7.1 and 4.1 in the pre- and post-crisis periods, respectively. This difference suggests that exposed insurers are more levered: the average leverage ratio, calculated as liabilities divided by surplus, is 3 and 2.1 times the average leverage of non-exposed insurers in the pre- and post-crisis periods, respectively. When weighting insurers' leverage by assets, exposed insurers still have nearly 12.2\% higher leverage in the pre-crisis period and 31.6\% higher leverage in the post-crisis period.

Consistent with our definition of variable annuity exposure, exposed insurers have substantially higher variable annuity liabilities as a share of total liabilities.\footnote{Note that insurers' liabilities are calculated differently than variable annuity liabilities. The share reported in the table and used for our classification is merely meant to separate those with high exposure from those with low exposure relative to their size.} This is not surprising, as the majority of non-exposed insurers do not issue variable annuities at all. That being said, certain life insurance products are also recorded as interest-sensitive and may be exposed to the low-rate environment in the post-crisis period. The table suggests that insurers exposed to variable annuities also have a substantially higher exposure to interest-sensitive life insurance policies.

Table \ref{tab:sum stats} also previews our findings across product markets. In the pre-crisis period, exposed insurers, despite being small in number, accounted for 40\% of total ordinary life insurance issuance. The remaining 90\% of insurance groups accounted for 58\% of the issuance, with the remainder being issued by small non-group companies. The numbers for group life issuance are similar. However, in the post-crisis period, exposed insurers only issued 28\% of new ordinary life insurance coverage, with non-exposed insurers increasing their share to 62\%. On the other hand, group life issuance shares remained relatively stable.

The decline in ordinary life issuance is echoed when considering life insurance in force. Exposed insurers' market share of life insurance coverage in force fell from 37\% to 29\% between the two periods, while non-exposed insurers' market share increased from 39\% to 40\%. Group life insurance in force again remained relatively stable. Note that the numbers for ordinary life only add up to 76\% in the pre-crisis period and 68\% in the post-crisis period; the majority of the remaining insurance was held by reinsurers, and within reinsurers, it was largely held by RGA, a prior subsidiary of MetLife until their split in 2008.

We further report summary statistics on RS and non-RS insurers in Appendix Table \ref{tab:sum stats rs}. Although insurers that issued RS guarantees are larger on average relative to those that did not, they had similar leverage, variable annuity liability shares, and interest-sensitive reserve shares in the pre-crisis period. To ensure that our tests are not driven by balancing issues across our exposed insurers, many of our analyses also consider the share of RS liabilities, a continuous measure of RS variable annuity exposure, as a robustness check. Importantly, among insurers with RS variable annuities, their RS liability share is uncorrelated with their asset size.\footnote{The measured correlation is 0.18 and statistically insignificant with a $p$-value of 0.577.}

\subsection{The Evolution of Duration Gaps}
\label{sec:empirics:duration}

\begin{figure}[t!]
    \centering
    \includegraphics[width=0.9\linewidth]{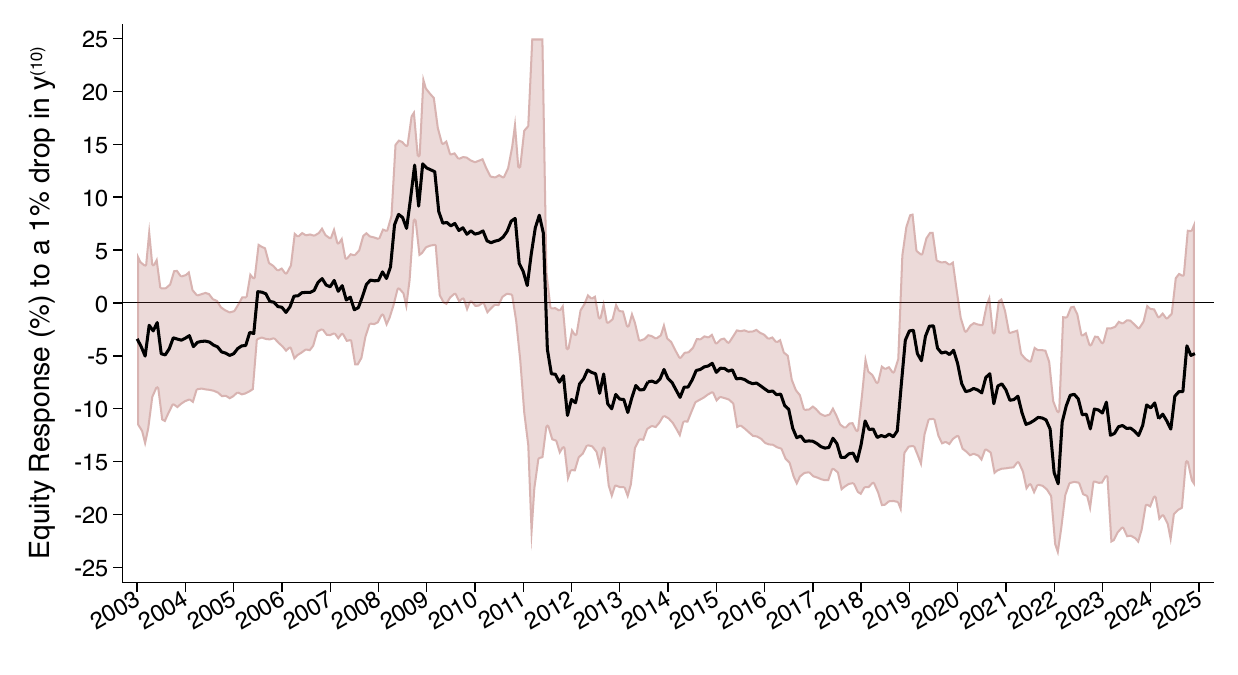}
    \caption{Rolling Estimates of Life Insurers' Duration Gaps}
    \label{fig:duration stocks}
    \floatfoot{Note: This figure reports rolling regression estimates of equation (\ref{eq:duration regression}). The black line plots estimates of the interest rate sensitivity, $\gamma$, for two-year (24-month) rolling windows. Red bands report 95\% confidence intervals using heteroscedasticity-robust standard errors.}
\end{figure}

We begin our analysis by documenting the widening of life insurers' duration gaps following the financial crisis. Many studies (e.g., \citealp{berends2013sensitivity, ozdagli2019interest, huber2022, li2024}) provide evidence of larger duration gaps using rolling estimates of interest rate betas for a portfolio of public insurers' returns. We replicate this finding in Figure \ref{fig:duration stocks}, where we estimate
    \begin{equation}\label{eq:duration regression}
        \text{Insurance Portfolio Returns}_{t} = \alpha + \beta \times \text{Market Returns}_{t} - \gamma \times \Delta\text{10-Year Yield}_t + \veps_{t}.
    \end{equation}
\nin We use monthly returns and consider rolling two-year intervals to calculate $\beta$ and $\gamma$. We plot the results, along with 95\% confidence bands, in Figure \ref{fig:duration stocks}. As in the aforementioned studies, we find that after approximately 2011, the life insurance portfolio exhibits a consistently negative duration gap for all time periods except 2019. 

However, our theory predicts that certain insurance companies were more exposed to interest rate risk than others. In particular, as we highlighted in the previous section and as shown by \cite{sen2023regulatory}, insurers that are heavily exposed to variable annuities should have faced even more intense interest rate risk exposure. It is difficult to show this using stock returns since most public companies issued variable annuities prior to the crisis. Therefore, we instead rely on a more direct estimation strategy following \cite{huber2022}. Recall that the duration gap of insurer $j$ at time $t$ can be written as
    \begin{align}\label{eq:duration gap empirical}
        D^{K}_{jt} \equiv D_{jt}^{A} + \frac{L_{jt}}{K_{jt}} \times (D_{jt}^{A} - D_{jt}^{L}).
    \end{align}
\nin Therefore, in order to estimate insurers' duration gaps directly, we need three objects. First, we use liabilities $L_{jt}$ and surplus capital $K_{jt}$ from insurers' statutory filings. Second, we use liability duration estimates, $D_{jt}^{L}$, from \cite{huber2022}, which run from 2005-2020.\footnote{We would like to thank Max Huber for making these data publicly available to researchers.} Finally, we approximate $D_{jt}^{A}$ using the duration of insurers' corporate bond portfolios, the main asset class held by insurers, which we calculate directly using information on their corporate bond holdings (Schedule D in the statutory filings).

We would like to highlight two nuances with this measure of duration gaps. First, the liability duration measure from \cite{huber2022} actually excludes reserves for variable annuity minimum return guarantees. Instead, the convexity of variable annuity guarantees enters the duration gap measure through its impact on insurers' leverage. Changes in this measure over time should therefore be interpreted as suggestive evidence of insurers' differential exposure to interest rate risk rather than as precise magnitudes. However, including minimum return guarantee reserves would likely exacerbate differences between exposed and non-exposed groups, further validating our findings.

Second, the magnitudes of the changes in duration gaps using this measure can be large in some years. This is primarily due to our use of surplus capital (rather than economic capital) when measuring leverage ratios. Hence, our measure more closely captures regulatory duration gaps than economic duration gaps, which is consistent with the discussions in Section~\ref{sec:inst setting: regulation and hedging}: the 2009 regulatory reform primarily affects the incentives to hedge regulatory risk exposures. As we highlighted in Table \ref{tab:sum stats}, exposed insurers experienced an increase in their (asset-weighted) surplus-based leverage ratio from roughly 20.16 to 21.3 ($\approx 5.7\%$), while non-exposed insurers experienced a decline in their leverage ratio from 17.97 to 16.19 ($\approx -9\%$). Therefore, small mismeasurements in asset and liability durations can be exacerbated by differences in leverage across insurers and across time.\footnote{We explore a decomposition of duration gap changes in Appendix \ref{appendix:duration gap drivers} and show in Figure \ref{fig:dur gap components} that leverage played an important role in exacerbating the difference in duration gaps across groups.} We partially address this by trimming our duration gap measures at the 1\% and 99\% level, though Appendix Figure \ref{fig:duration gaps untrimmed} shows that this does not affect our results qualitatively. 

To further address mismeasurement concerns, we test for differences in the changes in duration gaps across exposed and non-exposed insurers by estimating the regression
    \begin{equation} \label{eq:duration gap regression}
        D_{jt}^{K} = \sum_{\tau \neq \text{2008-10}} \beta_\tau \left(\mathbf{1}\{t \in \tau\} \times \text{Exposed}_j\right) + \alpha_j + \delta_t + \veps_{jt},
    \end{equation}
\nin This specification allows us to control for mismeasurement in the average level of duration gaps due to our inclusion of time fixed effects as well as time invariant mismeasurement at the insurer-level due to our inclusion of insurer fixed effects. We bin years into time periods ($\tau$) of approximately 3 years to reduce idiosyncratic noise, akin to a difference-in-differences approach. The precise time periods are $\mathcal{T} = $ \{2005-07, 2008-10, 2011-13, 2014-16, 2017-20\}. The estimates $\{\beta_\tau\}_{\tau \in \mathcal{T}}$ are therefore a more robust measure of how duration gaps changed across our two groups over time. We weight observations by insurers' assets and set 2008-10 to be our omitted period, so the estimates are relative to the years immediately post-crisis. 

In addition to the baseline results, we also estimate a specification that accounts for differences in the risk sensitivity within our exposed group of insurers,
    \begin{align} \label{eq:duration gap regression RS} \nonumber
        D_{jt}^{K} & = \sum_{\tau \neq \text{2008-10}} \beta_\tau \left(\mathbf{1}\{t \in \tau\} \times \text{Exposed}_j\right) \\ 
        & + \sum_{\tau \neq \text{2008-10}} \beta_\tau^{\text{RS}} \left(\mathbf{1}\{t \in \tau\} \times \text{Exposed}_j \times \text{RS}_j\right) + \alpha_j + \delta_t + \veps_{jt},
    \end{align}

\nin where $\text{RS}_j$ is an indicator for whether insurer $j$ carried risk-sensitive variable annuity liabilities between 2005 and 2007. 

\begin{figure}[t!]
	\begin{subfigure}{0.49\textwidth}
		\includegraphics[width = \textwidth]{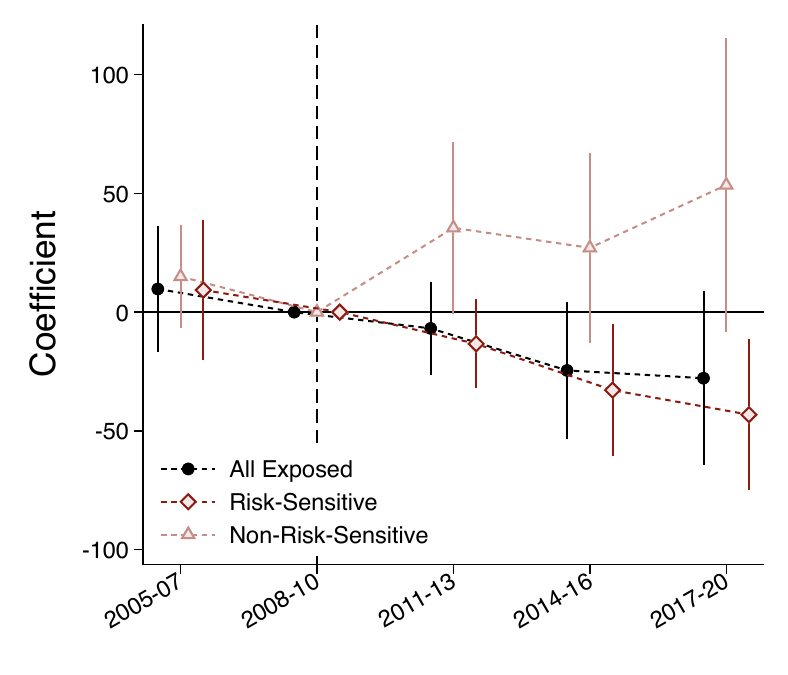}
		\caption{Total Effects}
	\end{subfigure}
	\begin{subfigure}{0.49\textwidth}
		\includegraphics[width = \textwidth]{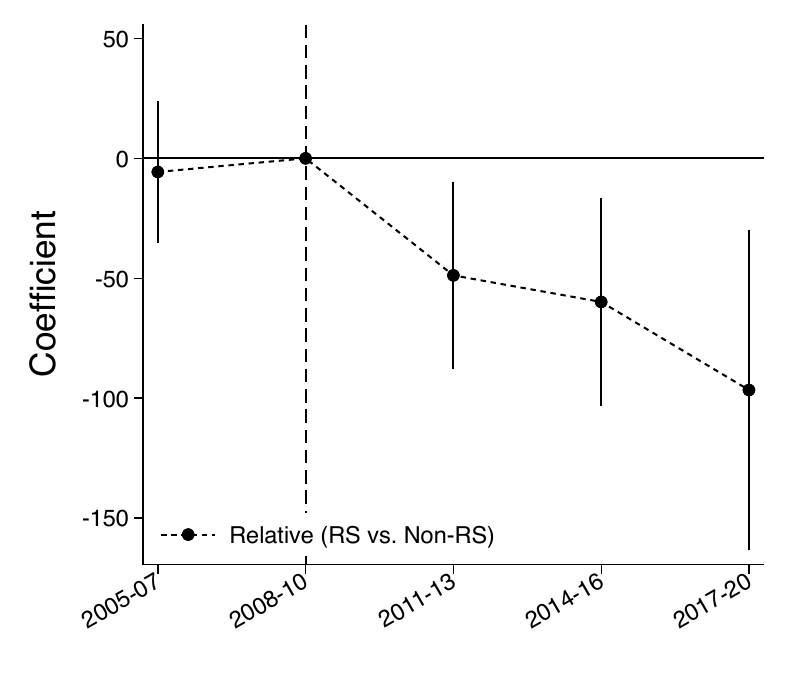}
		\caption{Relative Effects}
	\end{subfigure}
    \vspace{-1em}
	\caption{Changes in Duration Gaps by Exposure}
	\label{fig:duration gaps}
	\floatfoot{Note: This figure presents estimates from regressions (\ref{eq:duration gap regression}) and (\ref{eq:duration gap regression RS}). Duration gaps are constructed as in equation (\ref{eq:duration gap empirical}). In panel (a), estimates are presented as relative to non-exposed insurers; black circles represent the effects of all exposed insurers, red diamonds represent the effects of exposed RS insurers, and pink triangles represent the effects of exposed non-RS insurers. In panel (b), the black circles represent the difference between exposed RS and exposed non-RS insurers. The regressions are weighted by insurers' assets. Vertical lines represent 90\% confidence intervals using standard errors clustered at the insurer level.}

\end{figure}

We present our estimation results for equations (\ref{eq:duration gap regression}) and (\ref{eq:duration gap regression RS}) graphically in Figure \ref{fig:duration gaps}. There is a clear negative trend that begins after the 2011-2013 period, although for the broad group of exposed insurers, the effects are not significant at the 10\% level. However, once we split the effects between risk-sensitive and non-risk-sensitive exposed insurers, we find that the decline stems from the risk-sensitive insurers. In particular, risk-sensitive insurers faced a relative decline in their duration gaps by the 2014-16 period that was statistically significant at the 5\% level, while non-risk-sensitive insurers did not. Panel (b) also shows that the relative effects between risk-sensitive and non-risk-sensitive insurers (the coefficients $\{\beta_{\tau}^{\text{RS}}\}$) are statistically significant at the 5\% level for all post-crisis periods, but was near zero and insignificant in the pre-crisis period.

To ensure that the effects we document are not driven by systematic differences in insurers that issued risk-sensitive variable annuities and insurers that did not, we consider an additional specification in Appendix Figure \ref{fig:duration gaps rs} that replaces the indicator $\text{RS}_j$ with $\text{RSshare}_j$, the share of insurer $j$'s net variable annuity liabilities that are risk-sensitive. We find similar, albeit slightly noisier, results: the cross-term is not significantly different from zero in the pre-crisis period, but is negative and eventually significant in the post-crisis periods. In contrast, the effect on non-risk-sensitive insurers is not economically or statistically different from zero.

Overall, the evidence in this section points to an increase in interest rate risk exposure by insurance companies that were exposed to (risk-sensitive) variable annuities. Having validated this, we now turn to an analysis of insurers' product market behavior.

\subsection{The Effect of Interest Rate Risk on Insurance Pricing} \label{sec:empirics:pricing}

\nin Our theory predicts that the changes in relative duration gaps across insurers should manifest in their pricing decisions. This section tests this formally using the model as a guide. Consider two products, $\ell$ and $s$, in which $D_{\ell t} > D_{st}$. \lemmaref{lemma:rmm} suggests that to first order, we have that\footnote{See Section \ref{appendix:relative markup derivation} for a complete derivation.}
    \begin{multline}\label{eq:markup approx double diff}
                \overbrace{\underbrace{\ee_{\text{Ex}}\Bigg[\log \frac{P_{\ell jt}/V_{\ell t}}{P_{sjt}/V_{st}}\Bigg]}_{\substack{\text{long-short markup} \\ \text{for exposed insurers}}} - \underbrace{\ee_{\text{NonEx}}\Bigg[\log \frac{P_{\ell jt}/V_{\ell t}}{P_{sjt}/V_{st}}\Bigg]}_{\substack{\text{long-short markup} \\ \text{for non-exposed insurers}}}}^{\text{long-short markup spread}}  \\[2ex] \approx \ \sigma_{t+1}^{2} \times  \Big(\ee_{\text{Ex}}\Big[\bar\lambda_{jt}'D_{jt}^{K}\Big] - \ee_{\text{NonEx}}\Big[\bar\lambda_{jt}'D_{jt}^{K}\Big]\Big) \times (D_{\ell t} - D_{st}).
    \end{multline}

\nin We refer to the left-hand-side of equation (\ref{eq:markup approx double diff}) as the ($\ell,s$) markup spread. It measures how long-duration products are priced relative to short-duration products, comparing exposed insurers to non-exposed insurers. Our theory makes two predictions. First, it predicts that, due to their larger interest rate risk exposures, exposed insurers will have a higher long-short markup spread. This implies that (\ref{eq:markup approx double diff}) should be positive. Second, our theory predicts that this markup spread should increase when interest rates are low, as exposed insurers have more convex capital. Therefore, (\ref{eq:markup approx double diff}) should co-move negatively with long-term yields.

\begin{figure}
    \centering
    \includegraphics[width=0.9\linewidth]{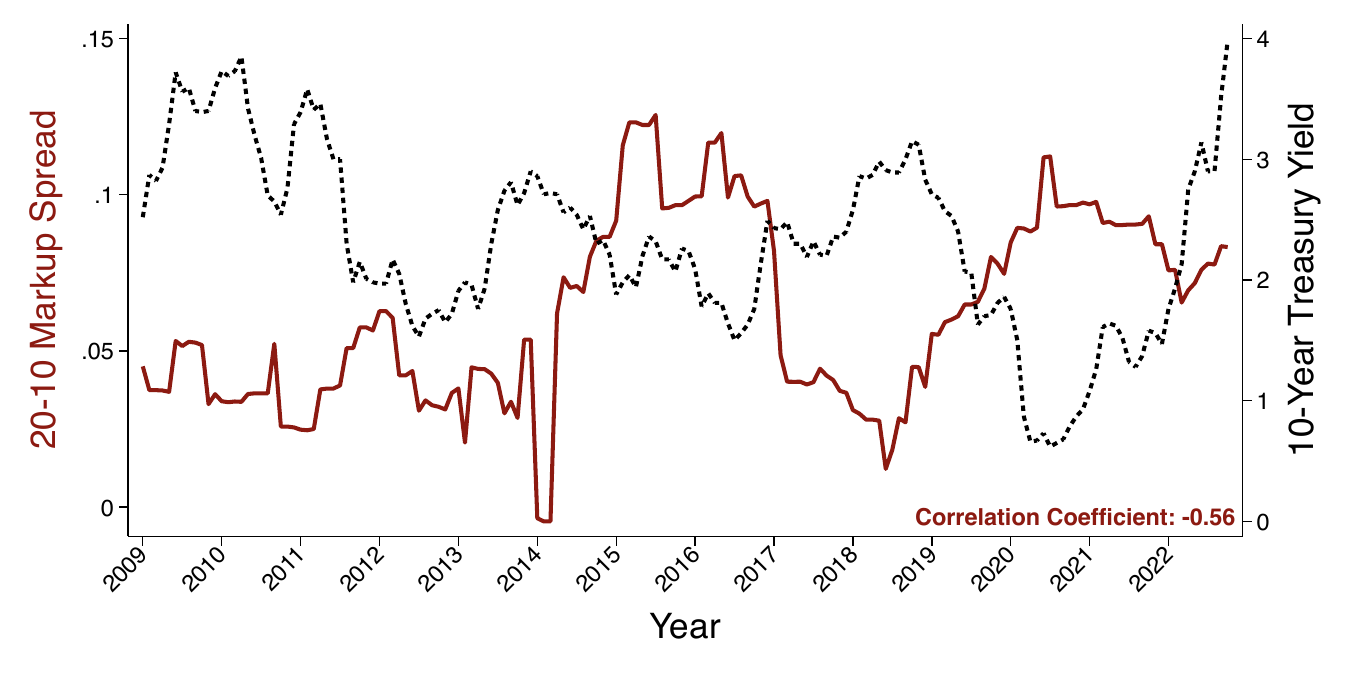}
    \caption{The Interaction between Product Pricing and Interest Rate Risk}
    \label{fig:markup spread}
    \floatfoot{Note: This figure plots the (20,10) markup spread (red, left axis) and the 10-year Treasury yield (black dotted, right axis) for each month between January 2009 and December 2022. When calculating the markup spread, averages are weighted by assets.}
\end{figure}

We plot the (20,10) markup spread for each month between January 2009 and December 2022 in Figure \ref{fig:markup spread}.\footnote{As \cite{koijen2015cost} show, the late months of 2008 displayed some extraordinary pricing behavior due to regulatory accounting practices. In particular, more distressed insurers reduced prices of long-term products relative to short-term products as a result of the differences between actuarially fair and statutory reserve values. We therefore start our sample in 2009 to avoid this episode and focus on time periods where interest rate risk was more pronounced.} When calculating averages as in equation (\ref{eq:markup approx double diff}), we weight insurers by their total assets to capture the most important insurers within each group.\footnote{Note that using assets as weights may put too much emphasis on large insurers that are not very active in ordinary life insurance markets. We show in Appendix Figure \ref{fig:markup spread infc} that our results hold (and in fact, become slightly stronger) when weighting by ordinary life insurance in force as well.} We also overlay monthly 10-year Treasury yields. Both conditions highlighted above hold: the markup spread is positive each month, is generally higher during periods of low interest rates, and co-moves negatively with 10-year yields.\footnote{We verify in Appendix Figure \ref{fig:markup spread vol} that this result is not driven by changes in interest rate uncertainty, as equation (\ref{eq:markup approx double diff}) suggests could be the case.} Over the full sample, the correlation coefficient between the relative markup spread and yields is $-0.56$. 

There could be other explanations for why the relative markup spread co-moves negatively with rates. For example, perhaps exposed insurers are simply larger and more active in long-duration term insurance markets, giving them more pricing power over time. It is also possible that, because exposed insurers are more likely to issue complex products, product-specific unobservables may be driving the results. As such, we formally test our results using a regression framework, which allows us to control for such differences. We begin with a triple difference specification in which we compare two product categories at a time,
    \begin{equation}\label{eq:price reg}
        \log \text{Price}_{ijt} = \beta \left(y_t^{(10)} \times \text{Exposed}_j \times \text{Long}_i\right) + \delta_{jt} + \delta_{it} + \delta_{ij} + \veps_{ijt},
    \end{equation}
\nin where $y_{t}^{(10)}$ is the monthly 10-year Treasury yield, $\text{Exposed}_j$ is an indicator for whether insurer $j$ is in the exposed group and $\text{Long}_i$ is an indicator for whether product $i$ has the longer duration of the two product categories. We interpret $\beta < 0$ as evidence that exposed insurers (relative to non-exposed insurers) set higher prices on longer-maturity products (relative to shorter-maturity products) when interest rates are low. As in Figure \ref{fig:markup spread}, we use 10- and 20-year term policies as our main specification and present results for other product combinations in the Appendix.

We include granular fixed effects to alleviate several endogeneity concerns. First, we include insurer $\times$ month fixed effects, which remove insurer-specific time variation (e.g., differences in size). Second, we include insurer $\times$ product fixed effects, which removes unobservable, time-invariant differences in product characteristics (e.g., convertibility clauses or renewal benefits). Third, we include month $\times$ product fixed effects, which absorb differences in demand over time for different product categories.

To further validate the prediction of Proposition \ref{prop:iss risk} (i.e., interest rate uncertainty increases insurers' interest rate risk exposure and relative maturity markups), we add a similar triple interaction term where the 10-year Treasury yield is replaced with a measure of interest rate uncertainty $\sigma_{t+1}^2$. Specifically, we use the policy rate uncertainty (PRU) index constructed by the Kansas City Fed, a market-based measure of uncertainty about where short-term US interest rates will be in one year \citep{bundick2024introducing}. Proposition \ref{prop:iss risk} predicts that the coefficient $\beta_{\text{PRU}}$ on the triple interaction term,
    \begin{equation*}
        \beta_{\text{PRU}}\left(\text{PRU}_t \times \text{Exposed}_j \times \text{Long}_i\right),
    \end{equation*}
should be positive, meaning that exposed insurers (relative to non-exposed insurers) set higher prices on longer-maturity products (relative to shorter-maturity products) when interest rate uncertainty is high. 

We also compare risk-sensitive (RS) to non-risk-sensitive (non-RS) insurers. As we documented in Section \ref{sec:empirics:duration}, RS insurers faced increasingly negative duration gaps in the post-crisis period, while non-RS insurers did not. We therefore expect RS insurers to make larger price adjustments in response to interest rates relative to non-RS insurers. To formally test this hypothesis, we include two more interaction terms in the regression: 
\begin{align*}
    \beta^{\text{RS}} \left(y_t^{(10)} \times \text{Exposed}_j \times \text{RS}_j \times \text{Long}_i\right), \quad \beta_{\text{PRU}}^{\text{RS}} \left(\mathrm{PRU}_t \times \text{Exposed}_j \times \text{RS}_j \times \text{Long}_i\right).
\end{align*}
\nin The $\text{Exposed}_j \times \text{RS}_j$ terms interact the baseline $\text{Exposed}_{j}$ indicator with an additional indicator for whether the insurer issued risk-sensitive variable annuities prior to the crisis, which allows us to inspect whether, \textit{within} the exposed group, insurers with more RS liabilities show stronger price responses. The coefficient $\beta^{\text{RS}}$ on this term should be interpreted relative to exposed non-RS insurers, so the total effects on exposed RS insurers relative to non-\textit{exposed} insurers are $\beta + \beta^{\text{RS}}$ for yields and $\beta_{\text{PRU}} + \beta^{\text{RS}}_{\text{PRU}}$ for rate uncertainty.

We report our results in Table \ref{table:price regression results baseline}. Column (1) finds that exposed insurers set higher markups on longer-duration products than non-exposed insurers when the 10-year Treasury yield is low. Further, after including the policy rate uncertainty measure in column (2), we see that exposed insurers increase markups for longer-duration products when facing greater monetary policy uncertainty. The within-exposed-insurer analysis further corroborates our mechanism: as shown in columns (3) and (4), exposed insurers with risk-sensitive liabilities adjust their markups more than exposed insurers with rate-insensitive liabilities, in response to movements in the 10-year yield and policy rate uncertainty.

\begingroup
\setlength{\tabcolsep}{3pt}
\renewcommand{\arraystretch}{1.4}
\begin{table}[t!]\small 
\sisetup{input-symbols = {( )}}
\begin{center}
    \begin{tabular}{@{}l *{4}{S[table-format = -1.3, table-space-text-post=$^{***}$, 
    table-column-width = 2.25cm]}@{}c}
        \toprule
        & \multicolumn{4}{c}{\textit{Dependent Variable:} $\log \text{Price}_{ijt}$} \\[1ex]  & {(1)} & {(2)} & {(3)} & {(4)} \\
        \midrule
        $ y_t^{(10)} \times \text{Exposed}_j \times \text{Long}_i$&      -0.025***&      -0.039***&      -0.011***&      -0.014*  \\
            &     (0.003)   &     (0.003)   &     (0.002)   &     (0.008)   \\
$ \mathrm{PRU}_t \times \text{Exposed}_j \times \text{Long}_i$&               &       0.048***&               &       0.014   \\
            &               &     (0.006)   &               &     (0.014)   \\
$ y_t^{(10)} \times \text{Exposed}_j \times \text{RS}_j \times \text{Long}_i$&               &               &      -0.015***&      -0.026***\\
            &               &               &     (0.003)   &     (0.008)   \\
$ \mathrm{PRU}_t \times \text{Exposed}_j \times \text{RS}_j \times \text{Long}_i$&               &               &               &       0.033** \\
            &               &               &               &     (0.014)   \\
\midrule Insurer $\times$ Month FE & \checkmark & \checkmark & \checkmark & \checkmark \\ Insurer $\times$ Product FE & \checkmark & \checkmark & \checkmark & \checkmark \\ Month $\times$ Product FE & \checkmark & \checkmark & \checkmark & \checkmark \\[1ex]
\midrule Observations & {8956} & {8956} & {8956} & {8956}  \\ Within-$ R^2$ & {       0.023} & {       0.034} & {       0.024} & {       0.034} \\[1ex]
\bottomrule

    \end{tabular}
    \caption{Interest Rates and Prices --- Regression Results}
    \label{table:price regression results baseline}
    \vspace{-0.5cm}
    \floatfoot{Note: This table reports regression results for equation (\ref{eq:price reg}). The dependent variable is the log of the premium quote for product $i$ sold by insurer $j$ in month $t$. $y_{t}^{(10)}$ is the monthly 10-year Treasury yield, $\text{PRU}_t$ is the Kansas Fed policy rate uncertainty index \citep{bundick2024introducing}, $\text{Exposed}_j$ is an indicator equal to 1 if insurer $j$ is in the exposed group, $\text{RS}_j$ is an indicator equal to 1 if insurer $j$ is in the risk-sensitive group, and $\text{Long}_i$ is an indicator equal to 1 if product $i$ has the longer maturity of the two product categories in the regression. Standard errors clustered at the product-time level are reported in parentheses. Observations are weighted by insurer $j$'s assets in the year corresponding to month $t$. * $p < 0.1$ ** $p < 0.05$ *** $p < 0.01$.}
\end{center}
\end{table}
\endgroup

We next conduct a range of robustness checks for Table \ref{table:price regression results baseline}. Appendix Tables \ref{table:price regression results 15-10} and \ref{table:price regression results 20-15} present results for alternative product pairs involving 15-year term life products. To avoid potential issues from the
differences in samples across policies, our baseline results consider insurer-month observations in which the insurer sells all policy categories. We show in Table \ref{table:price regression results unbal} that our results are robust to the inclusion of the remaining observations. In Table \ref{table:price regression results baseline}, we weight the regression by insurers’ assets; however, the results are robust to using ordinary life insurance in force as weights, as presented in Table \ref{table:price regression results infc}. Next, in Table \ref{table:price regression results MOVE}, we consider an alternative measure of interest rate uncertainty: the Merrill Lynch Option Volatility Estimate (MOVE) index, a market-based measure of uncertainty in long-term interest rates. Finally, to address the concern that the results might be driven by differences in size between RS and non-RS insurers, we control for insurers' log assets in triple-interactions similar to our main specification in Table \ref{table:price regression results size ctrl}. The results remain robust, suggesting that the findings are not driven by heterogeneous responses across the size distribution. Overall, the results in Tables \ref{table:price regression results 15-10}-\ref{table:price regression results size ctrl} confirm the patterns observed in Table \ref{table:price regression results baseline}, where exposed insurers, especially those exposed to risk-sensitive variable annuities, increase their markup more when rates are low or when rate uncertainty is high. 

While we primarily rely on cross-sectional identification using exposures to RS liabilities, we further strengthen the evidence on product pricing by examining how insurance prices respond to higher-frequency movements in $y_t^{(10)}$ around FOMC meetings. Appendix Figure \ref{fig:product price irf} shows that exposed insurers with risk-sensitive VAs mark up long-term products more following negative innovations in long rates. 

We have therefore shown that insurance companies set prices at least in part based on their interest rate risk exposure. Longer-term products are marked up more when interest rate risk increases, such as in low-interest-rate and high-uncertainty periods. While this mechanism is telling, we need to verify that the pricing behavior we observe is consistent with their product issuance. Unfortunately, we do not observe term life issuance at the insurer-maturity level. However, instead of looking \textit{within} product categories, we can instead focus our analysis \textit{across} product categories to capture differences in product duration. We therefore turn to a comparison of ordinary and group life issuance.

\subsection{Liability Rebalancing} \label{sec:empirics:quantity}

We have now shown that insurers exposed to variable annuities before the crisis experienced larger declines in their duration gaps and, in the context of term life insurance, passed through the additional interest rate risk to their prices. But how did they adjust their entire portfolio of liabilities? Our theory suggests that exposed insurers may have an incentive to shift their product issuance toward low-duration policies in response to an increase in interest rate risk. Group life policies, which are typically renewable yearly, have low reserve valuations, and carry very little interest rate risk, provide a natural alternative to long-dated term or whole life insurance policies.\footnote{See Appendix \ref{appendix:reserve values} for a discussion on the differences in reserve values across these two product categories.} 

\begin{figure}[t!]
	\begin{subfigure}{0.49\textwidth}
		\includegraphics[width = \textwidth]{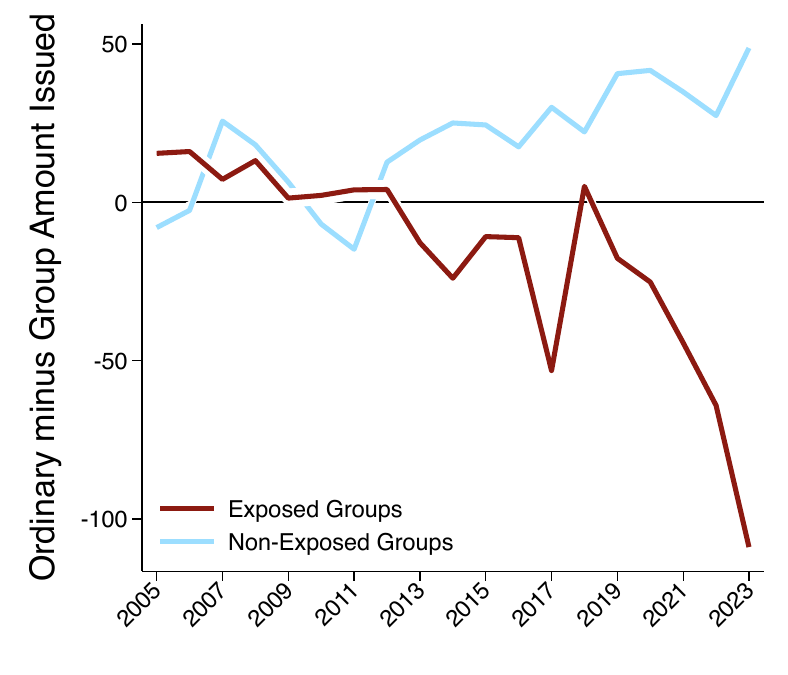}
		\caption{Relative Issuance}
	\end{subfigure}
	\begin{subfigure}{0.49\textwidth}
		\includegraphics[width = \textwidth]{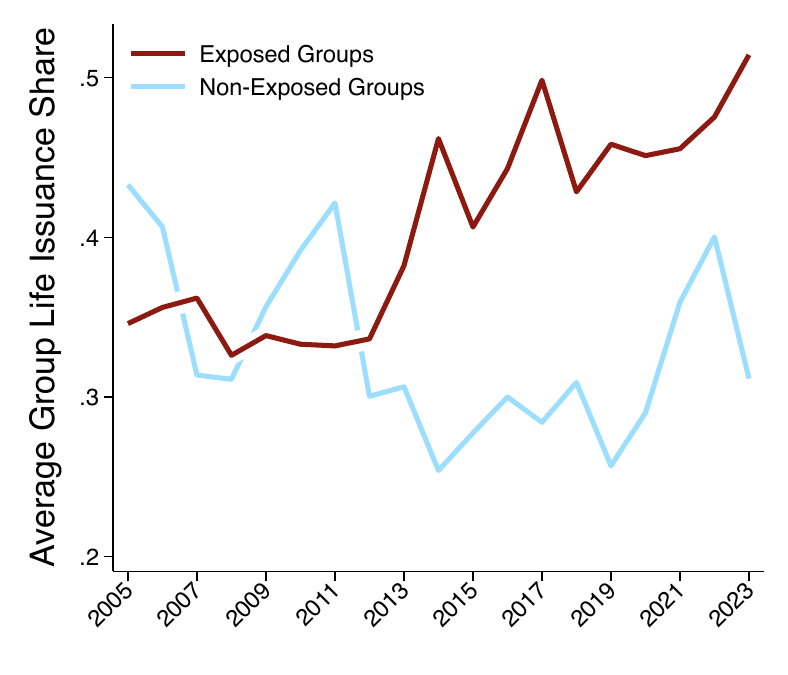}
		\caption{Issuance Shares}
	\end{subfigure}
    \vspace{-1em}
	\caption{Product Issuance Across Insurance Groups Over Time}
	\label{fig:product issuance avg}
	\floatfoot{Note: This figure reports average ordinary life insurance issuance relative to average group life issuance [panel (a)] and average group life issuance shares [panel (b)] for exposed (red) and non-exposed (blue) insurers over time. Averages are weighted by assets within each class of insurers. For panel (a), units are in billions of US dollars.}

\end{figure}

We therefore begin by exploring how the issuance of ordinary and group life products changed throughout the post-crisis period in Figure \ref{fig:product issuance avg}. Panel (a) reports the asset-weighted average difference between ordinary and group life issuance for each group of insurers, conditional on issuing both types of products. Units are in billions of nominal dollars. The figure strongly confirms the predictions of the theory. On average, exposed insurance groups reduced their relative issuance of ordinary life insurance coverage over the sample period. Notably, the decline begins after the financial crisis and accelerates after the drop in yields and the widening of duration gaps in 2011. At the same time, we see that non-exposed groups began to \textit{increase} their relative issuance after 2011. This is consistent with non-exposed groups capturing demand that was previously allocated to exposed groups.

The results on relative product issuance could be influenced by a few large insurers. For example, MetLife was a major force behind the strong growth in group life issuance. Panel (b) shows that this is not a driving factor: average group life issuance \textit{shares} also increased substantially for exposed groups while remaining stable for non-exposed groups. Since shares remove potentially large size differences across insurers, we interpret this as suggestive evidence of liability rebalancing. We further check for robustness in Appendix Figure \ref{app:fig:product issuance avg unweighted} by reproducing Figure \ref{fig:product issuance avg} using unweighted averages within exposed and non-exposed groups, as well as in Appendix Figure \ref{app:fig:product issuance avg no ML}, where we reproduce the figure after excluding MetLife from the sample. The trends are generally unchanged: relative issuance of ordinary life insurance declines for exposed groups, while increasing or remaining unchanged for non-exposed groups. Similarly, group life issuance shares increased from 35\% to over 50\% for exposed groups while remaining virtually unchanged for non-exposed groups.

We now turn to a more careful analysis of liability rebalancing. We estimate the following regression akin to equation (\ref{eq:duration gap regression}) in Section \ref{sec:empirics:duration}:
    \begin{equation}\label{eq:issuance reg}
        \text{Group Share}_{jt} = \sum_{\tau \neq \text{2008-10}} \beta_\tau \left(\mathbf{1}\{t \in \tau\} \times \text{Exposed}_j\right) + \alpha_j + \delta_t + \veps_{jt}
    \end{equation}
\nin where $\text{Group Share}_{jt}$ is the share of insurer $j$'s group life insurance coverage issued relative to total coverage at time $t$, $\text{Exposed}_j$ is an indicator equal to 1 if insurer $j$ was exposed to variable annuities in the pre-crisis period. As in Section \ref{sec:empirics:duration}, we divide our sample into roughly 3-year bins, $\mathcal{T} = $ \{2005-07, 2008-10, 2011-13, 2014-16, 2017-19, 2020-23\}, setting 2008-10 as our reference period. Observations are weighted by insurers' assets.

Our estimates of interest are the parameters $\{\beta_\tau\}_{\tau\in \mathcal{T}}$, which represent the difference in group issuance shares across exposed and non-exposed groups for each period. A positive coefficient for a given year indicates that, relative to non-exposed groups, exposed groups issue relatively more group life insurance than ordinary life insurance in that period relative to the reference period (2008-10). Given the evidence of interest rate risk differences across exposed and non-exposed insurers documented in the previous two sections, we interpret a positive $\beta_\tau$ as evidence of liability rebalancing as a risk management strategy.

To sharpen our identification, we estimate an additional specification that accounts for differences in variable annuity liabilities, akin to equation (\ref{eq:duration gap regression RS}):
        \begin{align} \label{eq:issuance reg rs} \nonumber
        \text{Group Share}_{jt} & = \sum_{\tau \neq \text{2008-10}} \beta_\tau \left(\mathbf{1}\{t \in \tau\} \times \text{Exposed}_j\right) \\ 
        & + \sum_{\tau \neq \text{2008-10}} \beta_\tau^{\text{RS}} \left(\mathbf{1}\{t \in \tau\} \times \text{Exposed}_j \times \text{RS}_j\right) + \alpha_j + \delta_t + \veps_{jt}.
    \end{align}
\nin Consistent with our duration gap results in Section \ref{sec:empirics:duration}, we expect $\beta_{\tau}^{RS}$ to be positive in the post-crisis period, reflecting stronger liability rebalancing by insurers more exposed to interest rate risk.

\begin{figure}[t!]
	\begin{subfigure}{0.49\textwidth}
		\includegraphics[width = \textwidth]{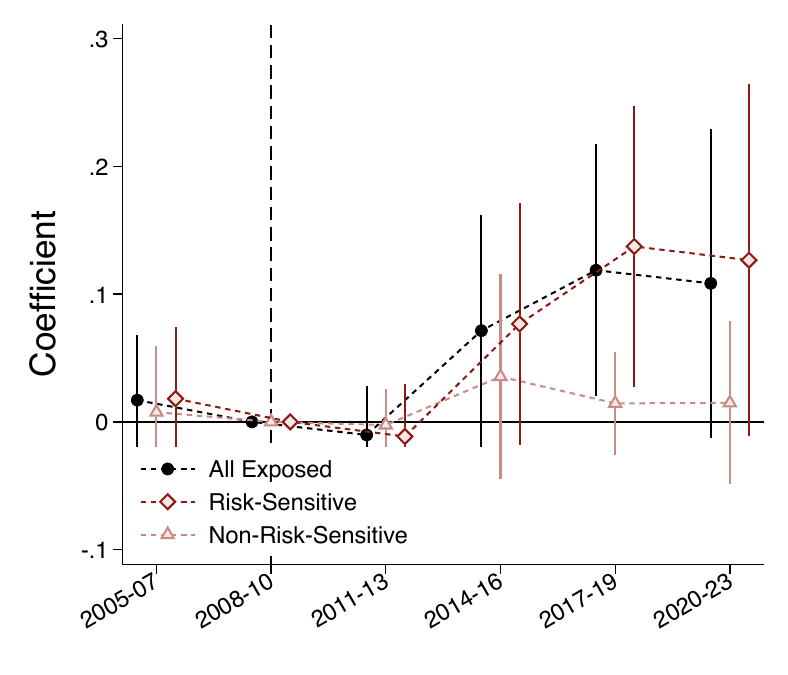}
		\caption{Total Effects}
	\end{subfigure}
	\begin{subfigure}{0.49\textwidth}
		\includegraphics[width = \textwidth]{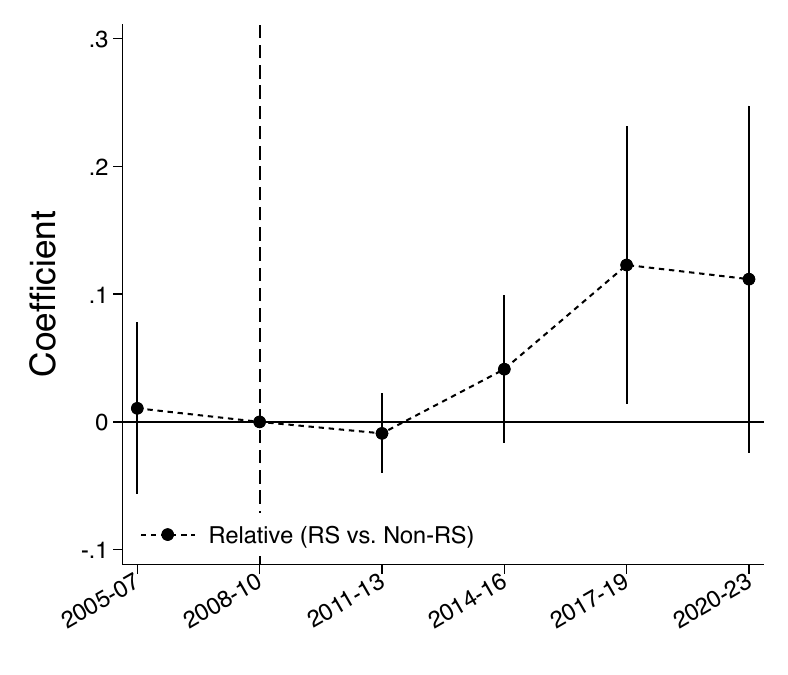}
		\caption{Relative Effects}
	\end{subfigure}
    \vspace{-1em}
	\caption{Liability Rebalancing --- Group Life Issuance Shares}
	\label{fig:liability rebalancing regs}
	\floatfoot{Note: This figure reports regression results for equations (\ref{eq:issuance reg}) and (\ref{eq:issuance reg rs}). The dependent variable is the share of insurer $j$'s insurance coverage issued in the form of group life insurance. In panel (a), estimates are presented as relative to non-exposed insurers; black circles represent the effects of all exposed insurers, red diamonds represent the effects of exposed RS insurers, and pink triangles represent the effects of exposed non-RS insurers. In panel (b), the black circles represent the difference between exposed RS and exposed non-RS insurers. The regressions are weighted by insurers' assets. Vertical lines represent 90\% confidence intervals using standard errors clustered at the insurer level.}
\end{figure}

We present our estimates in Figure \ref{fig:liability rebalancing regs}. In the pre-crisis period, exposed insurers did not appear to have significantly different group life issuance shares relative to non-exposed insurers and relative to the reference period. However, during the post-crisis period, our estimates become positive and significant at the 5\% level. These effects are driven entirely by risk-sensitive variable annuity issuers: the effect on non-risk-sensitive ticks up slightly in 2014-16, but is insignificant and declines thereafter. In contrast, the total effect of risk-sensitive insurers is positive and turns significant in 2017-19. The difference between the two, reflected in Panel (b), is also positive and significant at the 10\% by the 2017-19 period.

We conduct several robustness exercises to validate our claim. First, Appendix Figure \ref{app:fig:liability rebalancing regs cont} reports the results using our continuous measure of risk-sensitive variable annuity exposure. We find strong support for our mechanism, as the 2017-19 and 2020-23 coefficients are both positive and significant at the 1\% level. Second, Appendix Figure \ref{app:fig:liability rebalancing regs noML} excludes MetLife from the analysis to ensure that MetLife's dramatic shift toward group life issuance does not drive the results. We find that the results look similar, although the results become noisier due to the small number of exposed insurers. Third, we include time-varying size and leverage controls in Appendix Figure \ref{app:fig:liability rebalancing regs controls} to ensure that our estimates are not picking up differences in insurer fundamentals. Our results retain their sign and significance.

It is, of course, possible that the observed liability rebalancing is merely reflecting an expansion of group life issuance by exposed insurers rather than a contraction in ordinary life issuance. If this is the case, liability rebalancing would have implications for product composition in life insurance markets, but would not necessarily have implications for market size or participation. To test which of the two scenarios is active, we estimate equations (\ref{eq:issuance reg}) and (\ref{eq:issuance reg rs}) using the log of ordinary life insurance coverage issued as our dependent variable.

\begin{figure}[t!]
	\begin{subfigure}{0.49\textwidth}
		\includegraphics[width = \textwidth]{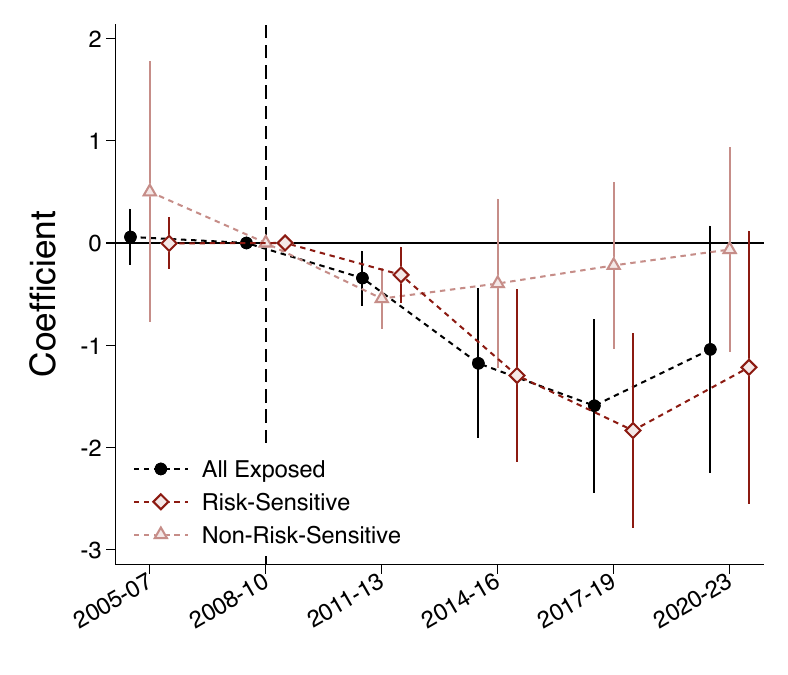}
		\caption{Total Effects}
	\end{subfigure}
	\begin{subfigure}{0.49\textwidth}
		\includegraphics[width = \textwidth]{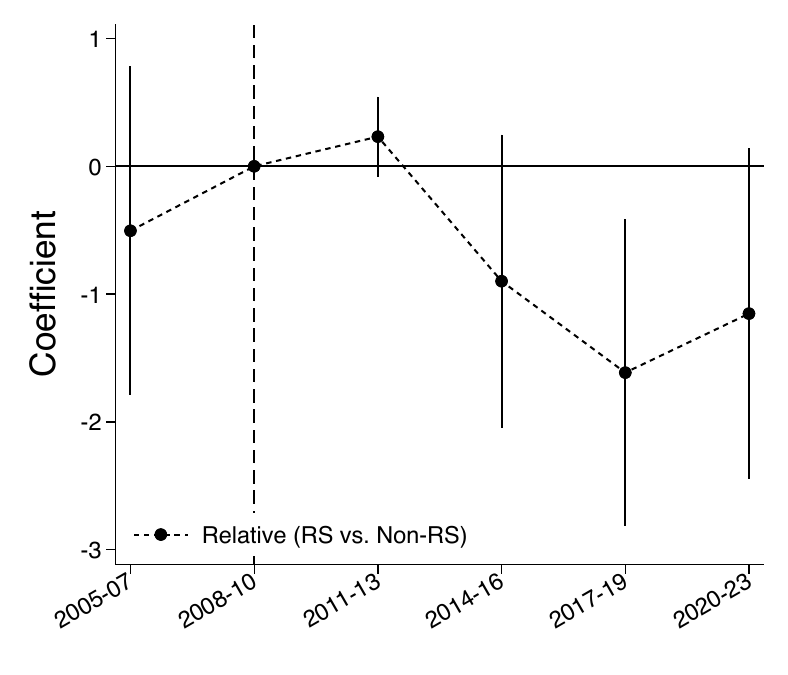}
		\caption{Relative Effects}
	\end{subfigure}
    \vspace{-1em}
	\caption{Liability Rebalancing --- Ordinary Life Issuance Contraction}
	\label{fig:liability rebalancing OL regs}
	\floatfoot{Note: This figure reports regression results for equations (\ref{eq:issuance reg}) and (\ref{eq:issuance reg rs}) where the dependent variable is the log of insurer $j$'s ordinary life insurance coverage issued in each year. In panel (a), estimates are presented as relative to non-exposed insurers; black circles represent the effects of all exposed insurers, red diamonds represent the effects of exposed RS insurers, and pink triangles represent the effects of exposed non-RS insurers. In panel (b), the black circles represent the difference between exposed RS and exposed non-RS insurers. The regressions are weighted by insurers' assets. Vertical lines represent 90\% confidence intervals using standard errors clustered at the insurer level.}
\end{figure}

We present the results in Figure \ref{fig:liability rebalancing OL regs}.\footnote{We also conduct robustness tests akin to the tests conducted for group issuance shares. These can be found in Appendix Figures \ref{app:fig:liability rebalancing regs cont OL}-\ref{app:fig:liability rebalancing regs controls OL}.} Our findings support the latter interpretation: exposed insurers --- and in particular, those exposed to risk-sensitive variable annuity liabilities --- sharply contracted their issuance of ordinary life insurance in the post-crisis period relative to non-exposed life insurers. Note that the contraction began in the 2011-13 period despite the fact that group life shares only began to increase in the 2014-16 period. This reflects the fact that exposed insurers also experienced shocks to their capital generally due to the more intensive treatment of variable annuity liabilities, which led them to reduce their issuance of all products; but, due to their sustained interest rate risk exposure, those exposed to risk-sensitive variable annuity liabilities reduced their ordinary issuance relatively more and for a relatively longer period of time. We demonstrate this in Appendix Figure \ref{app:fig:liability rebalancing regs both types}.\footnote{The figure displays estimated coefficients from our main regressions using the inverse hyperbolic sine transform of ordinary and group life insurance as dependent variables to allow for insurers that do not issue one of the two types of insurance. This allows for sample consistency when comparing across specifications.}

As we highlight in our theory, prices are just one lever that insurers can pull to distort their issuance. In addition, insurers can disincentivize their agents from selling or renewing policies by reducing the commission rates they offer on their products. We show in Appendix Figure \ref{app:fig:commissions}, and more formally using our regression specification in Appendix Figure \ref{app:fig:commissions regressions}, that average ordinary life commission rates are decreasing at a faster rate for exposed insurers relative to non-exposed insurers, despite being similar in the pre-crisis period.\footnote{We define an insurer-level commission rate as their total commissions paid divided by their total premium revenues collected.} This is also consistent with our theory: insurers pay lower commissions to attract fewer agents, which in turn leads to a contraction in their issuance. 

Interestingly, the decline is driven mostly by renewals rather than by new policies. As we have shown so far, insurers can reduce the duration exposure of newly issued policies by adjusting their duration mix, whereas they cannot alter the duration of existing policies. Hence, as insurers' duration mismatch amplifies, they might have stronger incentives to encourage lapsation of previously issued long-duration policies by cutting renewal commissions. We provide support for this channel in Appendix Figure \ref{app:fig:lapse rates regs}, which shows that exposed (and particularly, risk-sensitive) insurers experienced higher lapse rates than non-exposed insurers during the post-crisis period.

\subsection{Aggregate Product Market Dynamics}\label{sec:empirics:agg}

We now turn to the market-level effects of liability rebalancing. In addition to exposed (risk-sensitive and non-risk-sensitive) and non-exposed groups, we also consider four other categories of insurance companies for completeness. First, we include spin-offs, e.g., companies that were part of either an exposed or non-exposed insurance group in the pre-crisis period but later left their pre-crisis-period group (e.g., Brighthouse Financial departing from MetLife in 2017). Second, include insurers that entered the market after the pre-crisis period. Third, we include insurers that are not a part of an insurance group. Fourth, we include reinsurers.

\begin{figure}[t!]
    \centering
    \includegraphics[width=1\linewidth]{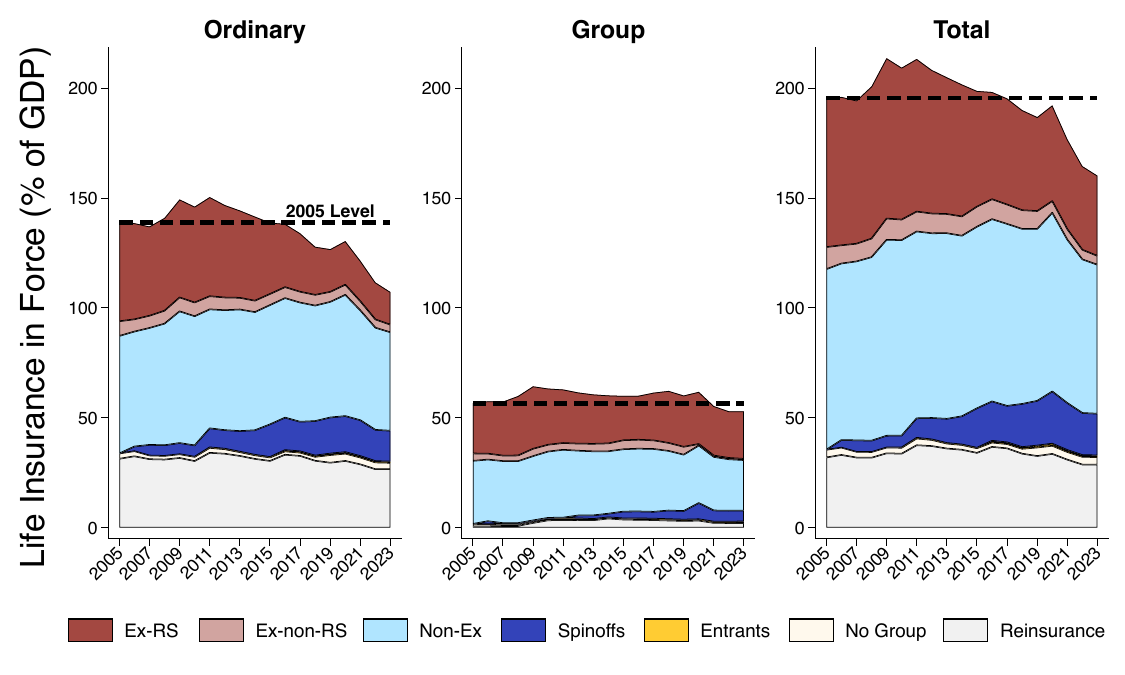}
    \caption{Aggregate Life Insurance Market Dynamics}
    \label{fig:aggregate infc dynamics}
    \floatfoot{Note: This figure reports real aggregate gross life insurance in force as a percentage of real GDP from 2005 to 2023. The first panel reflects ordinary life in force, the second panel reflects group life in force, and the third panel reflects the sum of ordinary and group life insurance. Red areas represent RS-exposed insurance groups, light red areas represent non-RS-exposed groups, light blue areas represent non-exposed insurance groups, dark blue areas reflect insurance companies that belonged to either the exposed or non-exposed insurance groups in the pre-crisis period but have since spun off, yellow areas represent new entrants relative to the pre-crisis period, white areas represent insurers not in a life insurance group, and gray areas reflect reinsurance companies. Dashed black lines represent aggregate insurance in force within each panel as of 2005.}
\end{figure}

We report market-level insurance coverage in force over our sample period in Figure \ref{fig:aggregate infc dynamics}, disaggregated by insurer categories.\footnote{See Appendix Figure \ref{fig:aggregate iss dynamics} for a corresponding figure on aggregate issuance, which largely follows the same pattern.} We express insurance coverage as a percentage of GDP, which measures the purchasing power of a dollar of insurance coverage. Although ordinary life insurance in force increased in the early part of the post-crisis period, peaking around 150.4\% of GDP, it ultimately fell to 107\% of GDP by 2023. While both exposed and non-exposed groups were responsible, the majority of the decline can be explained by exposed insurers: their life insurance in force fell from 51.7\% of GDP in 2005 to 20.3\% of GDP in 2023, accounting for three quarters of the decline. The majority of the decline was driven by RS-exposed insurers, whose insurance in force fell from 45\% to 14\% of GDP.

In contrast, group life insurance in force remained stable throughout most of the post-crisis period, only moderately declining relative to initial levels after the COVID-19 crisis. Consistent with our hypothesis, exposed insurers were a major difference between ordinary and group life market dynamics. Putting the two together, life insurance in force at the industry level fell from 213.2\% to 160\% of GDP.

In Appendix \ref{appendix:back of the envelope}, we provide a simple framework to isolate the role of interest rate risk and liability rebalancing in the contraction of the ordinary life insurance market. We estimate three different components: a common component that applies to all insurers, which we interpret as a demand shock; a VA-specific component to applies to all exposed insurers, which we interpret either as a shock to VA issuers' reputation or as a shock to their capital due to generally stricter regulation; and an interest rate risk component that applies only to RS-exposed insurers, which we interpret as the consequence of liability rebalancing. The estimates suggest that the liability rebalancing response reduced ordinary life insurance coverage between 2005 and 2023 by 12.1\% of GDP, which accounts for 51\% of the decline in RS insurers' ordinary life insurance coverage and 38\% of the decline in aggregate ordinary life insurance coverage.

\section{Conclusion}

Interest rate risk is of the first order to many financial institutions. During the low interest rate period that accompanied the recovery from the financial crisis, exposure to interest rate risk grew for many of these institutions. In particular, due to the long-term nature of their liabilities and issues of market incompleteness and regulatory frictions, many life insurance companies had their equity squeezed by low rates.

We provide theory and evidence that insurers with especially convex liabilities, such as variable annuities, may retreat from long-duration product markets to reduce their exposure to interest rate risk. Such liability rebalancing patterns are further amplified by post-GFC regulations that have increased interest rate risk for some insurers. While insurers substitute toward short-duration products to an extent, the industry as a whole may not remain stable if there are substantive differences in product market characteristics. This appears to be the case for life insurers today: group life insurance markets did not grow enough to offset the decline in ordinary life insurance markets, resulting in a shrunken system.

{
\singlespacing

}

\begin{appendix}
\setcounter{table}{0}
\setcounter{figure}{0}
\renewcommand{\thetable}{\Alph{section}.\arabic{table}}
\renewcommand{\thefigure}{\Alph{section}.\arabic{figure}}

\section{Proof Sketches for the Model}

This appendix gives self-contained proof sketches for all theoretical results.
Online Appendix~\ref{app_sec:proofs:sub:full} supplies the longer algebra for
the equilibrium substitution result.

\medskip
\phantomsection\label{app_sec:proofs:decisions}
\nin\textit{Sketch for \lemmaref{lemma:decisions}.}
Let
\[
    a_{jt}\equiv 1+\ee_t[\lambda_{jt+1}
    (R_{jt+1}^{A}-\overline R_{jt+1}^{A})],\qquad
    b_{ijt}\equiv 1+\ee_t[\lambda_{jt+1}
    (R_{it+1}-\overline R_{it+1})].
\]
Product $i$ contributes
$[a_{jt}P_{ijt}-b_{ijt}V_{it}]Q_{ijt}-a_{jt}\eta_{it}T_{ijt}$
to the objective. Since
$\partial Q_{ijt}/\partial P_{ijt}=-\veps_{it}Q_{ijt}/P_{ijt}$, the price
first-order condition gives
\[
    \frac{P_{ijt}}{V_{it}}
    =\frac{\veps_{it}}{\veps_{it}-1}\frac{b_{ijt}}{a_{jt}}
    =\frac{\veps_{it}}{\veps_{it}-1}\mathcal M_{ijt}.
\]
At this price, the terms that depend on $T_{ijt}$ are proportional to
$\mathcal E_{it}\overline Q_{ijt}V_{it}^{1-\veps_{it}}
\mathcal M_{ijt}^{1-\veps_{it}}\kappa(T_{ijt})-\eta_{it}T_{ijt}$.
Its Kuhn--Tucker condition gives the stated agent choice. \QED

\medskip
\phantomsection\label{app_sec:proofs:rmm}
\nin\textit{Sketch for \lemmaref{lemma:rmm}.}
Using $R_{it+1}-\overline R_{it+1}=-D_{it}\Delta R_{t+1}$ in
(\ref{eq:rm approx 2}) yields
\begin{align*}
    \ee_t[\lambda_{jt+1}(R_{it+1}-\overline R_{it+1})]
    &\approx \ee_t[(\bar\lambda_{jt+1}
    -\bar\lambda_{jt+1}'D_{jt}^{K}\Delta R_{t+1})
    (-D_{it}\Delta R_{t+1})] \\
    &=\bar\lambda_{jt+1}'D_{jt}^{K}D_{it}\sigma_{t+1}^{2}.
\end{align*}
The martingale property eliminates the linear term. Replacing $D_{it}$ by
$D_{jt}^{A}$ gives the denominator of $\mathcal M_{ijt}$. \QED

\medskip
\phantomsection\label{app_sec:proofs:iss risk}
\nin\textit{Sketch for \propref{prop:iss risk}.}
Set $c_{jt}=\bar\lambda_{jt+1}'D_{jt}^{K}>0$. Then
\[
    \frac{\partial\mathcal M_{ijt}}{\partial\sigma_{t+1}^{2}}
    =\frac{c_{jt}(D_{it}-D_{jt}^{A})}
    {(1+c_{jt}\sigma_{t+1}^{2}D_{jt}^{A})^{2}}.
\]
Thus uncertainty raises (lowers) the markup when product duration is above
(below) asset duration. By \lemmaref{lemma:decisions}, price moves with the
markup, while agent distribution and demand move against it, proving both
inequalities. \QED

\medskip
\phantomsection\label{app_sec:proofs:iss convex}
\nin\textit{Sketch for \propref{prop:iss convex}.}
The rate decline and the ordering of capital convexity imply
\[
    D_{j't}^{K,2}-D_{j't}^{K,1}
    \approx |\gamma_{j't}^{K}|(R_t^2-R_t^1)
    <|\gamma_{jt}^{K}|(R_t^2-R_t^1)
    \approx D_{jt}^{K,2}-D_{jt}^{K,1}.
\]
Consequently $c_{j't}^{2}>c_{jt}^{2}>0$, where
$c_{jt}=\bar\lambda_{jt+1}'D_{jt}^{K}$. Since
\[
    \frac{\partial\mathcal M_{ijt}}{\partial c_{jt}}
    =\frac{\sigma_{t+1}^{2}(D_{it}-D_{jt}^{A})}
    {(1+c_{jt}\sigma_{t+1}^{2}D_{jt}^{A})^{2}},
\]
the markup of the more-convex insurer changes more. Applying
\lemmaref{lemma:decisions} gives the two issuance-ratio orderings. \QED

\medskip
\phantomsection\label{app_sec:proofs:sub}
\nin\textit{Sketch for \propref{prop:sub}.}
Suppress $i,t$, and write
$z_j=\alpha_j\mathcal M_j^{1-\veps}$ and
$x=\mathcal P^{1-\veps}$. \lemmaref{lemma:decisions} and
$\kappa(T)=1-e^{-T}$ imply
\begin{equation}\label{eq:substitution fixed point}
    \kappa_j=\left[1-\frac{\eta x}{\mathcal E Yz_j}\right]_+,
    \qquad
    x=\alpha^0+\veps\sum_j
    \left[\mathcal E z_j-\frac{\eta x}{Y}\right]_+ .
\end{equation}
The right side of the second equation is continuous and decreasing in $x$,
so equilibrium is unique; the positive parts incorporate entry and exit.
Quantity can be written
\begin{equation}\label{eq:substitution quantity}
    Q_j=\frac{\veps-1}{V\mathcal M_j}
    \left[\frac{\mathcal E Yz_j}{x}-\eta\right]_+.
\end{equation}
Let $\rho=x^2/x^1$. Under the symmetric-initial-penetration assumption,
$\kappa_j^1=\kappa^1$ for every initially active insurer. Using
$z_j^2=z_j^1\psi_j^{1-\veps}$ in
(\ref{eq:substitution quantity}) gives
\[
 Q_j^2\gtrless Q_j^1
 \quad\Longleftrightarrow\quad
 \rho\lessgtr
 H(\psi_j;\kappa^1)
 \equiv\frac{\psi_j^{1-\veps}}
 {1+\kappa^1(\psi_j-1)}.
\]
The positive part covers exit in environment 2. The function $H$ is
continuous and strictly decreasing from infinity to zero, so the common
price-index ratio $\rho$ determines a unique common cutoff
$H(\overline\psi;\kappa^1)=\rho$. The same cutoff inequalities apply to
long and short products; only the effect of exposure on $\psi_j$ reverses.
Online Appendix~\ref{app_sec:proofs:sub:full} gives the full derivation. \QED

\medskip
\phantomsection\label{app_sec:proofs:iss market}
\nin\textit{Sketch for \propref{prop:iss market}.}
Retain $z_j$ and $x$ from (\ref{eq:substitution fixed point}). For a long
product, higher markups lower every $z_j$ and hence lower the unique fixed
point $x$. The outside-option expenditure share $\alpha^0/x$ rises, so
inside expenditure $X_i^2=\sum_jP_{ij}^2Q_{ij}^2$ is below $X_i^1$.
Common initial markups imply $P_{ij}^1=P_i^1$ for every $j$, while
$P_{ij}^2\geq P_i^1$. Therefore
\[
    Q_i^2\leq\frac{X_i^2}{P_i^1}
    <\frac{X_i^1}{P_i^1}=Q_i^1.
\]
For a short product, markups and prices fall, $x$ and inside expenditure
rise, and the inequalities reverse. Strictness follows from the assumed
change for an active insurer. \QED

\clearpage

\begin{center}
 {\centering \LARGE \bf Online Appendix \\[1em] ``From Long to Short:
How Interest Rates Shape Life Insurance Markets''}  

 \;

 {\centering \large Ziang Li and Derek Wenning}  

\end{center}

\section{Additional Analyses}

\subsection{The Drivers of Duration Gaps}\label{appendix:duration gap drivers}

\begin{figure}[h!]
	\begin{subfigure}{\textwidth}
            \centering 
		\includegraphics[width = 0.9\textwidth]{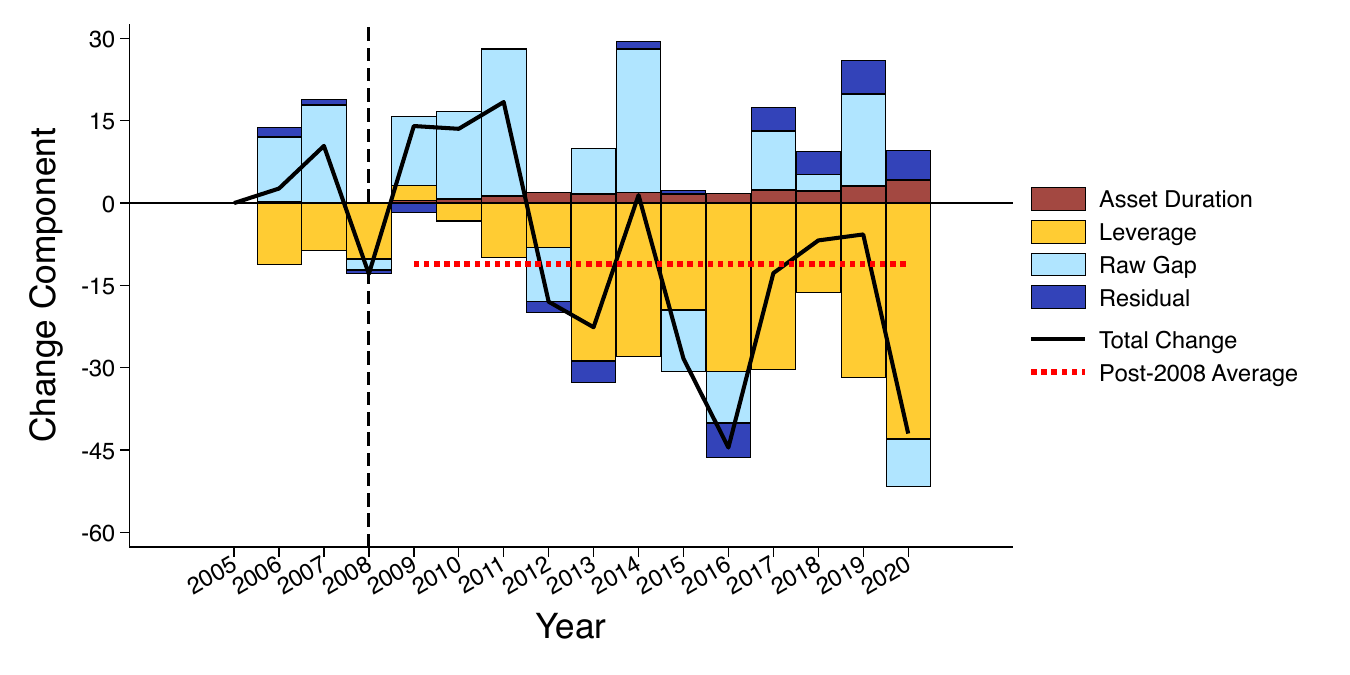}
		\caption{Exposed Insurers}
	\end{subfigure}
	\begin{subfigure}{\textwidth}
            \centering 
		\includegraphics[width = 0.9\textwidth]{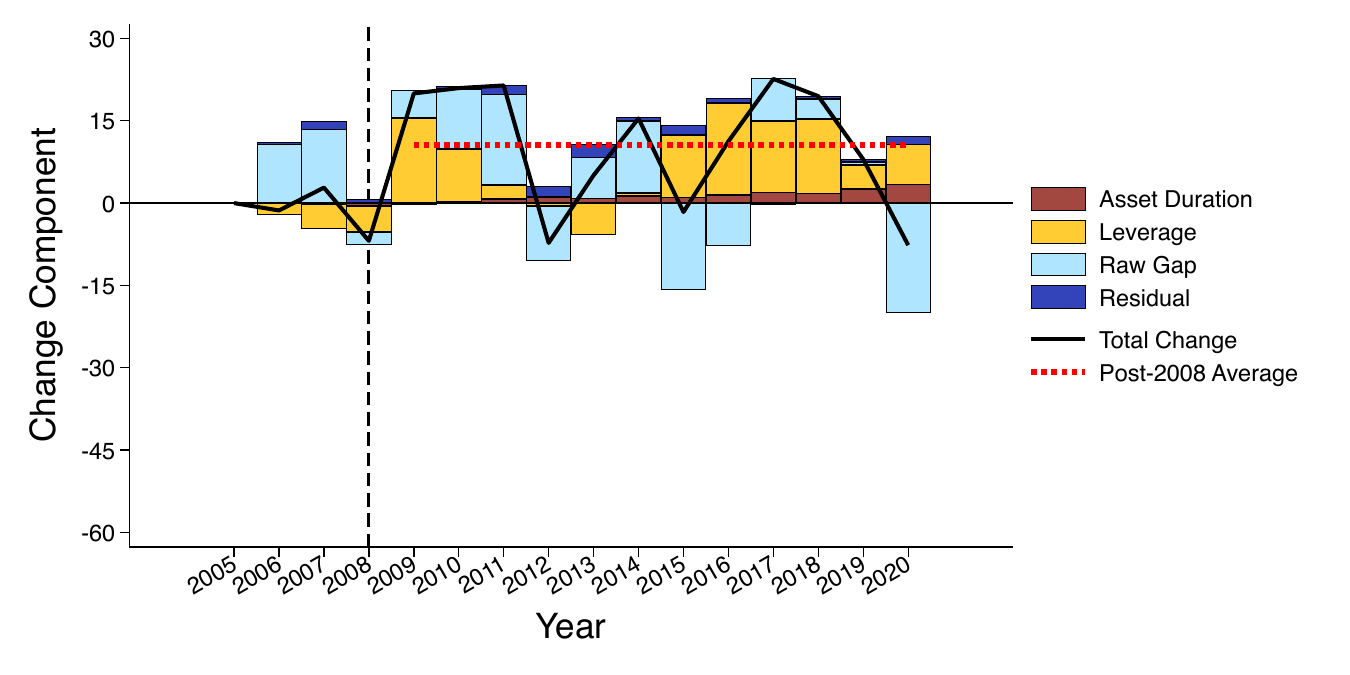}
		\caption{Non-Exposed Insurers}
	\end{subfigure}
    \vspace{-1em}
	\caption{Change in Duration Gap Components}
	\label{fig:dur gap components}
	\floatfoot{Note: This figure reports the decomposition of average duration gap changes as in equation (\ref{eq:dur gap components}) for exposed [panel (a)] and non-exposed [panel (b)] insurers. Red bars reflect changes in asset duration, light blue bars reflect changes in the duration mismatch component, yellow bars reflect changes in the leverage component, and dark blue bars reflect the residual component. Black lines represent the average change in duration gaps, while dotted gray lines reflect the average duration gap change relative to 2005 between 2008 and 2023. Averages are weighted by insurers' assets within exposed and non-exposed groups.}

\end{figure}

This section decomposes changes in duration gaps across the exposed and non-exposed groups into several components. To do so, define $\text{Lev}_{jt} = L_{jt} / K_{jt}$ to be insurer $j$'s leverage ratio, and let $G_{jt} = D_{jt}^{A} - D_{jt}^{L}$ be the difference between their asset duration and liability duration (but not their duration gap, which is the duration of their capital). Note that the change in the duration gap can be written
    \begin{align}
        \Delta D_{jt}^{K} & = \Delta D_{jt}^{A} + \Delta \Big[\text{Lev}_{jt} \times G_{jt}\Big] \nonumber\\
        & = \underbrace{\Delta D_{jt}^{A}}_{\substack{\text{Asset Duration} \\ \text{Component}}} + \underbrace{\Delta \text{Lev}_{jt} \times G_{jt}}_{\substack{\text{Leverage} \\ \text{Component}}} + \underbrace{\text{Lev}_{jt} \times \Delta G_{jt}}_{\substack{\text{Duration Mismatch} \\ \text{Component}}} + \underbrace{\Delta\text{Lev}_{jt} \times\Delta G_{jt}}_{\text{Residual}} \label{eq:dur gap components}
    \end{align}
\nin Figure \ref{fig:dur gap components} plots the four components cumulatively for exposed [panel (a)] and non-exposed [panel (b)] insurers, using 2005 as the base year. A striking pattern that emerges is that leverage is the primary driver of the decline in duration gaps for exposed insurers. This is consistent with variable annuities receiving higher capital requirements in the aftermath of the financial crisis \citep{koijen2022fragility}, thereby exacerbating these insurers' leverage. Therefore, any pre-existing duration mismatch would be amplified. This amplification would be even more dramatic if the liability duration measure of \cite{huber2022} included variable annuities and other interest-sensitive liabilities.

\subsection{Derivation for the Relative Markup Spread}\label{appendix:relative markup derivation}

This section derives equation (\ref{eq:markup approx double diff}) using the result in \lemmaref{lemma:rmm}. To first order, we can write the log markup over fair value as
    \begin{equation}\label{eq:markup approx empirical}
        \log \frac{P_{ijt}}{V_{it}} \approx \log \Bigg(\frac{\veps_{it}}{\veps_{it} - 1}\Bigg) + \bar\lambda_{jt+1}'D_{jt}^{K}\sigma_{t+1}^2\Big(D_{it} - D_{jt}^{A}\Big).
    \end{equation}
\nin Consider two products $\ell$ and $s$ in which $D_{\ell t} > D_{st}$. Differencing across products implies that
    \begin{equation}\label{eq:markup approx diff}
        \log \frac{P_{\ell jt}/V_{\ell t}}{P_{sjt}/V_{st}} \approx \log \frac{\veps_{\ell t}(\veps_{\ell t}-1)^{-1}}{\veps_{st}(\veps_{st}-1)^{-1}} + \bar\lambda_{jt+1}'D_{jt}^{K}\sigma_{t+1}^2\Big(D_{\ell t} - D_{st}\Big).
    \end{equation}
\nin We can then take averages across exposed and non-exposed insurers separately and take the difference between the two. This final step gives the expression in the text.

\subsection{Reserve Valuation Across Products}\label{appendix:reserve values}

We begin by exploring how the product-level reserve values of exposed insurers changed after the financial crisis. As we showed in Table \ref{tab:sum stats}, exposed groups had substantially more exposure to interest-sensitive life insurance policies in addition to their variable annuities, so we should expect their ordinary life insurance reserves to be sensitive to interest rate changes. Group life insurance, on the other hand, is yearly renewable, so its valuation should not systematically change with interest rates.

\begin{figure}
	\begin{subfigure}{0.49\textwidth}
		\includegraphics[width = \textwidth]{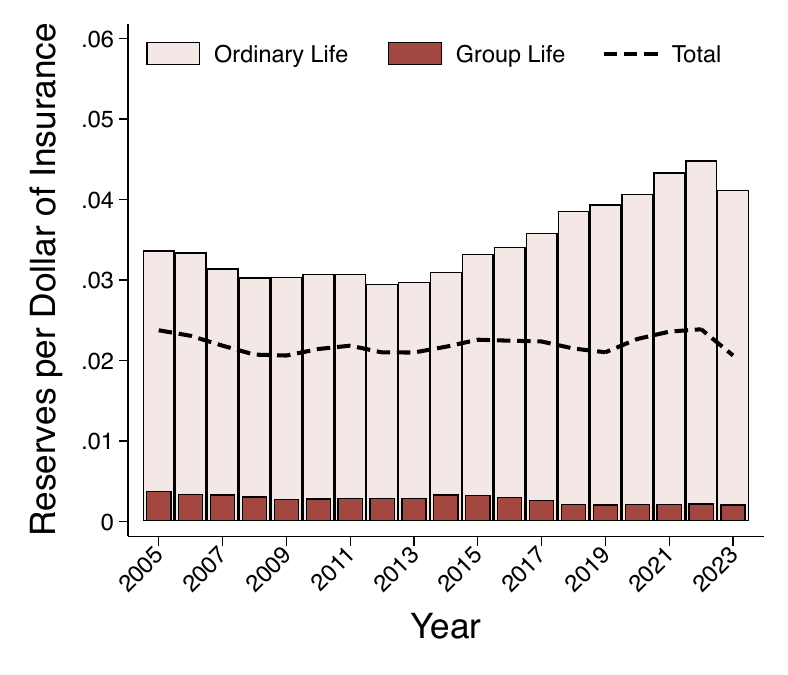}
		\caption{Exposed}
	\end{subfigure}
	\begin{subfigure}{0.49\textwidth}
		\includegraphics[width = \textwidth]{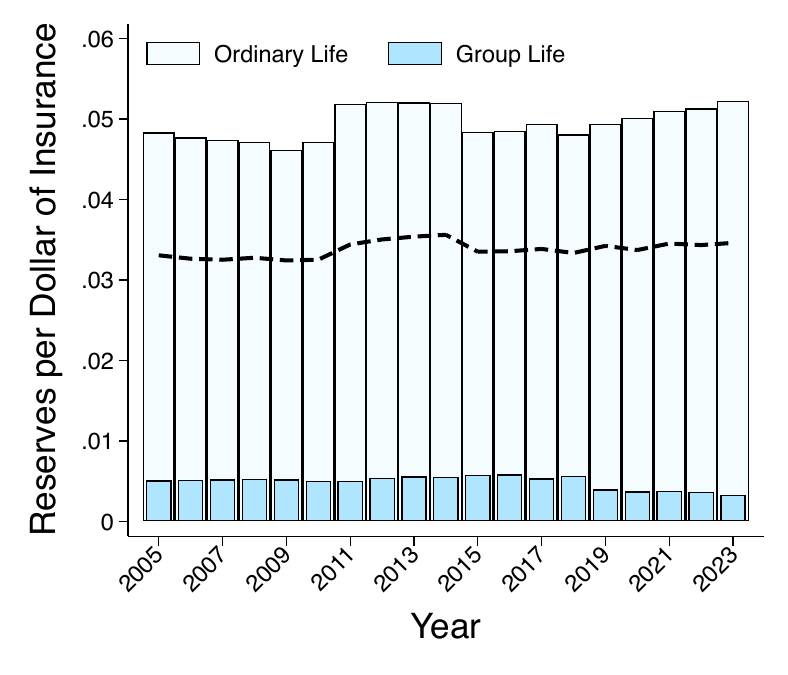}
		\caption{Non-Exposed}
	\end{subfigure}
    \vspace{-1em}
	\caption{Reserve Value Across Products Over Time}
	\label{fig:res val VA}
	\floatfoot{Note: This figure reports average reserve values for ordinary, group, and combined life insurance among exposed and non-exposed insurance groups. Reserve value is calculated as gross reserves divided by life insurance coverage in force. Panel (a) reports reserve values for exposed insurance groups, while panel (b) reports reserve values for non-exposed groups. Dark bars represent average group life reserve values, light bars represent average ordinary life reserve values, and the dashed black line represents the average of the total. Reserve values are weighted by life insurance in force within each category of insurance groups to avoid outliers.}

\end{figure}

Figure \ref{fig:res val VA} confirms this finding. Panel (a) plots the average reserve value of ordinary and group life policies separately for each year in our sample.\footnote{Note that these averages are weighted by the total amount of insurance in force for each insurer. Insurers who have small positions in a particular category tend to have high reserve values due to a lack of diversification. Additionally, reserve values are inflated when life insurance in force is close to 0, which creates outliers.} Three patterns emerge. First, group life policies require substantially fewer reserves than ordinary life policies. This is due to their shorter maturities. Second, average ordinary life reserve values for exposed insurers increased by 34\% (0.031 to 0.043) between 2010 and 2023, consistent with the decline in yields and the sensitivity of their reserves to interest rates. Ordinary reserve values also increased over the same time period for non-exposed insurers, but only by 11\% (0.047 to 0.052). Third, despite the increase in ordinary life reserve values over the post-crisis period, exposed insurers' total reserve value remained stable. This is suggestive of liability rebalancing: as reserve values increase for ordinary life insurance, the threat of future rate changes incentivizes exposed insurers to shift their issuance away from long-duration policies and toward short-duration policies. 

\subsection{A Back-of-the-Envelope Calculation}\label{appendix:back of the envelope}

\nin The goal of this exercise is to estimate how much interest rate risk --- and, as a result, liability rebalancing --- mattered for the decline in aggregate life insurance in force as a share of GDP. We focus on three categories of insurers: non-exposed insurance groups ($N$), non-RS-exposed insurance groups ($NRS$), and RS-exposed groups ($RS$). We begin by discussing insurance group accounting prerequisites and then follow with the derivation of our accounting exercise.

\subsubsection{Accounting for Spin-Offs}

\begin{figure}[t!]
    \centering
    \includegraphics[width=1\linewidth]{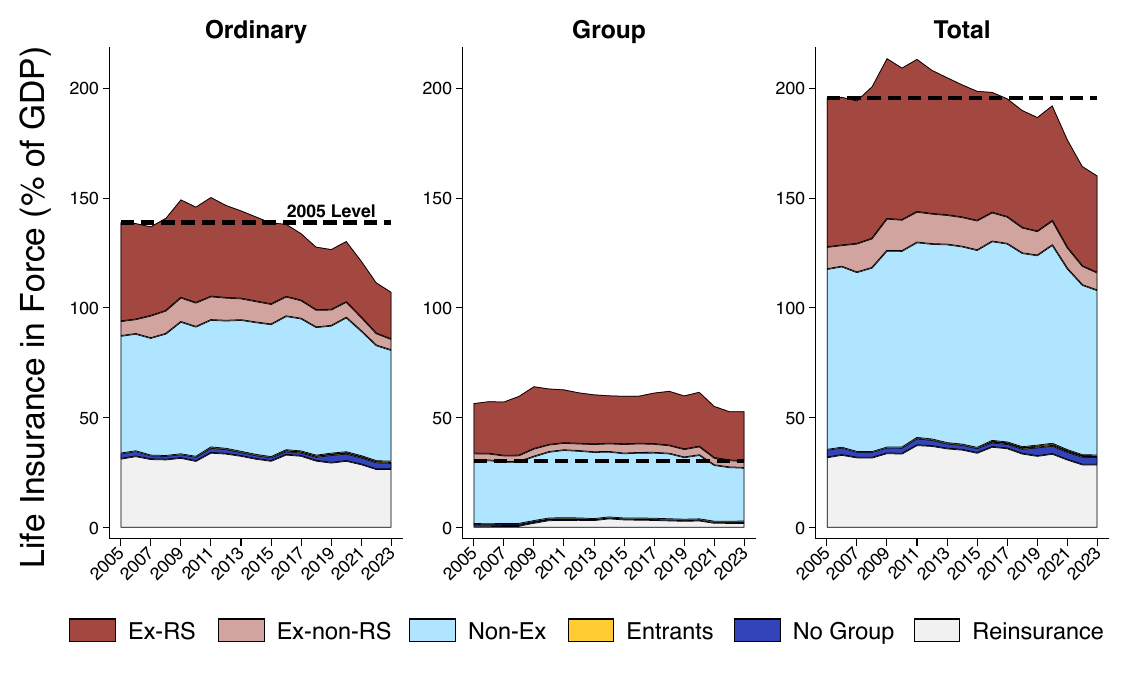}
    \caption{Aggregate Market Dynamics -- Reassigning Spinoffs}
    \label{fig:aggregate infc dynamics spinoffs}
    \floatfoot{Note: This figure reports real aggregate gross life insurance in force as a percentage of real GDP from 2005 to 2023. The first panel reflects ordinary life in force, the second panel reflects group life in force, and the third panel reflects the sum of ordinary and group life insurance. Red areas represent RS-exposed insurance groups, light red areas represent non-RS-exposed groups, light blue areas represent non-exposed insurance groups, yellow areas represent new entrants relative to the pre-crisis period, dark blue areas represent insurers not in a life insurance group, and gray areas reflect reinsurance companies. Dashed black lines represent aggregate insurance in force within each panel as of 2005.}
\end{figure}

A key challenge in keeping track of our three categories is accounting for the insurance \textit{companies} that leave (or are spun-off from) a given insurance \textit{group}. Tracking only the group identifier (and not the individual companies) risks interpreting a decline in aggregate life insurance in force \textit{of a given category} (e.g., RS-exposed insurance groups) as a decline in aggregate life insurance in force \textit{across the entire industry}. This creates two empirical problems. First, spin-offs may merely reflect a change in accounting that nets out in the aggregate. Second, there may be important economic effects of spin-offs that could be missed if we do not include the spun-off companies in our calculations (e.g., changes in perceived brand or reputation value or changes in capital requirements and financial frictions). Therefore, we perform our accounting exercise holding the \textit{category} (e.g., RS-exposed) of each \textit{company} fixed at their pre-crisis level. Appendix Figure \ref{fig:aggregate infc dynamics spinoffs} reproduces Figure \ref{fig:aggregate infc dynamics} from the main text using this reassignment and shows that the dynamics across categories do not materially change.

\subsubsection{Estimating the Aggregate Effects of Liability Rebalancing}

Let $Y_{kt}$ denote aggregate life insurance coverage in force as a percent of GDP for type $k$ insurers in year $t$. Suppose that between 2005 and 2023, we can express the log growth in these values as
    \begin{equation*}
        \Delta \log Y_{k} \equiv \log Y_{k,2023} - \log Y_{k,2005} = - \veps_D -\veps_{k,VA} - \veps_{k,LB}
    \end{equation*}
\nin where $\veps_D \geq 0$ is a demand shock, $\veps_{k,VA} \geq0$ is a shock associated with all variable annuities, and $\veps_{k,LB}\geq0$ is an interest rate risk exposure shock that captures insurers' liability rebalancing behavior. We assume that the demand shock is common across all insurers, that $\veps_{RS,VA} = \veps_{NRS,VA} = \veps_{VA} > 0 = \veps_{N,VA}$, and that $\veps_{RS,LB} = \veps_{LB} > 0 = \veps_{NRS,LB} = \veps_{N,LB}$. 

We can recover the demand shock directly using the decline in non-exposed insurers' life insurance in force, which implies that
    \begin{equation*}
       e^{-\veps_D} = \frac{e^{-\veps_D}Y_{N,2005}}{Y_{N,2005}} = \frac{Y_{N,2023}}{Y_{N,2005}} = \frac{50.69\% \text{ of GDP}}{53.49\% \text{ of GDP}} =  0.948.
    \end{equation*}
\nin Next, we can now recover the VA shock from the non-RS insurers' coverage decline:
    \begin{equation*}
        e^{-\veps_{VA}} = \frac{e^{-\veps_{VA}} e^{-\veps_D} Y_{NRS,2005}}{e^{-\veps_D}Y_{NRS,2005}} = e^{\veps_D} \frac{Y_{NRS,2023}}{Y_{NRS,2005}} = \frac{1}{0.948} \times \frac{4.98\%\text{ of GDP}}{6.69\%\text{ of GDP}} = 0.786.
    \end{equation*}
\nin Finally, it follows that we can recover the liability rebalancing component from the RS-exposed insurers:
    \begin{equation*}
        e^{-\veps_{LB}} = \frac{e^{-\veps_{LB}} e^{-\veps_D-\veps_{VA}} Y_{RS,2005}}{e^{-\veps_D-\veps_{VA}}Y_{RS,2005}} = e^{\veps_D+\veps_{VA}} \frac{Y_{RS,2023}}{Y_{RS,2005}} = \frac{1}{0.948\times 0.786} \times \frac{21.45\%\text{ of GDP}}{45.04\%\text{ of GDP}} = 0.640.
    \end{equation*}
\nin With the estimated the structural parameters, we can estimate counterfactual insurance in force. If RS-exposed insurers did not rebalance their liabilities to avoid interest rate risk, our estimates suggest that their counterfactual ($CF$) insurance in force would have been
    \begin{equation*}
        Y_{RS,2023}^{CF} = e^{-\veps_D-\veps_{VA}} Y_{RS,2005} = 0.948 \times 0.786 \times 45.04\% = 33.53\% \text{ of GDP}.
    \end{equation*}
\nin Therefore, the decline in RS-exposed insurers' insurance in force as a percentage of GDP attributed to their liability rebalancing is $Y_{RS,2023}^{CF} - Y_{RS,2023} = 33.53\% - 21.45\% = 12.08\% \text{ of GDP}$, or $12.08/(45.04-21.45) = 51.2\%$ of their total decline. This is the first set of numbers reported in the main text. The final, aggregate number can simply be calculated as $12.08 / (139-107.2) = 38\%$.

\subsection{Detailed Derivation for Insurer Substitution}
\label{app_sec:proofs:sub:full}

This section completes the argument behind \propref{prop:sub}, including
changes in the set of active insurers. Suppress the product and time
subscripts, and define
\[
    z_j\equiv\alpha_j\mathcal M_j^{1-\veps},
    \qquad x\equiv\mathcal P^{1-\veps}.
\]
Equilibrium demand has the partial-equilibrium form\newline
$Q_j=\overline Q_j\kappa(T_j)P_j^{-\veps}$, where
$\overline Q_j=\alpha_jYV^{\veps-1}/x$.
\lemmaref{lemma:decisions} and
$\kappa'(T)=e^{-T}=1-\kappa(T)$ therefore give
\begin{equation}\label{eq:oa penetration}
    \kappa_j=\left[1-\frac{\eta x}{\mathcal E Yz_j}\right]_+.
\end{equation}
Moreover,
$(\veps/(\veps-1))^{1-\veps}=\veps\mathcal E$. Substituting the optimal
price and (\ref{eq:oa penetration}) into the price index yields the global
fixed-point representation
\begin{equation}\label{eq:oa price fixed point}
    x=\alpha^0+\veps\sum_j
    \left[\mathcal E z_j-\frac{\eta x}{Y}\right]_+.
\end{equation}
The right side is continuous and weakly decreasing in $x$, whereas the left
side is strictly increasing, so (\ref{eq:oa price fixed point}) has a unique
solution. This representation does not presume that the active set remains
fixed.

For completeness, order insurers so that
$z_{(1)}\geq\cdots\geq z_{(J)}$. If the first $n$ insurers are active, the
candidate solution is
\begin{equation}\label{eq:oa active price}
    x_n=\frac{\alpha^0+\veps\mathcal E
    \sum_{m=1}^{n}z_{(m)}}{1+\veps\eta n/Y}.
\end{equation}
The equilibrium $n$ is the unique cutoff satisfying
$\mathcal E z_{(n)}>\eta x_n/Y\geq\mathcal E z_{(n+1)}$, with the natural
endpoint conventions. Hence every insurer with a larger $z_j$ than an
active insurer is also active.

Using the demand curve and optimal price once more gives
\begin{equation}\label{eq:oa equilibrium quantity}
    P_jQ_j=\veps\left[\frac{\mathcal E Yz_j}{x}-\eta\right]_+,
    \qquad
    Q_j=\frac{\veps-1}{V\mathcal M_j}
    \left[\frac{\mathcal E Yz_j}{x}-\eta\right]_+.
\end{equation}

Now impose the symmetric-initial-penetration assumption in
\propref{prop:sub}. For every insurer active in environment 1,
$\kappa_j^1=\kappa^1\in(0,1)$. Equation (\ref{eq:oa penetration}) then
implies that
\[
    a\equiv\frac{\mathcal E Yz_j^1}{x^1}
    =\frac{\eta}{1-\kappa^1}
\]
is common across those insurers. Let $\rho\equiv x^2/x^1$. Because
$z_j^2=z_j^1\psi_j^{1-\veps}$ and
$\mathcal M_j^2=\mathcal M_j^1\psi_j$, equation
(\ref{eq:oa equilibrium quantity}) gives
\[
 Q_j^1=\frac{\veps-1}{V\mathcal M_j^1}(a-\eta),
 \qquad
 Q_j^2=\frac{\veps-1}{V\mathcal M_j^1\psi_j}
 \left[\frac{a\psi_j^{1-\veps}}{\rho}-\eta\right]_+.
\]
Using $\eta=a(1-\kappa^1)$, for every initially active insurer,
\begin{equation}\label{eq:oa common cutoff comparison}
 Q_j^2\gtrless Q_j^1
 \quad\Longleftrightarrow\quad
 \rho\lessgtr H(\psi_j;\kappa^1),
 \qquad
 H(\psi;\kappa)\equiv
 \frac{\psi^{1-\veps}}{1+\kappa(\psi-1)}.
\end{equation}
If insurer $j$ exits in environment 2, the positive part is zero and the
decline inequality in (\ref{eq:oa common cutoff comparison}) continues to
hold. Moreover,
\[
 \frac{\partial\log H(\psi;\kappa)}{\partial\psi}
 =\frac{1-\veps}{\psi}
 -\frac{\kappa}{1+\kappa(\psi-1)}<0,
\]
and $H$ decreases continuously from infinity to zero. Since $\rho$ is the
common equilibrium price-index ratio determined by
(\ref{eq:oa price fixed point}), including any entry or exit, there is a
unique common cutoff $\overline\psi$ satisfying
$H(\overline\psi;\kappa^1)=\rho$. Equation
(\ref{eq:oa common cutoff comparison}) gives the stated inequalities. They
are the same for long and short products; only the effect of exposure on
$\psi_j$ reverses. \QED

\clearpage

\clearpage
\section{Additional Tables and Figures}

\begingroup
\setlength{\tabcolsep}{16pt}
\renewcommand{\arraystretch}{1.3}
\sisetup{input-symbols = {( )}}
\begin{table}[h!]
    \begin{center}
        \vspace{-0.5em}

        \begin{tabular}{@{}l *{4}{S[table-format = -1.1, table-space-text-post=$^{***}$]}@{}c}
            \hline
             & \multicolumn{2}{c}{Non-RS Insurers} & \multicolumn{2}{c}{RS Insurers} \\  \cmidrule(l){2-3} \cmidrule(l){4-5}
             & {2005-2008} & {2009-2023} & {2005-2008} & {2009-2023} \\
             \hline\\[-2.5ex]
             Number of Groups&&&&\\
\cmidrule{1-1} \hfill Full Sample&{9}&{9}&{18}&{17}\\
\hfill Compulife Sample&{0}&{3}&{11}&{12}\\
\\[-1em] Assets&36.31&44.33&116.96&125.91\\
Surplus&1.61&2.66&6.46&6.56\\
Leverage Ratio&18.77&18.33&20.02&19.06\\
Leverage Ratio (Weighted)&26.36&23.96&19.26&20.80\\
\\[-1em] VA Liability Share&0.66&0.45&0.49&0.41\\
IS Reserve Share&0.51&0.51&0.72&0.70\\
\\[-1em] Issuance Market Share&&&&\\
\cmidrule{1-1} \hfill Ordinary&0.06&0.05&0.33&0.23\\
\hfill Group&0.05&0.05&0.39&0.37\\
\\[-1em] In Force Market Share&&&&\\
\cmidrule{1-1} \hfill Ordinary&0.04&0.04&0.33&0.24\\
\hfill Group&0.05&0.04&0.43&0.39\\
\hline

        \end{tabular}
        \caption{Summary Statistics by Risk-sensitive VA Issuance}
        \label{tab:sum stats rs}
        \vspace{-3em}
        \floatfoot{Note: This table reports summary statistics for our sample of exposed insurers, broken down by those that did not issue risk-sensitive variable annuities (non-RS) and those that did (RS). Assets and surplus are reported in billions of dollars. All variables except market shares, the weighted leverage ratio, and the number of groups are unweighted averages across insurers. The weighted leverage ratio is weighted by insurer assets within each period. Market shares are calculated across all years within each period.}
    \end{center}
\end{table}
\endgroup

\begingroup
\setlength{\tabcolsep}{3pt}
\renewcommand{\arraystretch}{1.4}
\begin{table}[h!]
\sisetup{input-symbols = {( )}}
\begin{center}
    \begin{tabular}{@{}l *{4}{S[table-format = -1.3, table-space-text-post=$^{***}$, 
    table-column-width = 2.25cm]}@{}c}
        \toprule
        & \multicolumn{4}{c}{\textit{Dependent Variable:} $\log \text{Price}_{ijt}$} \\[1ex]
         & {(1)} &  {(2)} &  {(3)} & {(4)}  \\
        \midrule
        $ y_t^{(10)} \times \text{Exposed}_j \times \text{Long}_i$&      -0.005***&      -0.016***&      -0.001   &      -0.004   \\
            &     (0.002)   &     (0.002)   &     (0.001)   &     (0.006)   \\
$ \mathrm{PRU}_t \times \text{Exposed}_j \times \text{Long}_i$&               &       0.038***&               &       0.012   \\
            &               &     (0.004)   &               &     (0.011)   \\
$ y_t^{(10)} \times \text{Exposed}_j \times \text{RS}_j \times \text{Long}_i$&               &               &      -0.005** &      -0.013*  \\
            &               &               &     (0.002)   &     (0.007)   \\
$ \mathrm{PRU}_t \times \text{Exposed}_j \times \text{RS}_j \times \text{Long}_i$&               &               &               &       0.027** \\
            &               &               &               &     (0.012)   \\
\midrule Insurer $\times$ Month FE & \checkmark & \checkmark & \checkmark & \checkmark \\ Insurer $\times$ Product FE & \checkmark & \checkmark & \checkmark & \checkmark \\ Month $\times$ Product FE & \checkmark & \checkmark & \checkmark & \checkmark \\[1ex]
\midrule Observations & {8956} & {8956} & {8956} & {8956}  \\ Within-$ R^2$ & {       0.001} & {       0.010} & {       0.001} & {       0.010} \\[1ex]
\bottomrule

    \end{tabular}
    \caption{Interest Rates and Prices Regression --- 15- and 10-year products}
    \label{table:price regression results 15-10}
    \vspace{-0.5cm}
    \floatfoot{Note: This table reports regression results for equation (\ref{eq:price reg}), using prices of 15- and 10-year term life products. The dependent variable is the log of the premium quote for product $i$ sold by insurer $j$ in month $t$. $y_{t}^{(10)}$ is the monthly 10-year Treasury yield, $\text{PRU}_t$ is the Kansas Fed policy rate uncertainty index \citep{bundick2024introducing}, $\text{Exposed}_j$ is an indicator equal to 1 if insurer $j$ is in the exposed group, $\text{RS}_j$ is an indicator equal to 1 if insurer $j$ is in the risk-sensitive group, and $\text{Long}_i$ is an indicator equal to 1 if product $i$ has the longer maturity of the two product categories in the regression. Standard errors clustered at the product-time level are reported in parentheses. Observations are weighted by insurer $j$'s assets in the year corresponding to month $t$. * $p < 0.1$ ** $p < 0.05$ *** $p < 0.01$.}
\end{center}
\end{table}
\endgroup

\begingroup
\setlength{\tabcolsep}{3pt}
\renewcommand{\arraystretch}{1.4}
\begin{table}[h!]
\sisetup{input-symbols = {( )}}
\begin{center}
    \begin{tabular}{@{}l *{4}{S[table-format = -1.3, table-space-text-post=$^{***}$,      table-column-width = 2.25cm]}@{}c}
        \toprule
        & \multicolumn{4}{c}{\textit{Dependent Variable:} $\log \text{Price}_{ijt}$} \\[1ex]
         & {(1)} &  {(2)} &  {(3)} & {(4)}  \\
        \midrule
        $ y_t^{(10)} \times \text{Exposed}_j \times \text{Long}_i$&      -0.020***&      -0.023***&      -0.010***&      -0.010***\\
            &     (0.002)   &     (0.003)   &     (0.001)   &     (0.002)   \\
$ \mathrm{PRU}_t \times \text{Exposed}_j \times \text{Long}_i$&               &       0.010*  &               &       0.002   \\
            &               &     (0.006)   &               &     (0.005)   \\
$ y_t^{(10)} \times \text{Exposed}_j \times \text{RS}_j \times \text{Long}_i$&               &               &      -0.011***&      -0.013***\\
            &               &               &     (0.002)   &     (0.003)   \\
$ \mathrm{PRU}_t \times \text{Exposed}_j \times \text{RS}_j \times \text{Long}_i$&               &               &               &       0.007   \\
            &               &               &               &     (0.005)   \\
\midrule Insurer $\times$ Month FE & \checkmark & \checkmark & \checkmark & \checkmark \\ Insurer $\times$ Product FE & \checkmark & \checkmark & \checkmark & \checkmark \\ Month $\times$ Product FE & \checkmark & \checkmark & \checkmark & \checkmark \\[1ex]
\midrule Observations & {8956} & {8956} & {8956} & {8956}  \\ Within-$ R^2$ & {       0.030} & {       0.031} & {       0.031} & {       0.032} \\[1ex]
\bottomrule

    \end{tabular}
    \caption{Interest Rates and Prices Regression --- 20- and 15-year products}
    \label{table:price regression results 20-15}
    \vspace{-0.5cm}
    \floatfoot{Note: This table reports regression results for equation (\ref{eq:price reg}), using prices of 20- and 15-year term life products. The dependent variable is the log of the premium quote for product $i$ sold by insurer $j$ in month $t$. $y_{t}^{(10)}$ is the monthly 10-year Treasury yield, $\text{PRU}_t$ is the Kansas Fed policy rate uncertainty index \citep{bundick2024introducing}, $\text{Exposed}_j$ is an indicator equal to 1 if insurer $j$ is in the exposed group, $\text{RS}_j$ is an indicator equal to 1 if insurer $j$ is in the risk-sensitive group, and $\text{Long}_i$ is an indicator equal to 1 if product $i$ has the longer maturity of the two product categories in the regression. Standard errors clustered at the product-time level are reported in parentheses. Observations are weighted by insurer $j$'s assets in the year corresponding to month $t$. * $p < 0.1$ ** $p < 0.05$ *** $p < 0.01$.}
\end{center}
\end{table}
\endgroup

\begingroup
\setlength{\tabcolsep}{3pt}
\renewcommand{\arraystretch}{1.4}
\begin{table}[h!]
\sisetup{input-symbols = {( )}}
\begin{center}
    \begin{tabular}{@{}l *{4}{S[table-format = -1.3, table-space-text-post=$^{***}$,      table-column-width = 2.25cm]}@{}c}
        \toprule
        & \multicolumn{4}{c}{\textit{Dependent Variable:} $\log \text{Price}_{ijt}$} \\[1ex]
         & {(1)} &  {(2)} &  {(3)} & {(4)}  \\
        \midrule
        $ y_t^{(10)} \times \text{Exposed}_j \times \text{Long}_i$&      -0.019***&      -0.026***&       0.003   &       0.011** \\
            &     (0.002)   &     (0.003)   &     (0.006)   &     (0.005)   \\
$ \mathrm{PRU}_t \times \text{Exposed}_j \times \text{Long}_i$&               &       0.025***&               &      -0.039***\\
            &               &     (0.005)   &               &     (0.008)   \\
$ y_t^{(10)} \times \text{Exposed}_j \times \text{RS}_j \times \text{Long}_i$&               &               &      -0.025***&      -0.043***\\
            &               &               &     (0.007)   &     (0.005)   \\
$ \mathrm{PRU}_t \times \text{Exposed}_j \times \text{RS}_j \times \text{Long}_i$&               &               &               &       0.074***\\
            &               &               &               &     (0.010)   \\
\midrule Insurer $\times$ Month FE & \checkmark & \checkmark & \checkmark & \checkmark \\ Insurer $\times$ Product FE & \checkmark & \checkmark & \checkmark & \checkmark \\ Month $\times$ Product FE & \checkmark & \checkmark & \checkmark & \checkmark \\[1ex]
\midrule Observations & {11830} & {11830} & {11830} & {11830}  \\ Within-$ R^2$ & {       0.011} & {       0.014} & {       0.014} & {       0.021} \\[1ex]
\bottomrule

    \end{tabular}
    \caption{Interest Rates and Prices Regression --- Unbalanced}
    \label{table:price regression results unbal}
    \vspace{-0.5cm}
    \floatfoot{Note: This table reports regression results for equation (\ref{eq:price reg}), using the full sample of insurer-product observations. The dependent variable is the log of the premium quote for product $i$ sold by insurer $j$ in month $t$. $y_{t}^{(10)}$ is the monthly 10-year Treasury yield, $\text{PRU}_t$ is the Kansas Fed policy rate uncertainty index \citep{bundick2024introducing}, $\text{Exposed}_j$ is an indicator equal to 1 if insurer $j$ is in the exposed group, $\text{RS}_j$ is an indicator equal to 1 if insurer $j$ is in the risk-sensitive group, and $\text{Long}_i$ is an indicator equal to 1 if product $i$ has the longer maturity of the two product categories in the regression. Standard errors clustered at the product-time level are reported in parentheses. Observations are weighted by insurer $j$'s assets in the year corresponding to month $t$. * $p < 0.1$ ** $p < 0.05$ *** $p < 0.01$.}
\end{center}
\end{table}
\endgroup

\begingroup
\setlength{\tabcolsep}{3pt}
\renewcommand{\arraystretch}{1.4}
\begin{table}
\sisetup{input-symbols = {( )}}
\begin{center}
    \begin{tabular}{@{}l *{4}{S[table-format = -1.3, table-space-text-post=$^{***}$,      table-column-width = 2.25cm]}@{}c}
        \toprule
        & \multicolumn{4}{c}{\textit{Dependent Variable:} $\log \text{Price}_{ijt}$} \\[1ex]
         & {(1)} &  {(2)} &  {(3)} & {(4)}  \\
        \midrule
        $ y_t^{(10)} \times \text{Exposed}_j \times \text{Long}_i$&      -0.033***&      -0.052***&      -0.014***&      -0.018***\\
            &     (0.004)   &     (0.004)   &     (0.001)   &     (0.006)   \\
$ \mathrm{PRU}_t \times \text{Exposed}_j \times \text{Long}_i$&               &       0.065***&               &       0.018   \\
            &               &     (0.009)   &               &     (0.011)   \\
$ y_t^{(10)} \times \text{Exposed}_j \times \text{RS}_j \times \text{Long}_i$&               &               &      -0.020***&      -0.034***\\
            &               &               &     (0.003)   &     (0.006)   \\
$ \mathrm{PRU}_t \times \text{Exposed}_j \times \text{RS}_j \times \text{Long}_i$&               &               &               &       0.047***\\
            &               &               &               &     (0.012)   \\
\midrule Insurer $\times$ Month FE & \checkmark & \checkmark & \checkmark & \checkmark \\ Insurer $\times$ Product FE & \checkmark & \checkmark & \checkmark & \checkmark \\ Month $\times$ Product FE & \checkmark & \checkmark & \checkmark & \checkmark \\[1ex]
\midrule Observations & {8956} & {8956} & {8956} & {8956}  \\ Within-$ R^2$ & {       0.033} & {       0.049} & {       0.034} & {       0.049} \\[1ex]
\bottomrule

    \end{tabular}
    \caption{Interest Rates and Prices Regression --- In Force Weighted}
    \label{table:price regression results infc}
    \vspace{-0.5cm}
    \floatfoot{Note: This table reports regression results for equation (\ref{eq:price reg}). The dependent variable is the log of the premium quote for product $i$ sold by insurer $j$ in month $t$. $y_{t}^{(10)}$ is the monthly 10-year Treasury yield, $\text{PRU}_t$ is the Kansas Fed policy rate uncertainty index \citep{bundick2024introducing}, $\text{Exposed}_j$ is an indicator equal to 1 if insurer $j$ is in the exposed group, $\text{RS}_j$ is an indicator equal to 1 if insurer $j$ is in the risk-sensitive group, and $\text{Long}_i$ is an indicator equal to 1 if product $i$ has the longer maturity of the two product categories in the regression. Standard errors clustered at the product-time level are reported in parentheses. Observations are weighted by insurer $j$'s ordinary life insurance in force in the year corresponding to month $t$. * $p < 0.1$ ** $p < 0.05$ *** $p < 0.01$.}
\end{center}
\end{table}
\endgroup

\begingroup
\setlength{\tabcolsep}{3pt}
\renewcommand{\arraystretch}{1.4}
\begin{table}
\sisetup{input-symbols = {( )}}
\begin{center}
    \begin{tabular}{@{}l *{4}{S[table-format = -1.3, table-space-text-post=$^{***}$,      table-column-width = 2.25cm]}@{}c}
        \toprule
        & \multicolumn{4}{c}{\textit{Dependent Variable:} $\log \text{Price}_{ijt}$} \\[1ex]
         & {(1)} &  {(2)} &  {(3)} & {(4)}  \\
        \midrule
        $ y_t^{(10)} \times \text{Exposed}_j \times \text{Long}_i$&      -0.025***&      -0.032***&      -0.011***&      -0.019***\\
            &     (0.003)   &     (0.003)   &     (0.002)   &     (0.007)   \\
$ \mathrm{MOVE}_t \times \text{Exposed}_j \times \text{Long}_i$&               &       0.041***&               &       0.034   \\
            &               &     (0.010)   &               &     (0.022)   \\
$ y_t^{(10)} \times \text{Exposed}_j \times \text{RS}_j \times \text{Long}_i$&               &               &      -0.015***&      -0.013*  \\
            &               &               &     (0.003)   &     (0.008)   \\
$ \mathrm{MOVE}_t \times \text{Exposed}_j \times \text{RS}_j \times \text{Long}_i$&               &               &               &       0.005   \\
            &               &               &               &     (0.023)   \\
\midrule Insurer $\times$ Month FE & \checkmark & \checkmark & \checkmark & \checkmark \\ Insurer $\times$ Product FE & \checkmark & \checkmark & \checkmark & \checkmark \\ Month $\times$ Product FE & \checkmark & \checkmark & \checkmark & \checkmark \\[1ex]
\midrule Observations & {8956} & {8956} & {8956} & {8956}  \\ Within-$ R^2$ & {       0.023} & {       0.029} & {       0.024} & {       0.029} \\[1ex]
\bottomrule

    \end{tabular}
    \caption{Interest Rates and Prices Regression --- MOVE Index}
    \label{table:price regression results MOVE}
    \vspace{-0.5cm}
    \floatfoot{Note: This table reports regression results for equation (\ref{eq:price reg}), with $\text{PRU}_t$ replaced by $\text{MOVE}_t$. The dependent variable is the log of the premium quote for product $i$ sold by insurer $j$ in month $t$. $y_{t}^{(10)}$ is the monthly 10-year Treasury yield, $\text{MOVE}_t$ is the Merrill Lynch Option Volatility Estimate (MOVE) index, $\text{Exposed}_j$ is an indicator equal to 1 if insurer $j$ is in the exposed group, $\text{RS}_j$ is an indicator equal to 1 if insurer $j$ is in the risk-sensitive group, and $\text{Long}_i$ is an indicator equal to 1 if product $i$ has the longer maturity of the two product categories in the regression. Standard errors clustered at the product-time level are reported in parentheses. Observations are weighted by insurer $j$'s assets in the year corresponding to month $t$. * $p < 0.1$ ** $p < 0.05$ *** $p < 0.01$.}
\end{center}
\end{table}
\endgroup

\begingroup
\setlength{\tabcolsep}{3pt}
\renewcommand{\arraystretch}{1.4}
\begin{table}
\sisetup{input-symbols = {( )}}
\begin{center}
    \begin{tabular}{@{}l *{4}{S[table-format = -1.3, table-space-text-post=$^{***}$, table-column-width = 2.25cm]}@{}c}
        \toprule
        & \multicolumn{4}{c}{\textit{Dependent Variable:} $\log \text{Price}_{ijt}$} \\[1ex]
         & {(1)} &  {(2)} &  {(3)} & {(4)}  \\
        \midrule
        $ y_t^{(10)} \times \text{Exposed}_j \times \text{Long}_i$&      -0.019***&      -0.032***&      -0.007***&      -0.005   \\
            &     (0.003)   &     (0.003)   &     (0.002)   &     (0.008)   \\
$ \mathrm{PRU}_t \times \text{Exposed}_j \times \text{Long}_i$&               &       0.044***&               &       0.004   \\
            &               &     (0.005)   &               &     (0.015)   \\
$ y_t^{(10)} \times \text{Exposed}_j \times \text{RS}_j \times \text{Long}_i$&               &               &      -0.013***&      -0.028***\\
            &               &               &     (0.003)   &     (0.009)   \\
$ \mathrm{PRU}_t \times \text{Exposed}_j \times \text{RS}_j \times \text{Long}_i$&               &               &               &       0.040***\\
            &               &               &               &     (0.015)   \\
$ y_t^{(10)} \times \log \text{Assets}_{jt} \times \text{Long}_i$&      -0.015***&      -0.014***&      -0.015***&      -0.014***\\
            &     (0.002)   &     (0.002)   &     (0.002)   &     (0.002)   \\
$ \mathrm{PRU}_t \times \log \text{Assets}_{jt} \times \text{Long}_i$&       0.018***&       0.013** &       0.018***&       0.013** \\
            &     (0.005)   &     (0.005)   &     (0.005)   &     (0.005)   \\
\midrule Insurer $\times$ Month FE & \checkmark & \checkmark & \checkmark & \checkmark \\ Insurer $\times$ Product FE & \checkmark & \checkmark & \checkmark & \checkmark \\ Month $\times$ Product FE & \checkmark & \checkmark & \checkmark & \checkmark \\[1ex]
\midrule Observations & {8956} & {8956} & {8956} & {8956}  \\ Within-$ R^2$ & {       0.044} & {       0.052} & {       0.044} & {       0.053} \\[1ex]
\bottomrule

    \end{tabular}
    \caption{Interest Rates and Prices Regression --- Size Control}
    \label{table:price regression results size ctrl}
    \vspace{-0.5cm}
    \floatfoot{Note: This table reports regression results for equation (\ref{eq:price reg}). The dependent variable is the log of the premium quote for product $i$ sold by insurer $j$ in month $t$. $y_{t}^{(10)}$ is the monthly 10-year Treasury yield, $\text{PRU}_t$ is the Kansas Fed policy rate uncertainty index \citep{bundick2024introducing}, $\text{Exposed}_j$ is an indicator equal to 1 if insurer $j$ is in the exposed group, $\text{RS}_j$ is an indicator equal to 1 if insurer $j$ is in the risk-sensitive group, $\text{Long}_i$ is an indicator equal to 1 if product $i$ has the longer maturity of the two product categories in the regression, and $\log \text{Asset}_{jt}$ is the log total assets of insurer $j$ in period $t$. Standard errors clustered at the product-time level are reported in parentheses. Observations are weighted by insurer $j$'s assets in the year corresponding to month $t$. * $p < 0.1$ ** $p < 0.05$ *** $p < 0.01$.}
\end{center}
\end{table}
\endgroup

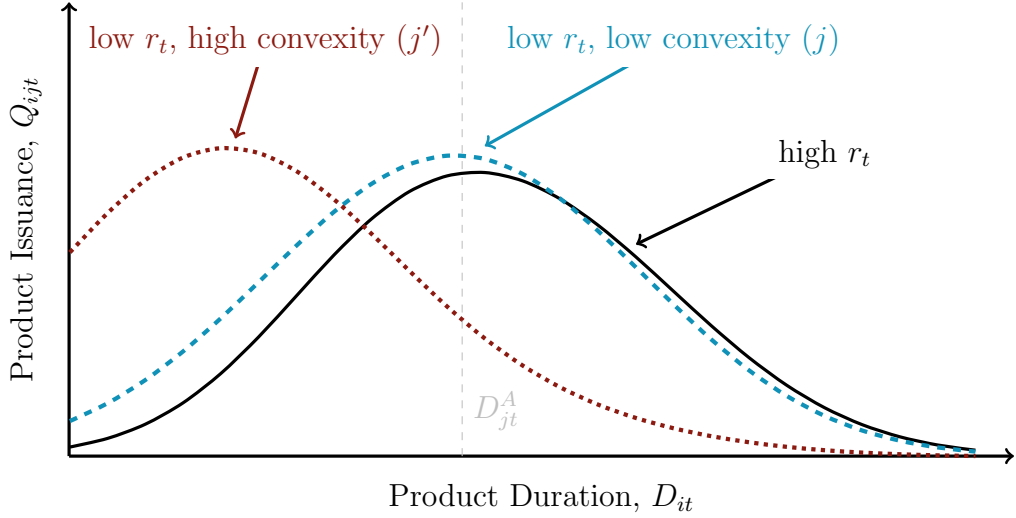
\begin{figure}[h!]
    \begin{center}
        \begin{tikzpicture}
            \draw[->, very thick] (0,0) -- (0,6) node[pos = 0.5, anchor = south, rotate = 90, yshift = 0.5em] {Product Issuance, $Q_{ijt}$};
            \draw[->, very thick] (0,0) -- (12.5,0) node[pos = 0.5, anchor = north, yshift = -0.5em] {Product Duration, $D_{it}$};

            \draw[very thick, black, domain = 0.01:11.99, variable = \x, smooth] plot ({\x}, {0.8*4.0*2.718^(-(\x - 6)^2 * 0.1) + 0.2*5.0*2.718^(-(\x - 4)^2 * 0.2) - 0.01});

            \draw[ultra thick, cyan!70!black, dashed, domain = 0.01:11.99, variable = \x, smooth] plot ({\x}, {0.6*4.0*2.718^(-(\x - 6)^2 * 0.1) + 0.4*5.0*2.718^(-(\x - 4)^2 * 0.1) - 0.01});

            \draw[ultra thick, mycolor, dotted, domain = 0.01:11.99, variable = \x, smooth] plot ({\x}, {0.1*4.0*2.718^(-(\x - 6)^2 * 0.1) + 0.8*5.0*2.718^(-(\x - 2)^2 * 0.1) - 0.01});

            \draw[->, black, very thick] (10,4) -- (7.5, 2.8) node[pos = 0, fill = white] {high $r_t$};

            \draw[->, cyan!70!black, very thick] (8,5.5) -- (5.5, 4.1) node[pos = 0, fill = white] {low $r_t$, low convexity ($j$)};

            \draw[->, mycolor, very thick] (2.6,5.5) -- (2.2, 4.2) node[pos = 0, fill = white] {low $r_t$, high convexity ($j'$)};

            \draw[dashed, thin, lightgray] (5.2,0) -- (5.2,6) node[pos = 0.1, anchor = west] {$D_{jt}^{A}$};

        \end{tikzpicture}
        \vspace{-2em}
        \caption{Interest Rate Risk, Capital Convexity, and Product Issuance}
        \label{fig:props 2-3}
        \floatfoot{Note: This figure presents hypothetical product issuance curves as a function of product duration. The black curve reflects the decisions of two insurers with identical duration gaps in a high interest rate environment. The red dotted and blue dashed line respectively reflect decisions of the more convex and less convex insurer when interest rates decline. The faint dashed gray line represents their shared asset duration.}
    \end{center}
\end{figure}

\begin{figure}
	\begin{subfigure}{0.49\textwidth}
		\includegraphics[width = \textwidth]{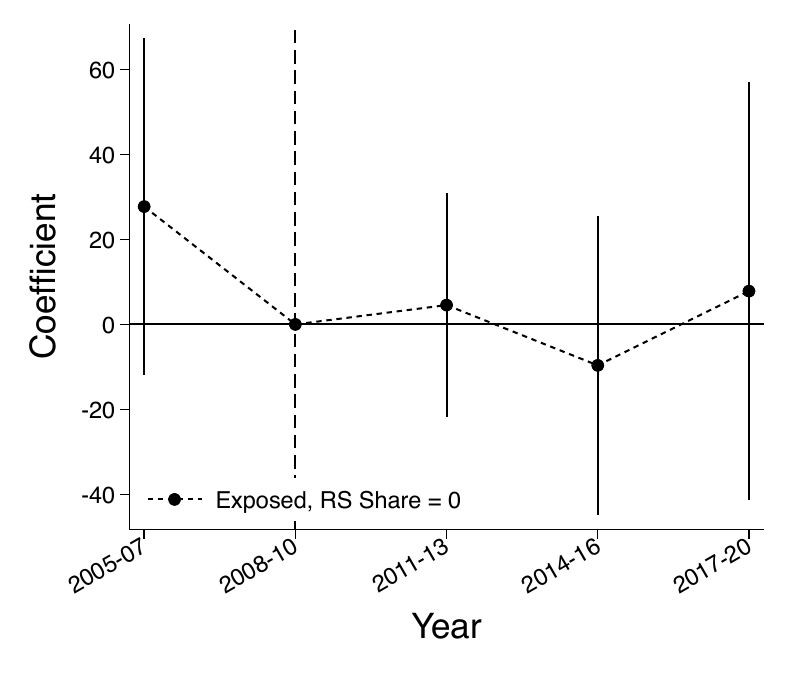}
		\caption{Non-RS Effects}
	\end{subfigure}
	\begin{subfigure}{0.49\textwidth}
		\includegraphics[width = \textwidth]{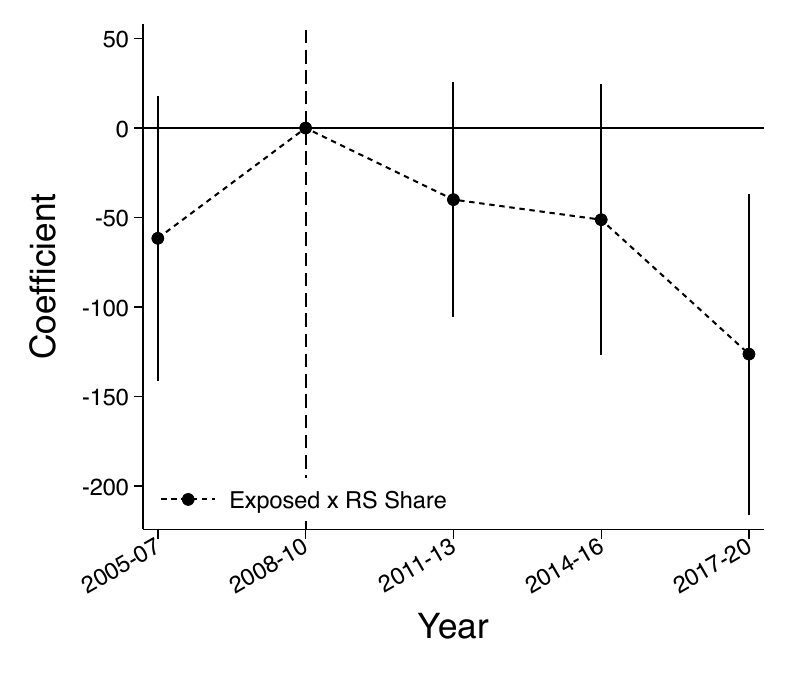}
		\caption{Cross-Term Coefficients}
	\end{subfigure}
    \vspace{-1em}
	\caption{Changes in Duration Gaps --- Continuous Exposure}
	\label{fig:duration gaps rs}
	\floatfoot{Note: This figure presents estimates from regressions (\ref{eq:duration gap regression}) and (\ref{eq:duration gap regression RS}) using $\text{RSshare}_j$ as a continuous measure of risk-sensitive variable annuity exposure. Duration gaps are constructed as in equation (\ref{eq:duration gap empirical}). Panel (a) plots the coefficient on $\text{Exposed}_j$, which captures the difference between exposed insurers with $\text{RSshare}_j=0$ and non-exposed insurers. Panel (b) plots the coefficient on $\text{Exposed}_j \times \text{RSshare}_j$, which captures how the exposed-insurer effect varies with the share of variable annuity reserves that are risk sensitive. For an exposed insurer with $\text{RSshare}_j=s$, the implied effect equals the coefficient in panel (a) plus $s$ times the coefficient in panel (b). The regressions are weighted by insurers' assets. Vertical lines represent 90\% confidence intervals using standard errors clustered at the insurer level.}

\end{figure}

\begin{figure}[t!]
	\begin{subfigure}{0.49\textwidth}
		\includegraphics[width = \textwidth]{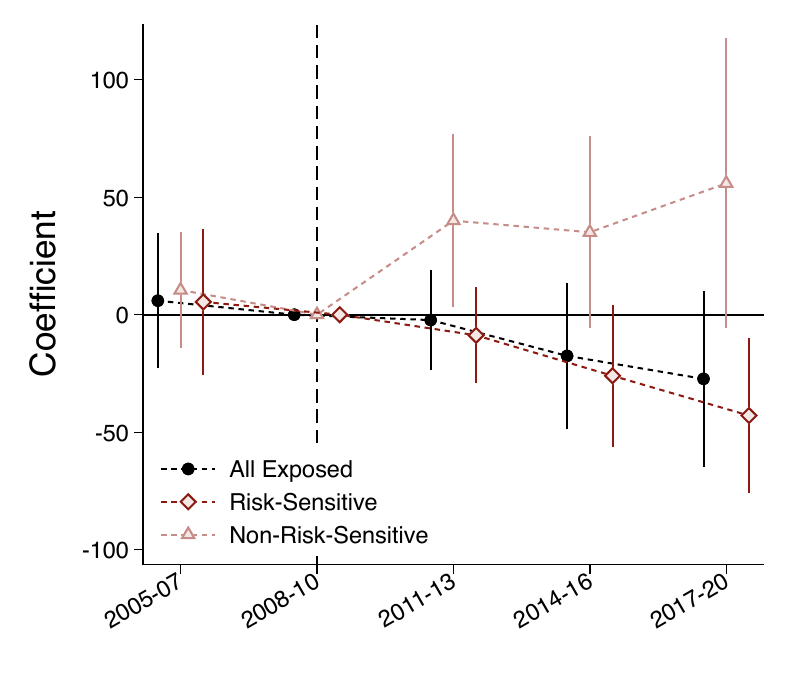}
		\caption{Total Effects}
	\end{subfigure}
	\begin{subfigure}{0.49\textwidth}
		\includegraphics[width = \textwidth]{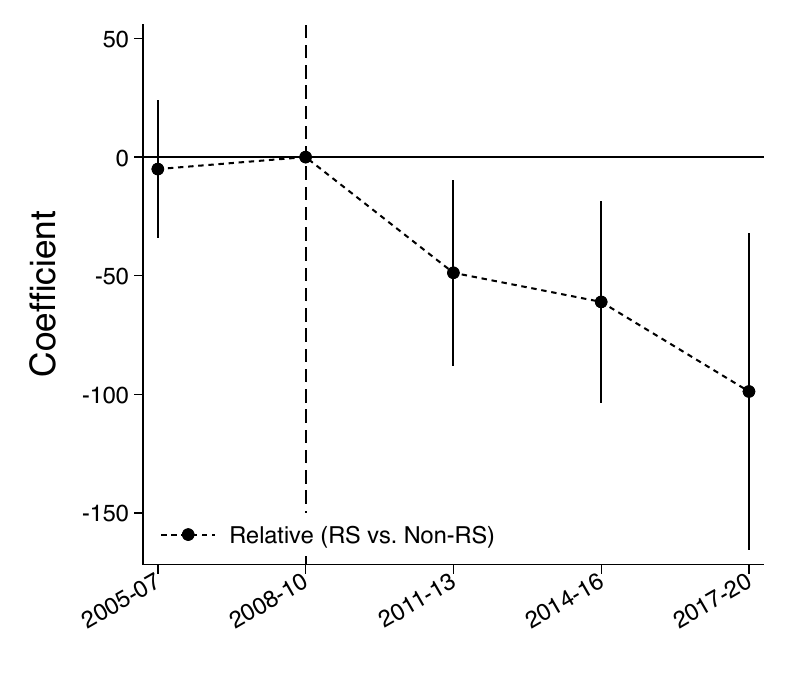}
		\caption{Relative Effects}
	\end{subfigure}
    \begin{subfigure}{0.49\textwidth}
		\includegraphics[width = \textwidth]{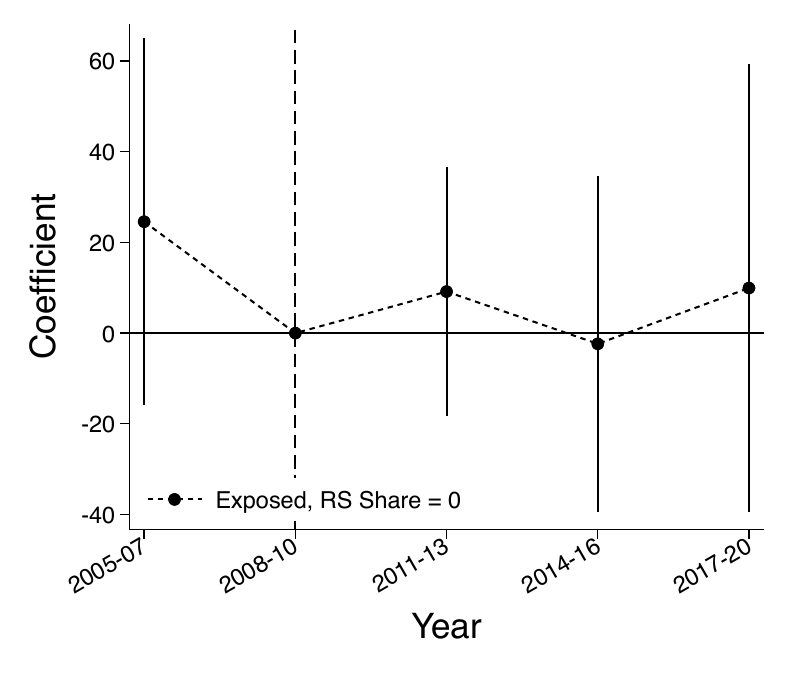}
		\caption{Non-RS Effects, Cont. Measure}
	\end{subfigure}
	\begin{subfigure}{0.49\textwidth}
		\includegraphics[width = \textwidth]{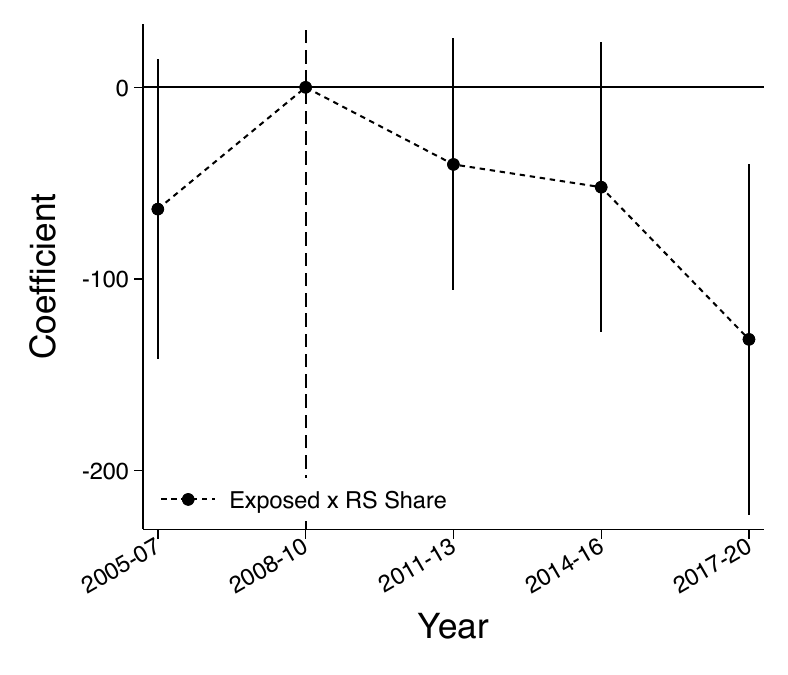}
		\caption{Cross-Term Coefficient, Cont. Measure}
	\end{subfigure}
    \vspace{-1em}
	\caption{Changes in Duration Gaps by Exposure --- Includes Outliers}
	\label{fig:duration gaps untrimmed}
	\floatfoot{Note: This figure reports regression results for equations (\ref{eq:duration gap regression}) and (\ref{eq:duration gap regression RS}) using an untrimmed version of the dependent variable. Duration gaps are constructed as in equation (\ref{eq:duration gap empirical}). In panel (a), estimates are presented as relative to non-exposed insurers; black circles represent the effects of all exposed insurers, red diamonds represent the effects of exposed RS insurers, and pink triangles represent the effects of exposed non-RS insurers. In panel (b), the black circles represent the difference between exposed RS and exposed non-RS insurers. In panel (c), estimates are presented as relative to non-exposed insurers for insurers with no RS exposure. In panel (d), the black circles represent the coefficient on the continuous RS exposure. The regressions are weighted by insurers' assets. Vertical lines represent 90\% confidence intervals using standard errors clustered at the insurer level.}
\end{figure}

\begin{figure}
    \centering
    \includegraphics[width=0.9\linewidth]{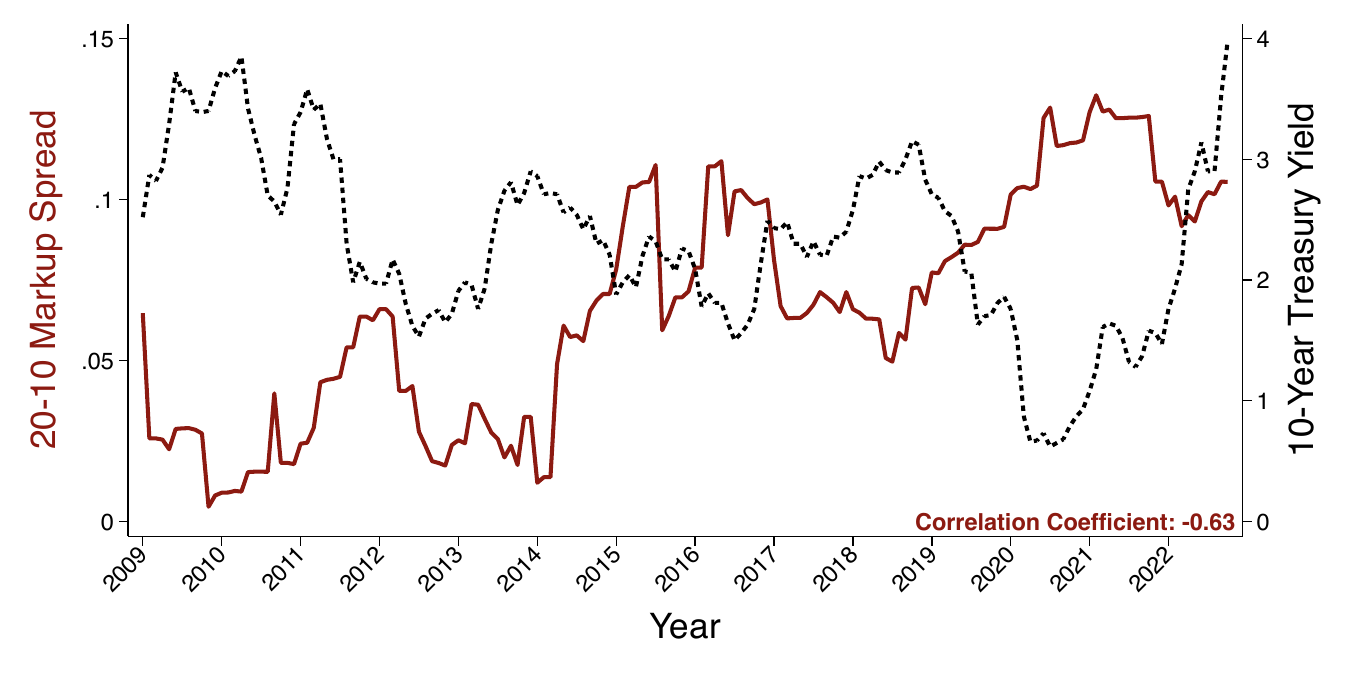}
    \caption{The Interaction between Product Pricing and Interest Rate Risk}
    \label{fig:markup spread infc}
    \floatfoot{Note: This figure plots the (20,10) markup spread (red, left axis) and the 10-year Treasury yield (black dotted, right axis) for each month between January 2009 and December 2022. When calculating the markup spread, averages are weighted by ordinary life insurance in force.}
\end{figure}

\begin{figure}
    \centering
    \includegraphics[width=0.9\linewidth]{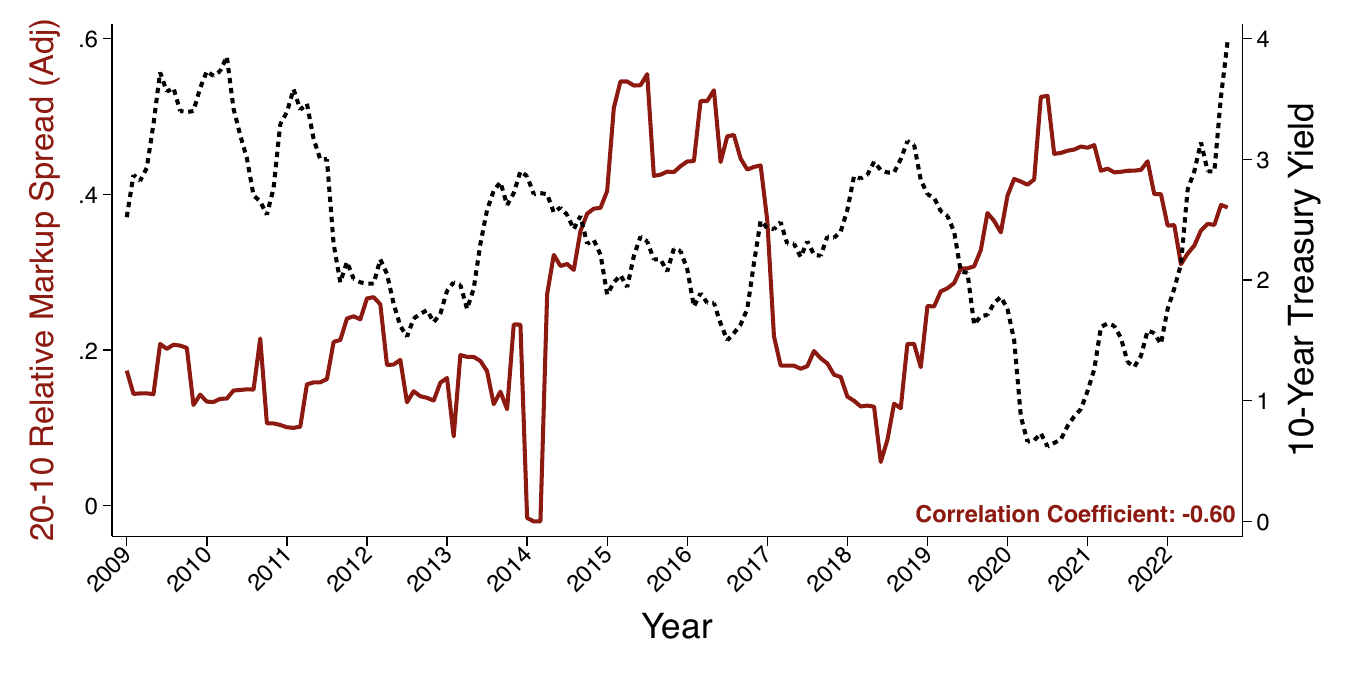}
    \caption{The Interaction between Product Pricing and Interest Rate Risk}
    \label{fig:markup spread vol}
    \floatfoot{Note: This figure plots the (20,10) markup spread (red, left axis) and the 10-year Treasury yield (black dotted, right axis) for each month between January 2009 and December 2022. When calculating the markup spread, averages are weighted by assets. We report the markup spread divided by interest rate volatility, which we estimate each year using a 24-month rolling window.}
\end{figure}

\begin{figure}[t!]
	\includegraphics[width = 0.8\textwidth]{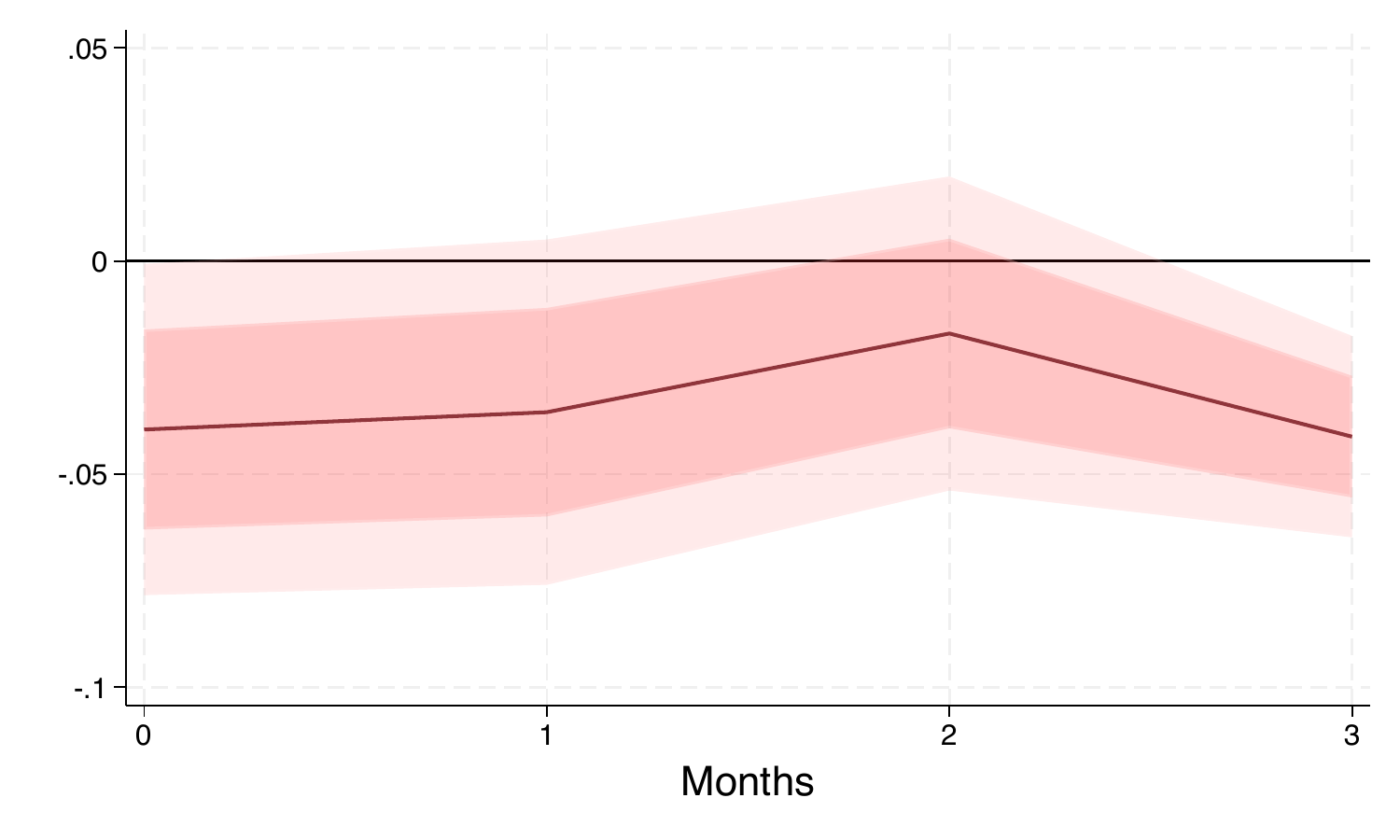} 
	\caption{Relative Price Responses to Long Rates Movements around FOMC Meetings}
	\label{fig:product price irf}
	\floatfoot{Note: This figure reports the $\beta_k^{\text{RS}}$ coefficients estimated from the local projection $\Delta_{3}^k \log \text{Price}_{ijt} = \beta_k \times \left(\Delta_{3} y^{(10)}_t\mid\text{FOMC}\right) \times \text{Exposed}_j \times \text{Long}_i + \beta_k^{\text{RS}} \times \left(\Delta_{3} y^{(10)}_t\mid\text{FOMC}\right) \times \text{Exposed}_j \times \text{RS}_j \times \text{Long}_i + \delta_{jt} + \delta_{it} + \delta_{ij} + \veps_{ijt}.$ The dependent variable $\Delta_{3}^k \log \text{Price}_{ijt}$ is the change in the log price of product $i$ issued by insurer $j$ from month $t-3$ to $t+k$. $\left(\Delta_{3} y^{(10)}_t\mid\text{FOMC}\right)$ is the total change in the 10-year Treasury yield over two-day windows around FOMC meetings as in \cite{li2024}, aggregated over the three-month period between month $t-3$ and month $t$. The coefficients for $k = 0$ capture the contemporaneous responses of quarterly log price changes to quarterly long rate shocks, while those for $k \in \{1, 2, 3\}$ capture longer-term price reactions. Standard errors are clustered at the product-time level, and the shaded areas indicate 68\% and 90\% confidence intervals. Consistent with Table \ref{table:price regression results baseline}, estimated $\{\beta_k^{\text{RS}}\}$ are negative, suggesting that exposed insurers with risk-sensitive VAs mark up long-term products more following negative innovations in the long rate. The responses persist for at least 3 months.}
\end{figure}

\begin{figure}
	\begin{subfigure}{0.49\textwidth}
		\includegraphics[width = \textwidth]{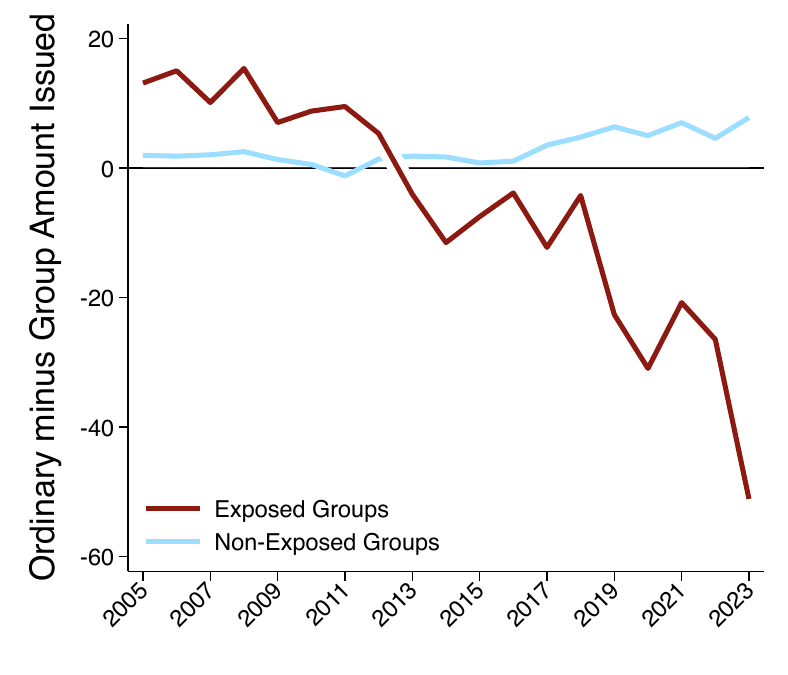}
		\caption{Ordinary Life}
	\end{subfigure}
	\begin{subfigure}{0.49\textwidth}
		\includegraphics[width = \textwidth]{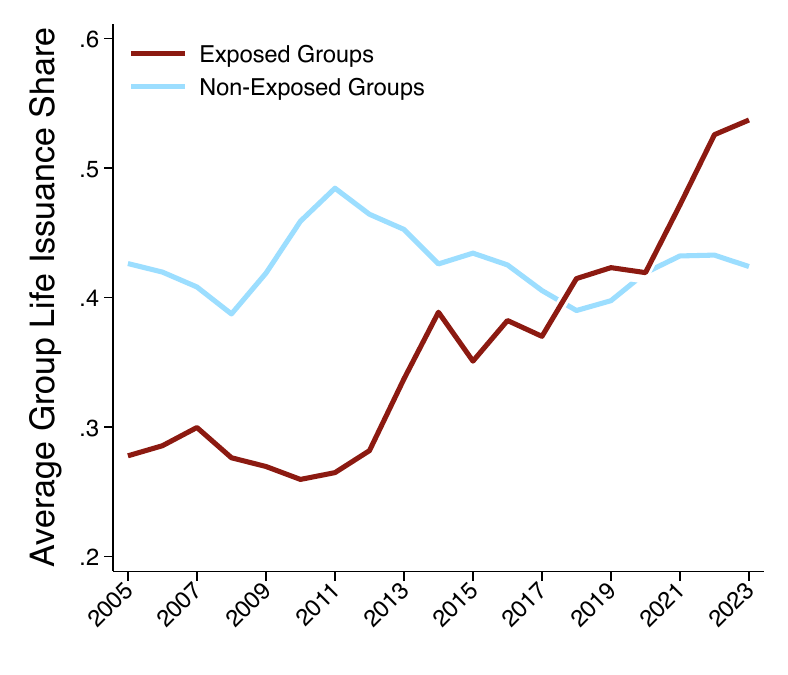}
		\caption{Group Life}
	\end{subfigure}
	\caption{Relative Product Issuance --- Unweighted}
	\label{app:fig:product issuance avg unweighted}
	\floatfoot{Note: This figure reports average ordinary life insurance issuance relative to average group life issuance [panel (a)] and average group life issuance shares [panel (b)] for exposed (red) and non-exposed (blue) insurers over time. Averages are unweighted. For panel (a), units are in billions of US dollars.}

\end{figure}

\begin{figure}
	\begin{subfigure}{0.49\textwidth}
		\includegraphics[width = \textwidth]{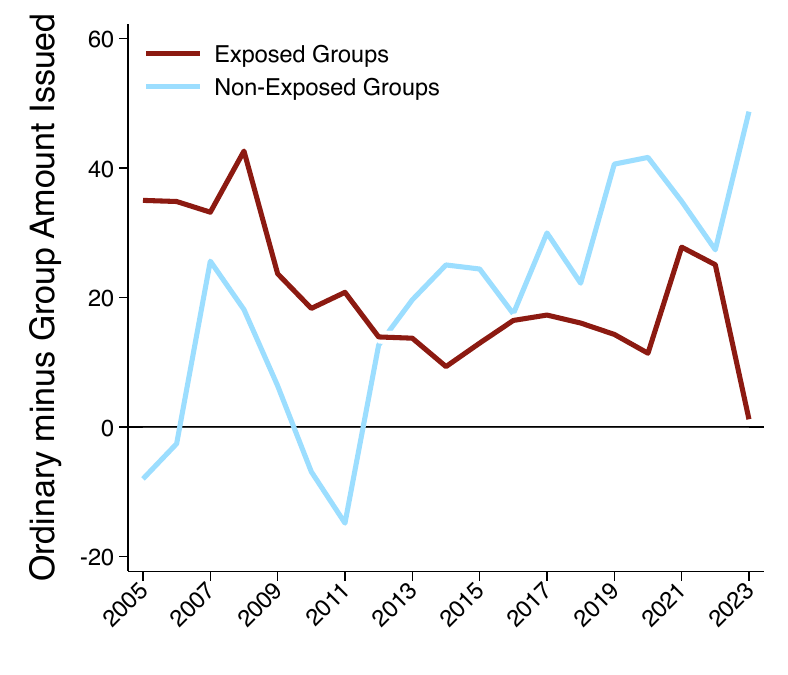}
		\caption{Ordinary Life}
	\end{subfigure}
	\begin{subfigure}{0.49\textwidth}
		\includegraphics[width = \textwidth]{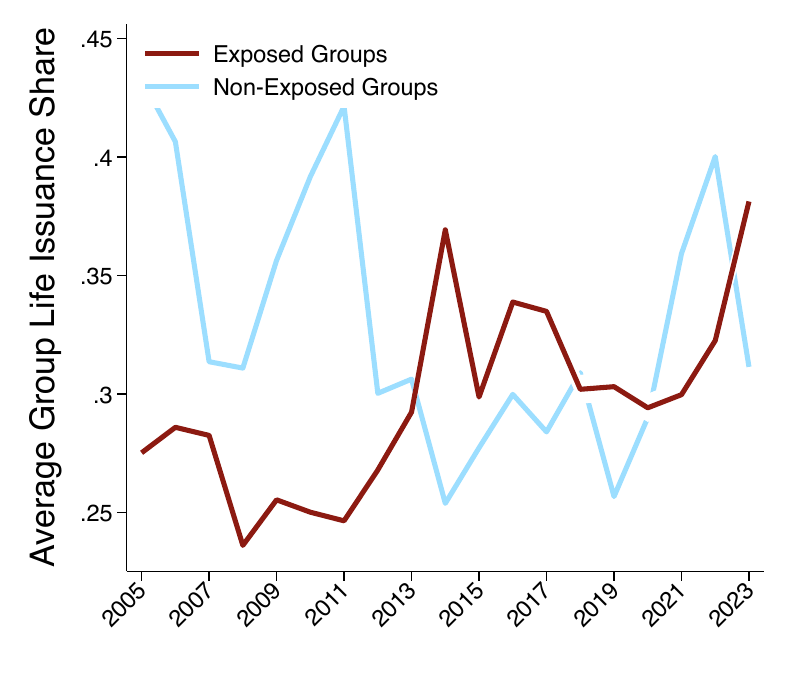}
		\caption{Group Life}
	\end{subfigure}
	\caption{Relative Product Issuance --- Excludes Metlife}
	\label{app:fig:product issuance avg no ML}
	\floatfoot{Note: This figure reports average ordinary life insurance issuance relative to average group life issuance [panel (a)] and average group life issuance shares [panel (b)] for exposed (red) and non-exposed (blue) insurers over time. Averages are weighted by assets within each class of insurers. For panel (a), units are in billions of US dollars. Metlife is excluded from calculations.}

\end{figure}

\begin{figure}[t!]
	\begin{subfigure}{0.45\textwidth}
		\includegraphics[width = \textwidth]{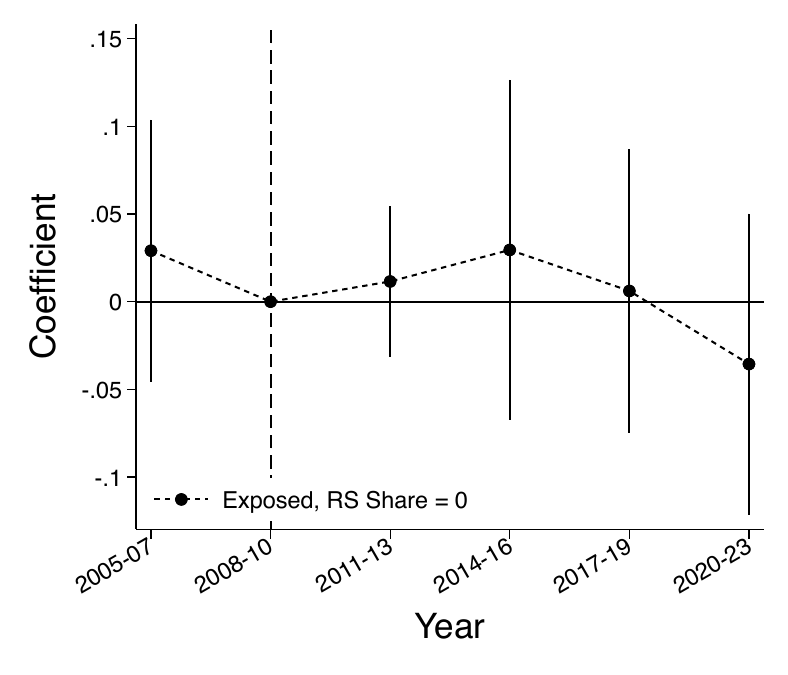}
		\caption{Non-RS Effects}
	\end{subfigure}
	\begin{subfigure}{0.45\textwidth}
		\includegraphics[width = \textwidth]{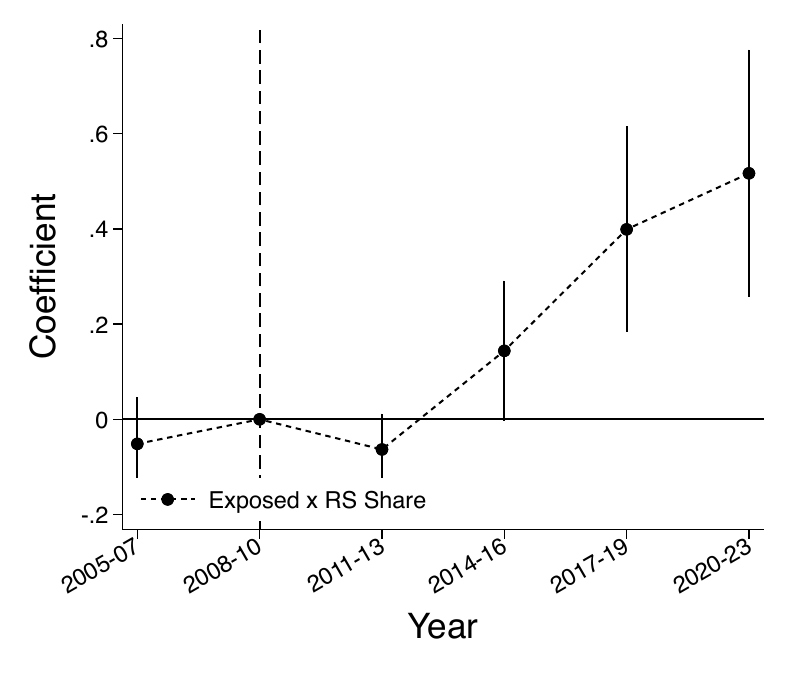}
		\caption{Cross-Term Coefficients}
	\end{subfigure}
    \vspace{-1em}
	\caption{Liability Rebalancing --- Continuous Exposure}
	\label{app:fig:liability rebalancing regs cont}
	\floatfoot{Note: This figure reports regression results for equations (\ref{eq:issuance reg}) and (\ref{eq:issuance reg rs}) using the continuous exposure measure $\text{RSshare}_j$. The dependent variable is the share of insurer $j$'s insurance coverage issued in the form of group life insurance. Panel (a) plots the coefficient on $\text{Exposed}_j$, which captures the difference between exposed insurers with $\text{RSshare}_j=0$ and non-exposed insurers. Panel (b) plots the coefficient on $\text{Exposed}_j \times \text{RSshare}_j$, which captures how the exposed-insurer effect varies with the share of variable annuity reserves that are risk sensitive. For an exposed insurer with $\text{RSshare}_j=s$, the implied effect equals the coefficient in panel (a) plus $s$ times the coefficient in panel (b). The regressions are weighted by insurers' assets. Vertical lines represent 90\% confidence intervals using standard errors clustered at the insurer level.}
\end{figure}

\begin{figure}[t!]
	\begin{subfigure}{0.45\textwidth}
		\includegraphics[width = \textwidth]{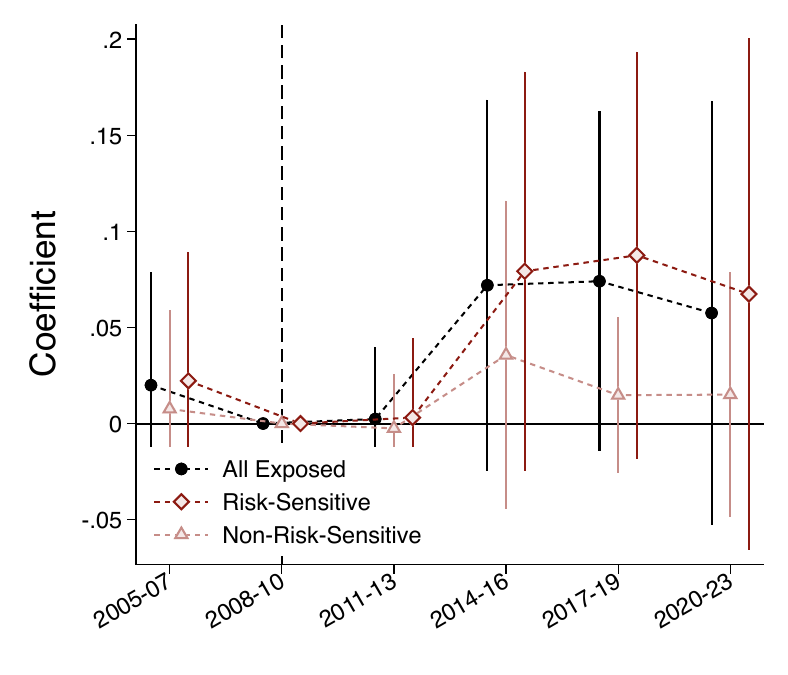}
		\caption{Total Effects}
	\end{subfigure}
	\begin{subfigure}{0.45\textwidth}
		\includegraphics[width = \textwidth]{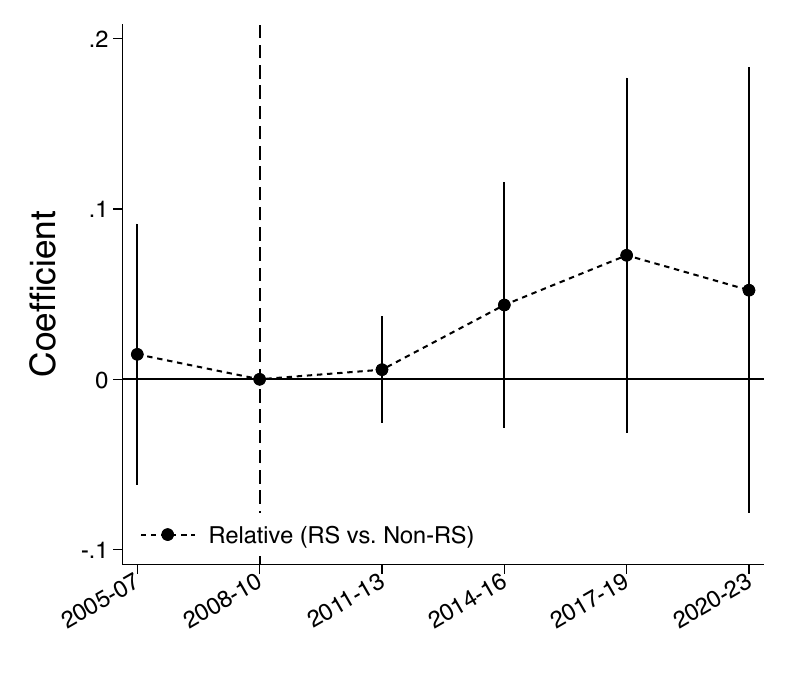}
		\caption{Relative Effects}
	\end{subfigure}
    \vspace{-1em}
	\caption{Liability Rebalancing --- No MetLife}
	\label{app:fig:liability rebalancing regs noML}
	\floatfoot{Note: This figure reports regression results for equations (\ref{eq:issuance reg}) and (\ref{eq:issuance reg rs}) excluding MetLife. The dependent variable is the share of insurer $j$'s insurance coverage issued in the form of group life insurance. In panel (a), estimates are presented as relative to non-exposed insurers; black circles represent the effects of all exposed insurers, red diamonds represent the effects of exposed RS insurers, and pink triangles represent the effects of exposed non-RS insurers. In panel (b), the black circles represent the difference between exposed RS and exposed non-RS insurers. The regressions are weighted by insurers' assets. Vertical lines represent 90\% confidence intervals using standard errors clustered at the insurer level.}
\end{figure}

\begin{figure}[t!]
	\begin{subfigure}{0.45\textwidth}
		\includegraphics[width = \textwidth]{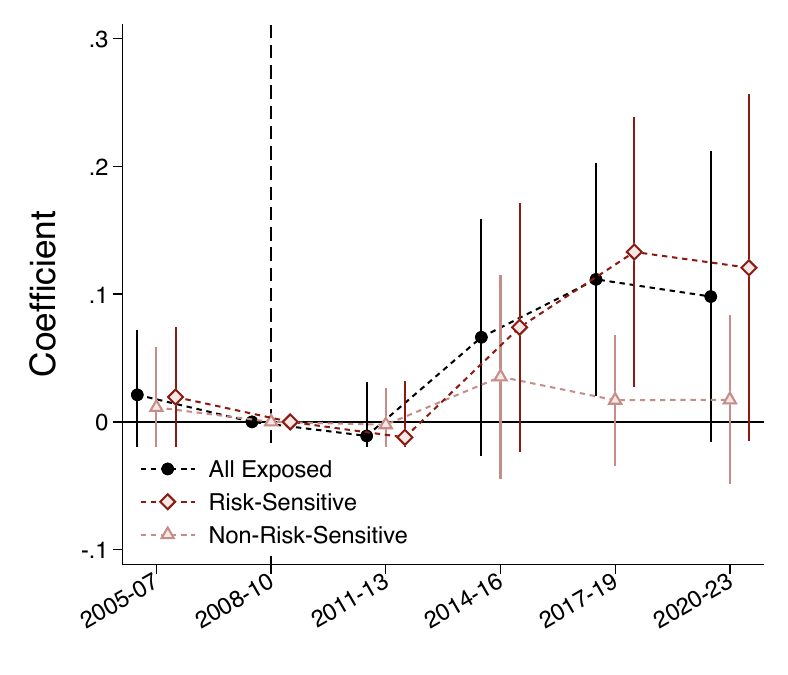}
		\caption{Total Effects}
	\end{subfigure}
	\begin{subfigure}{0.45\textwidth}
		\includegraphics[width = \textwidth]{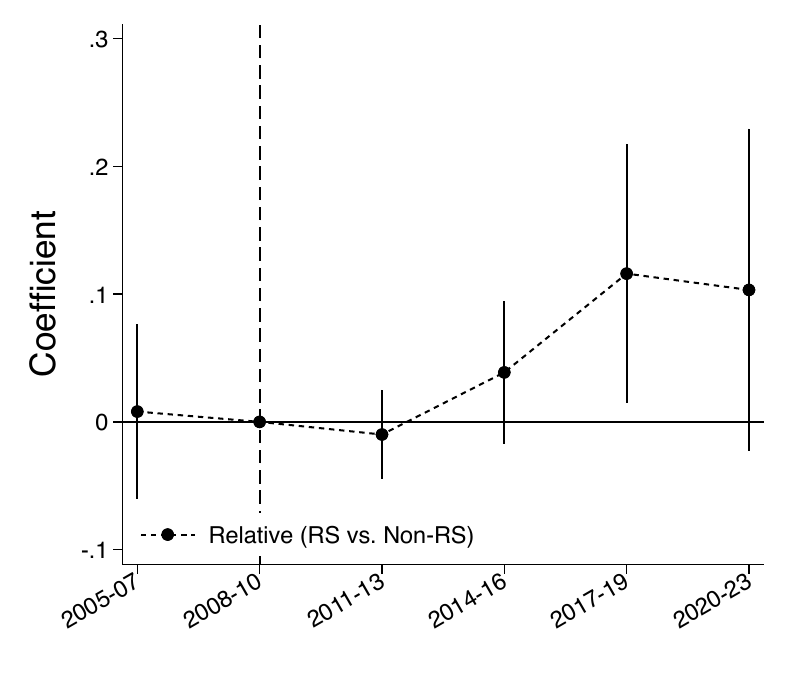}
		\caption{Relative Effects}
	\end{subfigure}
    \vspace{-1em}
	\caption{Liability Rebalancing --- Added Controls}
	\label{app:fig:liability rebalancing regs controls}
	\floatfoot{Note: This figure reports regression results for equations (\ref{eq:issuance reg}) and (\ref{eq:issuance reg rs}) adding time-varying size (log assets) and leverage controls. The dependent variable is the share of insurer $j$'s insurance coverage issued in the form of group life insurance. In panel (a), estimates are presented as relative to non-exposed insurers; black circles represent the effects of all exposed insurers, red diamonds represent the effects of exposed RS insurers, and pink triangles represent the effects of exposed non-RS insurers. In panel (b), the black circles represent the difference between exposed RS and exposed non-RS insurers. The regressions are weighted by insurers' assets. Vertical lines represent 90\% confidence intervals using standard errors clustered at the insurer level.}
\end{figure}

\begin{figure}[t!]
	\begin{subfigure}{0.45\textwidth}
		\includegraphics[width = \textwidth]{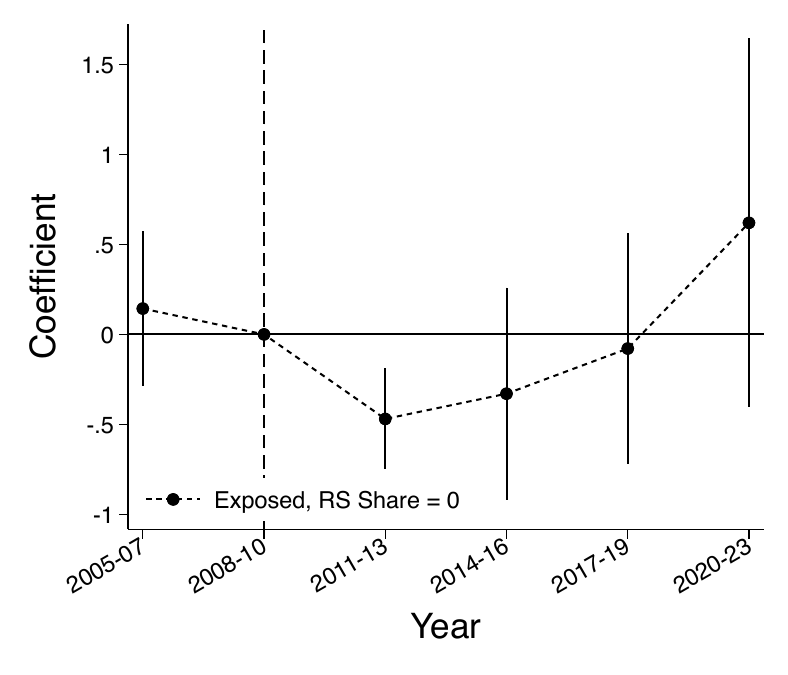}
		\caption{Non-RS Effects}
	\end{subfigure}
	\begin{subfigure}{0.45\textwidth}
		\includegraphics[width = \textwidth]{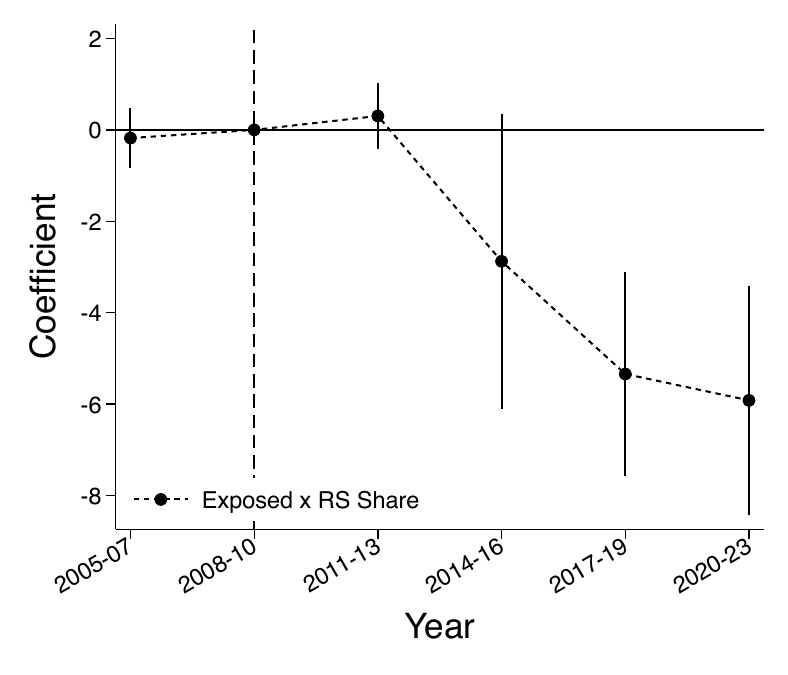}
		\caption{Cross-Term Coefficients}
	\end{subfigure}
    \vspace{-1em}
	\caption{Liability Rebalancing --- Ordinary Contraction, Continuous Exposure}
	\label{app:fig:liability rebalancing regs cont OL}
	\floatfoot{Note: This figure reports regression results for equations (\ref{eq:issuance reg}) and (\ref{eq:issuance reg rs}) using the continuous exposure measure $\text{RSshare}_j$ and log ordinary life insurance coverage issued as the dependent variable. In panel (a), estimates are presented as relative to non-exposed insurers for insurers with no RS exposure. In panel (b), the black circles represent the coefficient on the continuous RS exposure. The regressions are weighted by insurers' assets. Vertical lines represent 90\% confidence intervals using standard errors clustered at the insurer level.}
\end{figure}

\begin{figure}[t!]
	\begin{subfigure}{0.49\textwidth}
		\includegraphics[width = \textwidth]{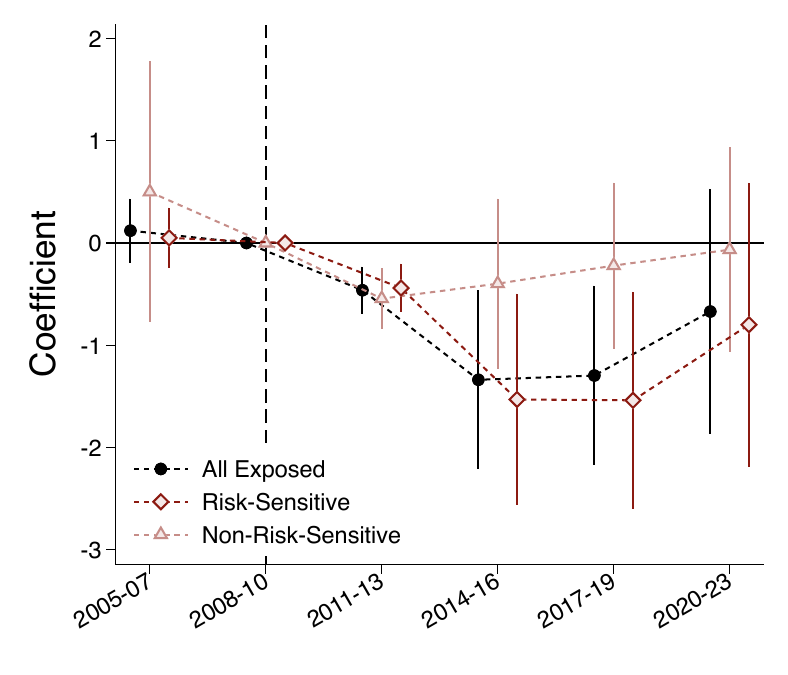}
		\caption{Total Effects}
	\end{subfigure}
	\begin{subfigure}{0.49\textwidth}
		\includegraphics[width = \textwidth]{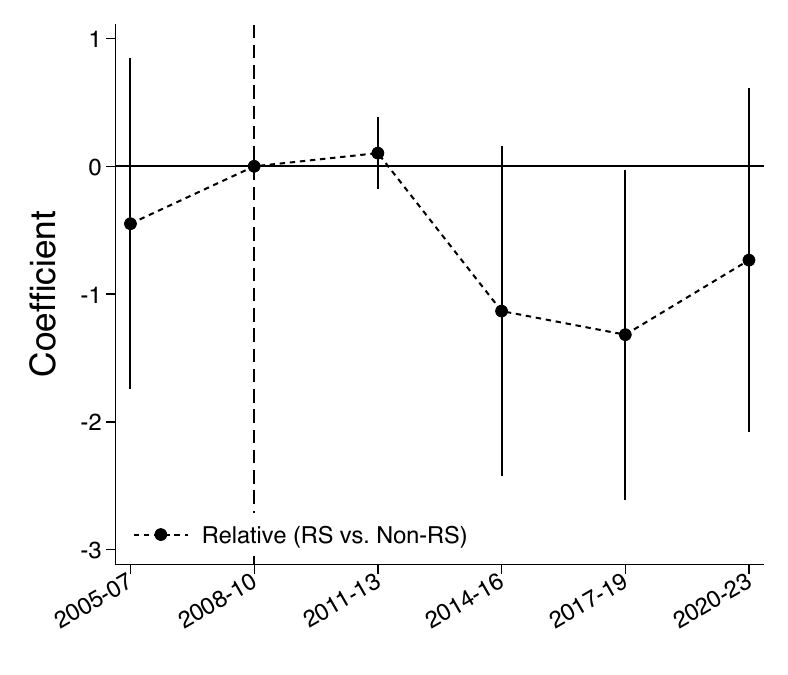}
		\caption{Relative Effects}
	\end{subfigure}
    \begin{subfigure}{0.49\textwidth}
		\includegraphics[width = \textwidth]{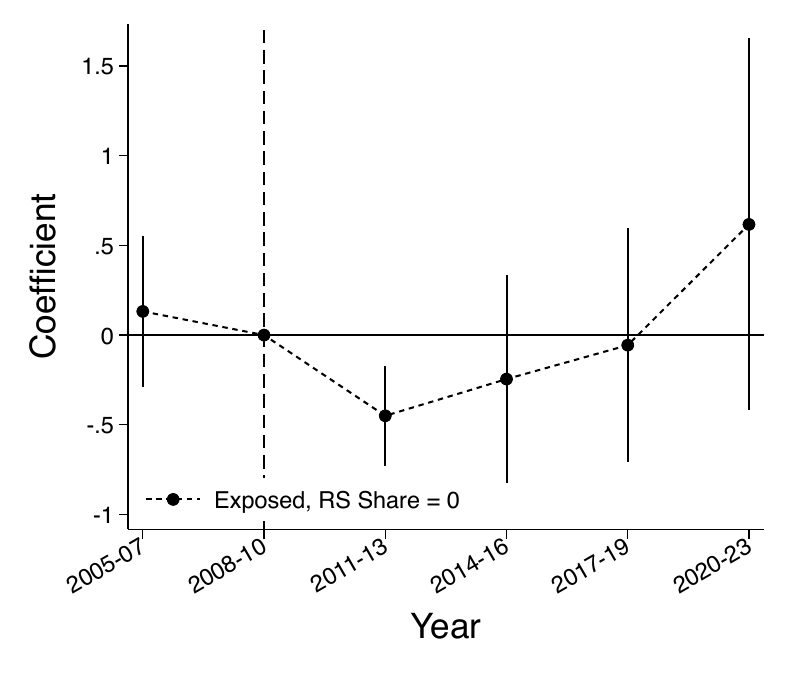}
		\caption{Non-RS Effects, Cont. Measure}
	\end{subfigure}
	\begin{subfigure}{0.49\textwidth}
		\includegraphics[width = \textwidth]{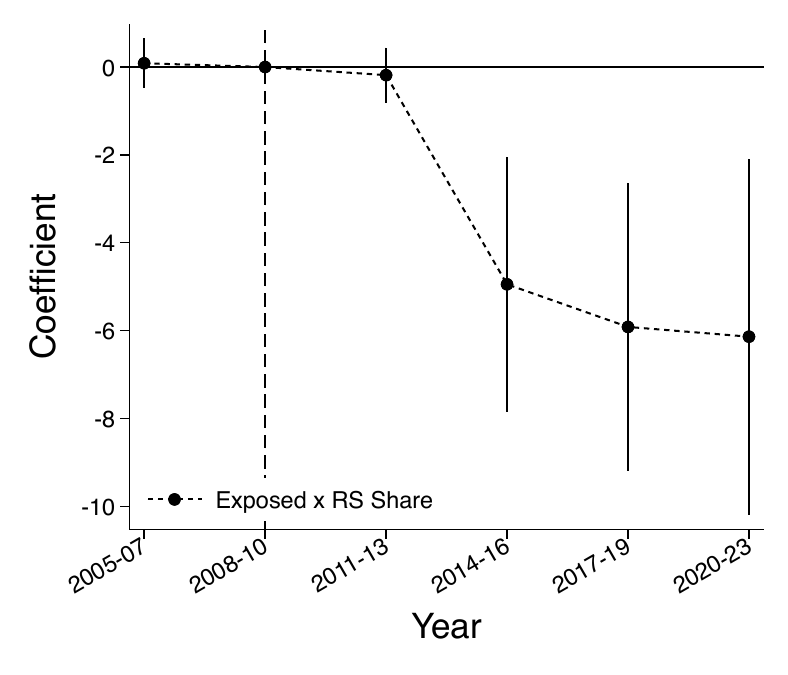}
		\caption{Cross-Term Coefficient, Cont. Measure}
	\end{subfigure}
    \vspace{-1em}
	\caption{Liability Rebalancing --- Ordinary Contraction, No MetLife}
	\label{app:fig:liability rebalancing regs noML OL}
	\floatfoot{Note: This figure reports regression results for equations (\ref{eq:issuance reg}) and (\ref{eq:issuance reg rs}) excluding MetLife and log ordinary life insurance coverage issued as the dependent variable. In panel (a), estimates are presented as relative to non-exposed insurers; black circles represent the effects of all exposed insurers, red diamonds represent the effects of exposed RS insurers, and pink triangles represent the effects of exposed non-RS insurers. In panel (b), the black circles represent the difference between exposed RS and exposed non-RS insurers. In panel (c), estimates are presented as relative to non-exposed insurers for insurers with no RS exposure. In panel (d), the black circles represent the coefficient on the continuous RS exposure. The regressions are weighted by insurers' assets. Vertical lines represent 90\% confidence intervals using standard errors clustered at the insurer level.}
\end{figure}

\begin{figure}[t!]
	\begin{subfigure}{0.49\textwidth}
		\includegraphics[width = \textwidth]{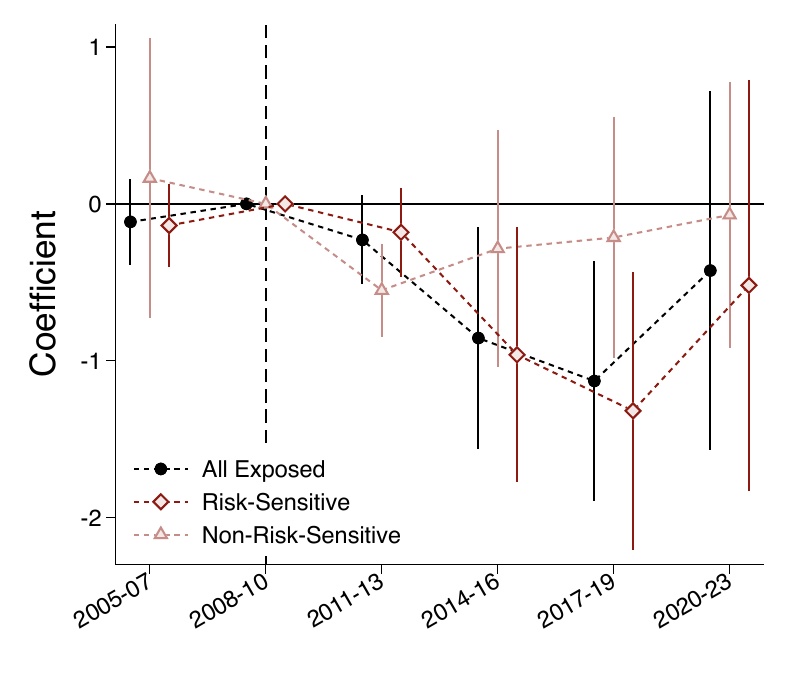}
		\caption{Total Effects}
	\end{subfigure}
	\begin{subfigure}{0.49\textwidth}
		\includegraphics[width = \textwidth]{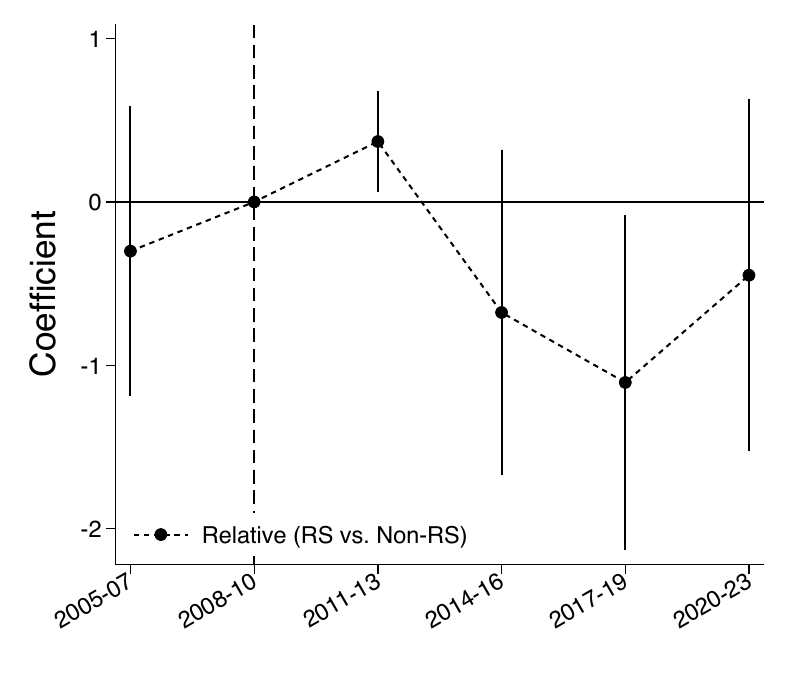}
		\caption{Relative Effects}
	\end{subfigure}
    \begin{subfigure}{0.49\textwidth}
		\includegraphics[width = \textwidth]{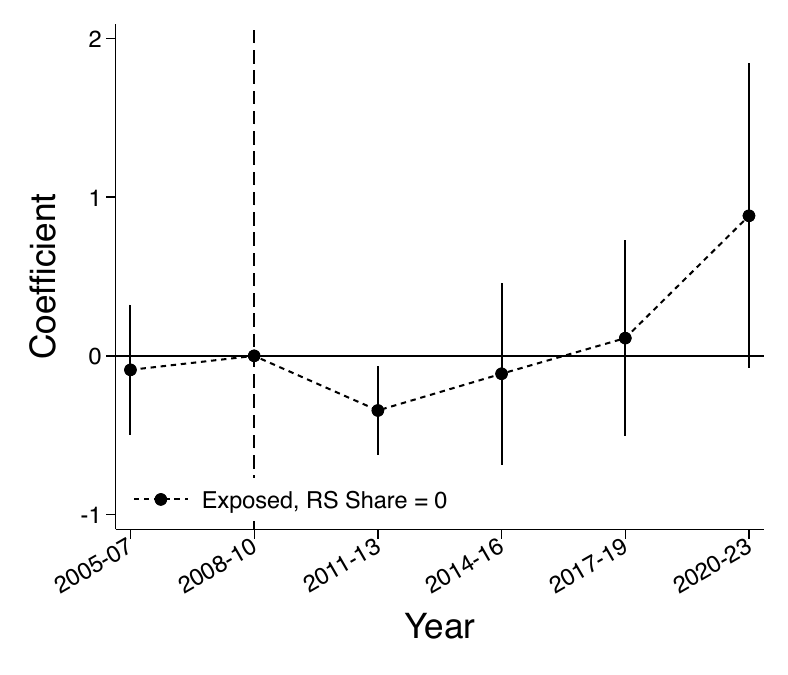}
		\caption{Non-RS Effects, Cont. Measure}
	\end{subfigure}
	\begin{subfigure}{0.49\textwidth}
		\includegraphics[width = \textwidth]{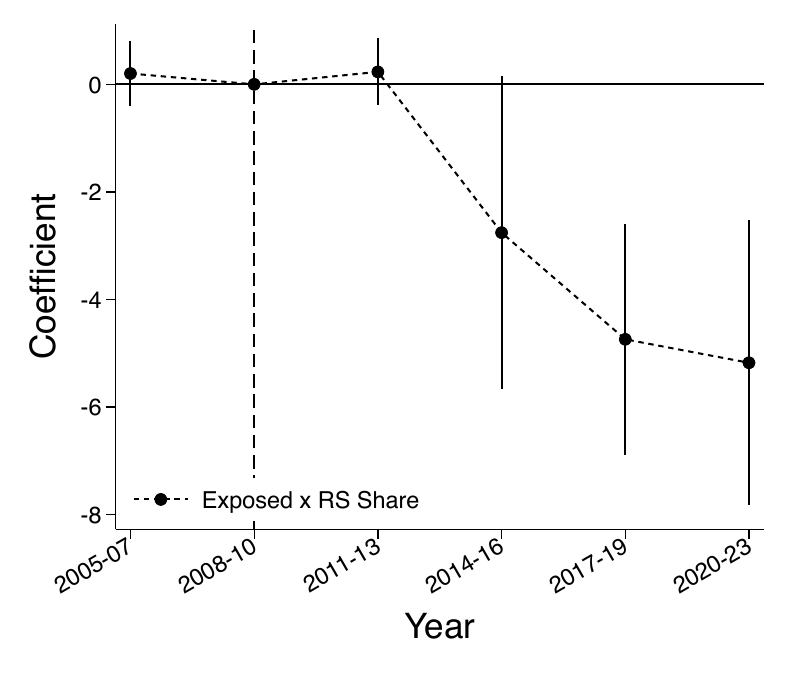}
		\caption{Cross-Term Coefficient, Cont. Measure}
	\end{subfigure}
    \vspace{-1em}
	\caption{Liability Rebalancing --- Ordinary Contraction, Added Controls}
	\label{app:fig:liability rebalancing regs controls OL}
	\floatfoot{Note: This figure reports regression results for equations (\ref{eq:issuance reg}) and (\ref{eq:issuance reg rs}) adding time-varying size (log assets) and leverage controls. The dependent variable is the log ordinary life insurance coverage issued. In panel (a), estimates are presented as relative to non-exposed insurers; black circles represent the effects of all exposed insurers, red diamonds represent the effects of exposed RS insurers, and pink triangles represent the effects of exposed non-RS insurers. In panel (b), the black circles represent the difference between exposed RS and exposed non-RS insurers. In panel (c), estimates are presented as relative to non-exposed insurers for insurers with no RS exposure. In panel (d), the black circles represent the coefficient on the continuous RS exposure. The regressions are weighted by insurers' assets. Vertical lines represent 90\% confidence intervals using standard errors clustered at the insurer level.}
\end{figure}

\begin{figure}[t!]
	\begin{subfigure}{0.49\textwidth}
		\includegraphics[width = \textwidth]{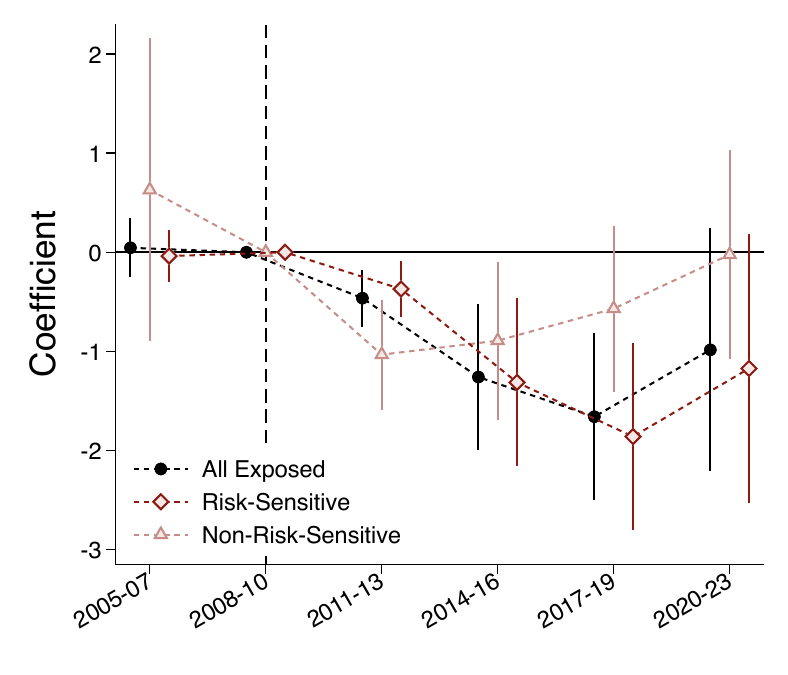}
		\caption{Ordinary -- Total Effects}
	\end{subfigure}
	\begin{subfigure}{0.49\textwidth}
		\includegraphics[width = \textwidth]{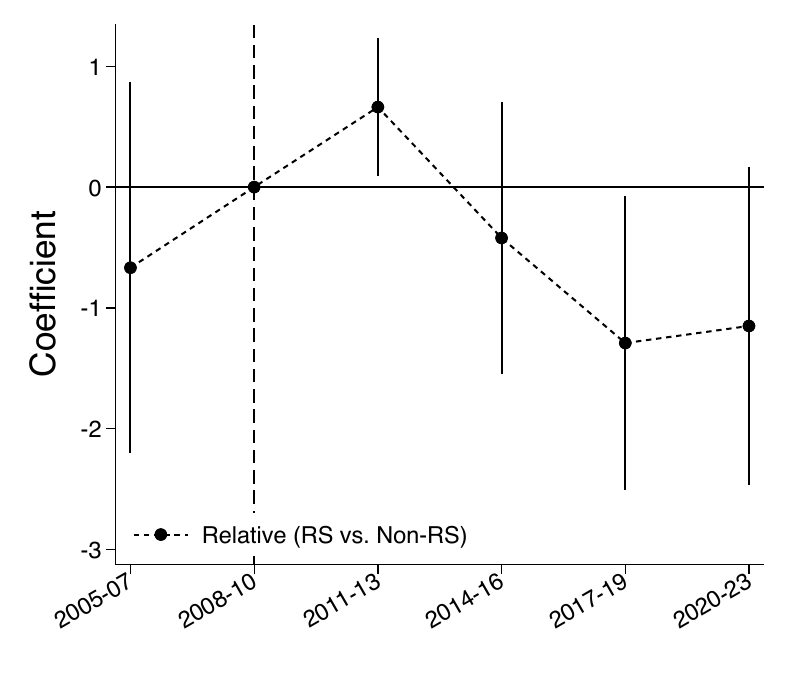}
		\caption{Ordinary -- Relative Effects}
	\end{subfigure}
    \begin{subfigure}{0.49\textwidth}
		\includegraphics[width = \textwidth]{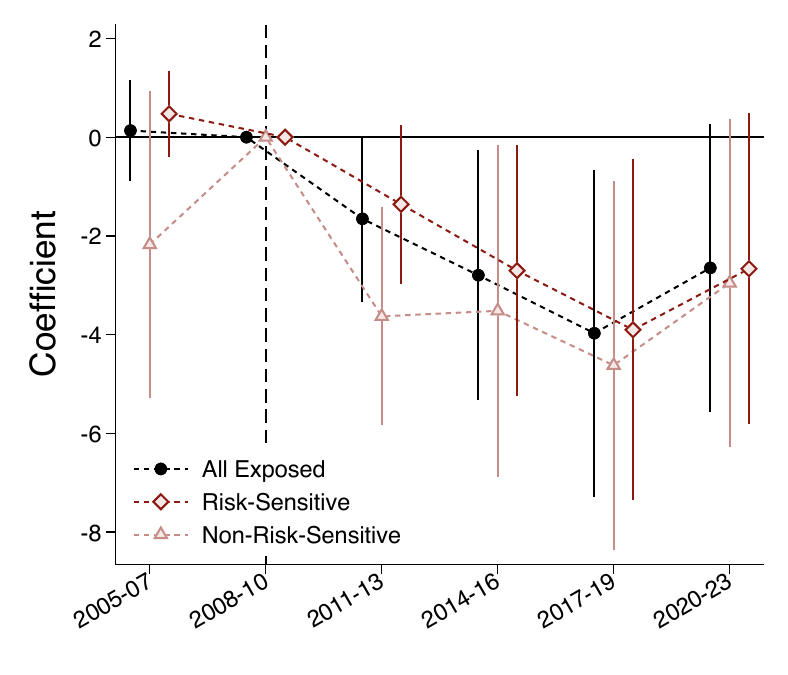}
		\caption{Group -- Total Effects}
	\end{subfigure}
	\begin{subfigure}{0.49\textwidth}
		\includegraphics[width = \textwidth]{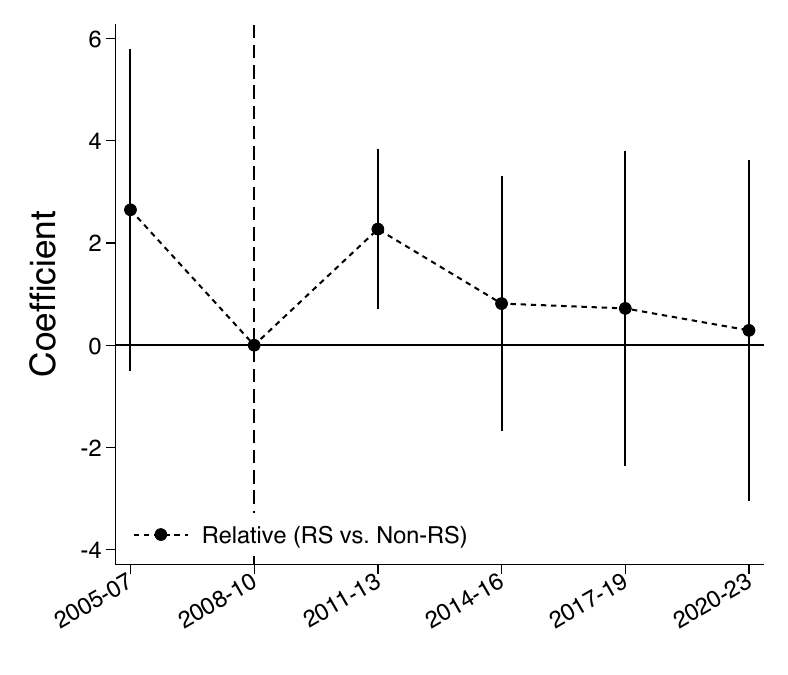}
		\caption{Group -- Relative Effects}
	\end{subfigure}
    \vspace{-1em}
	\caption{Liability Rebalancing --- Ordinary \& Group Comparison}
	\label{app:fig:liability rebalancing regs both types}
	\floatfoot{Note: This figure reports regression results for equations (\ref{eq:issuance reg}) and (\ref{eq:issuance reg rs}) using the inverse hyperbolic sine transform of ordinary [(a) and (b)] and group [(c) and (d)] life issuance. In panels (a) and (c), estimates are presented as relative to non-exposed insurers; black circles represent the effects of all exposed insurers, red diamonds represent the effects of exposed RS insurers, and pink triangles represent the effects of exposed non-RS insurers. In panels (b) and (d), the black circles represent the difference between exposed RS and exposed non-RS insurers. The regressions are weighted by insurers' assets. Vertical lines represent 90\% confidence intervals using standard errors clustered at the insurer level.}
\end{figure}

\begin{figure}
	\begin{subfigure}{0.32\textwidth}
		\includegraphics[width = \textwidth]{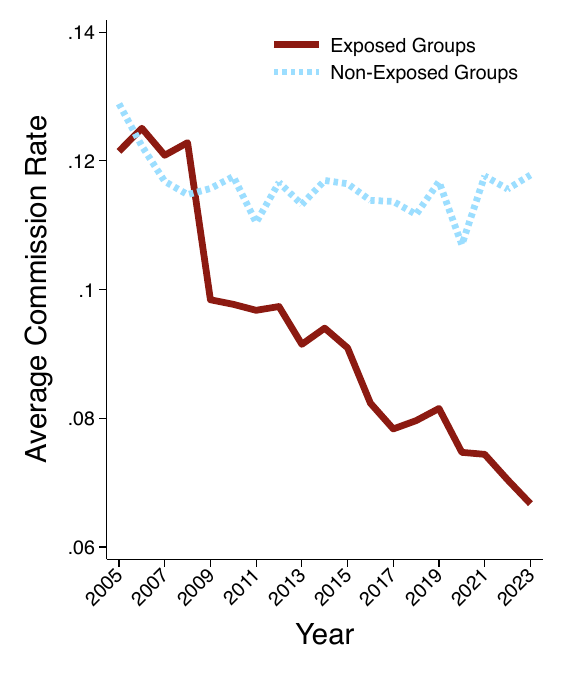}
		\caption{All Policies}
	\end{subfigure}
	\begin{subfigure}{0.32\textwidth}
		\includegraphics[width = \textwidth]{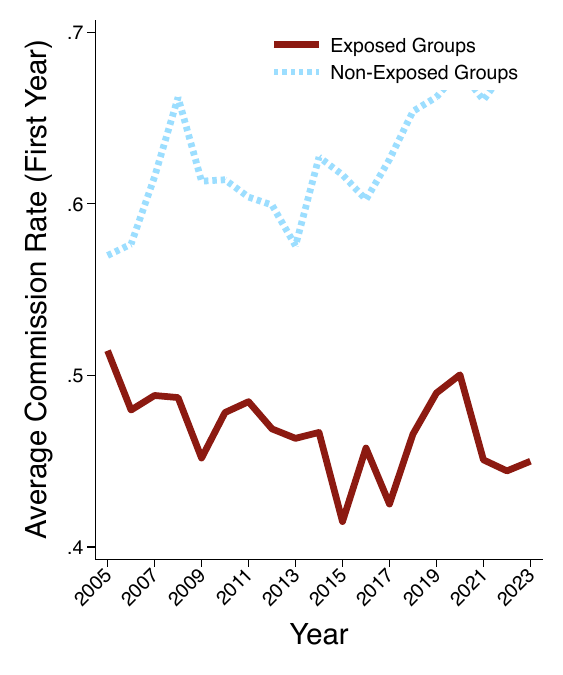}
		\caption{New Policies}
	\end{subfigure}
    \begin{subfigure}{0.32\textwidth}
		\includegraphics[width = \textwidth]{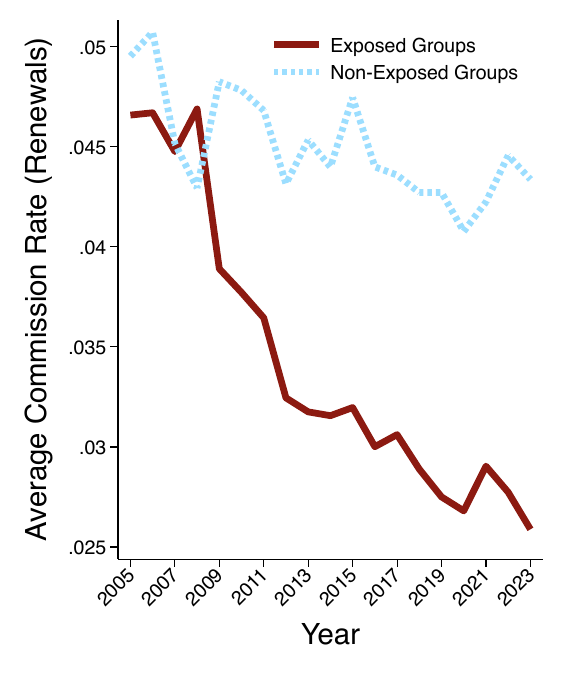}
		\caption{Renewals}
	\end{subfigure}

	\caption{Average Commission Rates by Exposure Group}
	\label{app:fig:commissions}
	\floatfoot{Note: This figure reports average commission rates for exposed (red) and non-exposed (blue) insurance groups from 2005 to 2023. Panel (a) reports total commission rates, panel (b) reports commissions on policies issued in the current year, and panel (c) reports commissions on policy renewals. Commission rates are calculated as direct commissions paid to agents divided by direct premium revenues. The data are winsorized at the 1\% and 99\% level to avoid outliers.}

\end{figure}

\begin{figure}
    \begin{subfigure}{0.32\textwidth}
		\includegraphics[width = \textwidth]{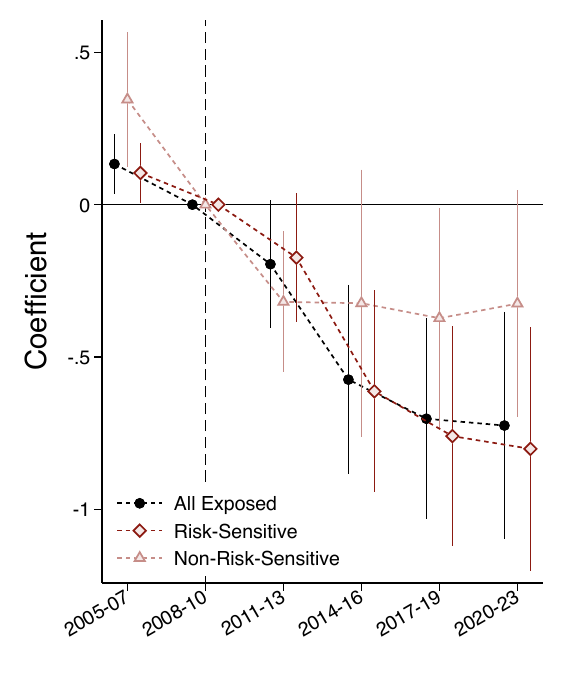}
		\caption{All (Total)}
	\end{subfigure}
	\begin{subfigure}{0.32\textwidth}
		\includegraphics[width = \textwidth]{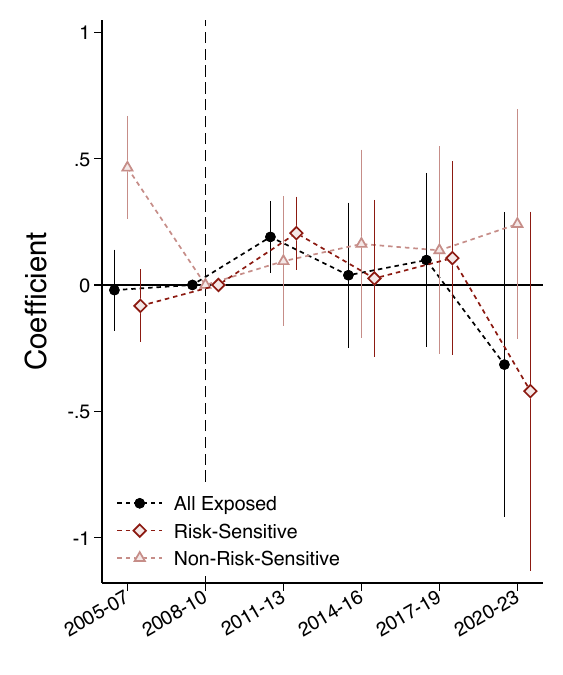}
		\caption{New (Total)}
	\end{subfigure}
    \begin{subfigure}{0.32\textwidth}
		\includegraphics[width = \textwidth]{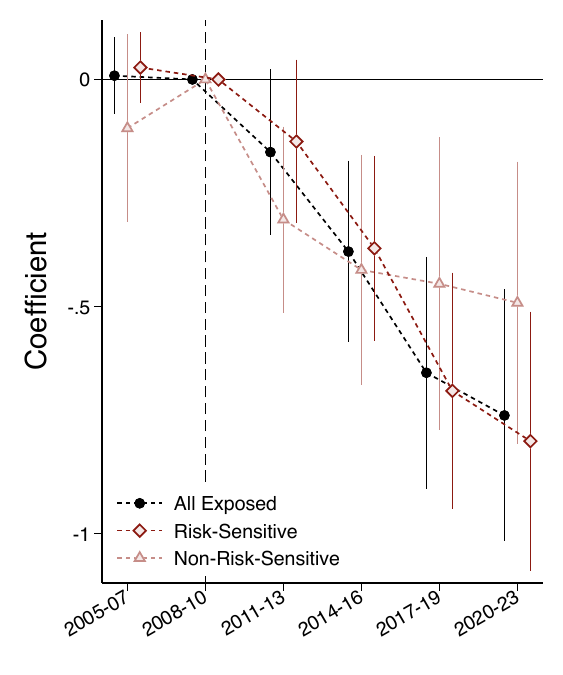}
		\caption{Renewals (Total)}
	\end{subfigure}
    \begin{subfigure}{0.32\textwidth}
		\includegraphics[width = \textwidth]{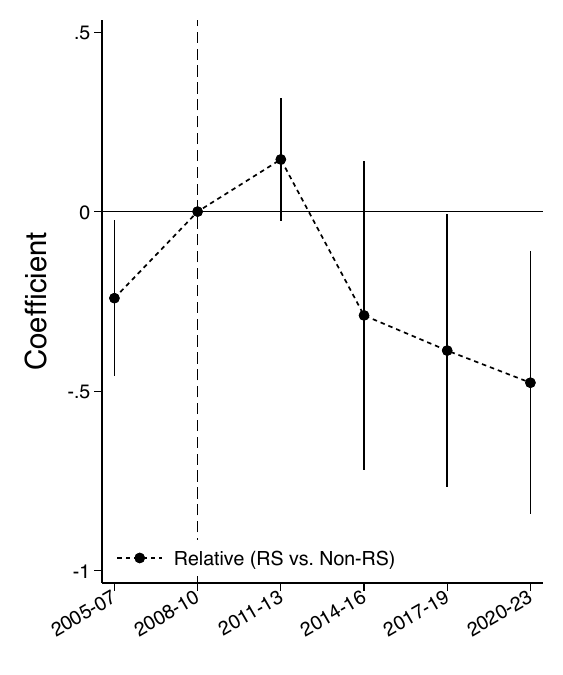}
		\caption{All (Relative)}
	\end{subfigure}
	\begin{subfigure}{0.32\textwidth}
		\includegraphics[width = \textwidth]{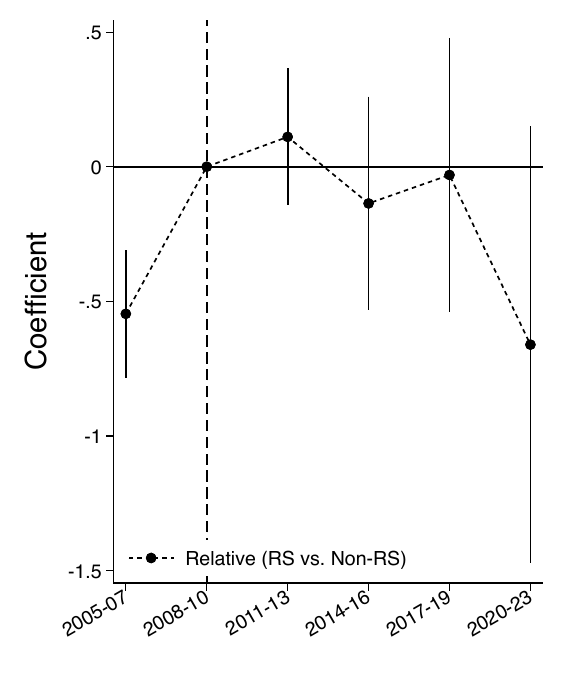}
		\caption{New (Relative)}
	\end{subfigure}
    \begin{subfigure}{0.32\textwidth}
		\includegraphics[width = \textwidth]{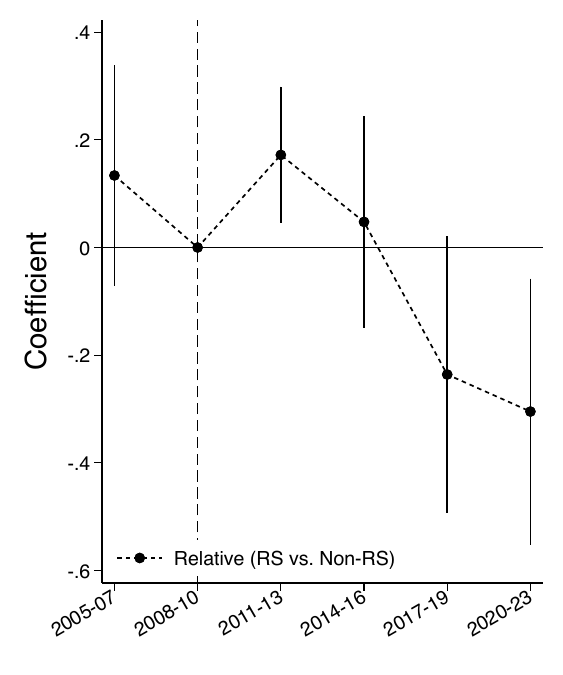}
		\caption{Renewals (Relative)}
	\end{subfigure}
	\caption{Regression Results by Commission Type}
	\label{app:fig:commissions regressions}
	\floatfoot{Note: This figure reports regression results for equations (\ref{eq:issuance reg}) and (\ref{eq:issuance reg rs}) using the log of total commission rates, new policy commission rates, and renewal rates as dependent variables. In panels (a)-(c), estimates are presented as relative to non-exposed insurers; black circles represent the effects of all exposed insurers, red diamonds represent the effects of exposed RS insurers, and pink triangles represent the effects of exposed non-RS insurers. In panels (d)-(f), the black circles represent the difference between exposed RS and exposed non-RS insurers. The regressions are weighted by insurers' assets. Vertical lines represent 90\% confidence intervals using standard errors clustered at the insurer level.}

\end{figure}

\begin{figure}[t!]
	\begin{subfigure}{0.49\textwidth}
		\includegraphics[width = \textwidth]{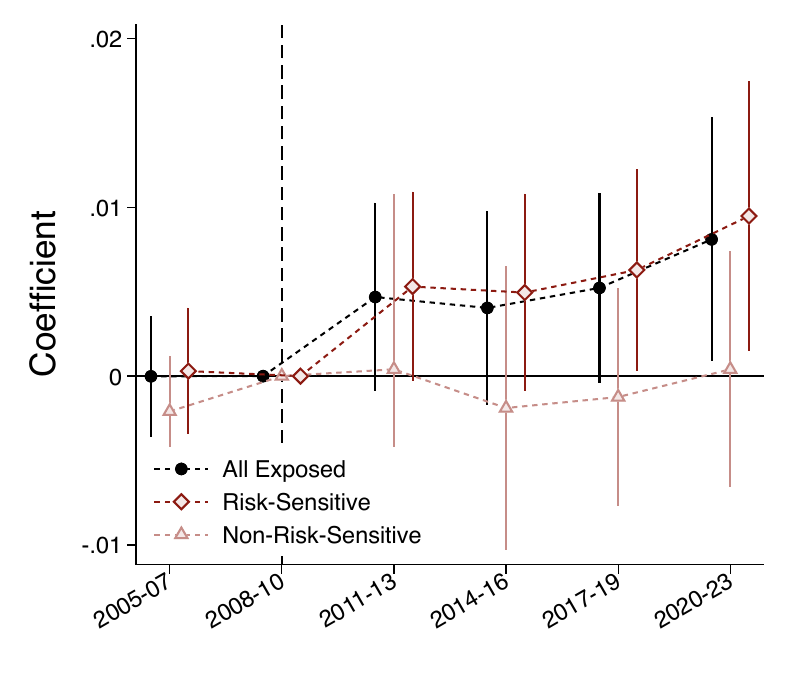}
		\caption{Total Effects}
	\end{subfigure}
	\begin{subfigure}{0.49\textwidth}
		\includegraphics[width = \textwidth]{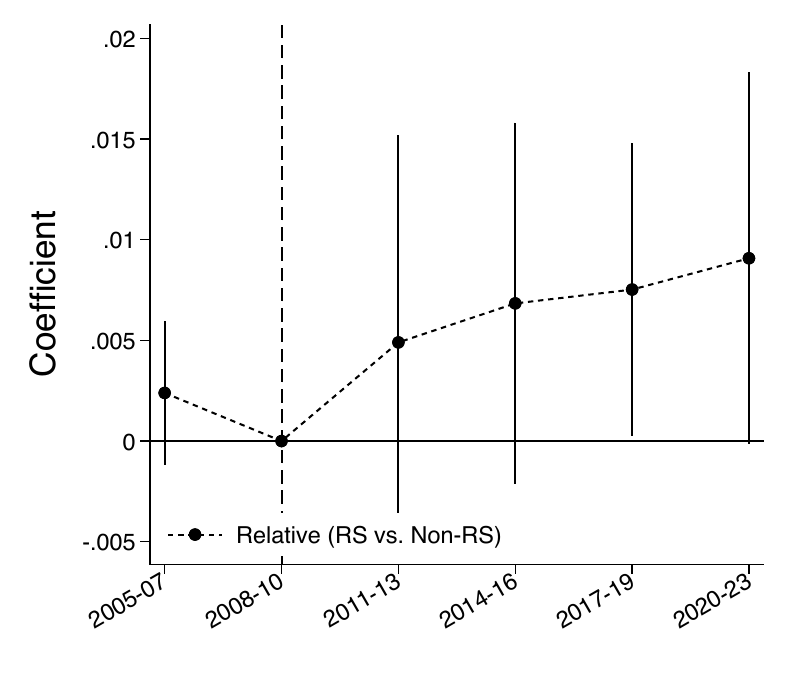}
		\caption{Relative Effects}
	\end{subfigure}
    \begin{subfigure}{0.49\textwidth}
		\includegraphics[width = \textwidth]{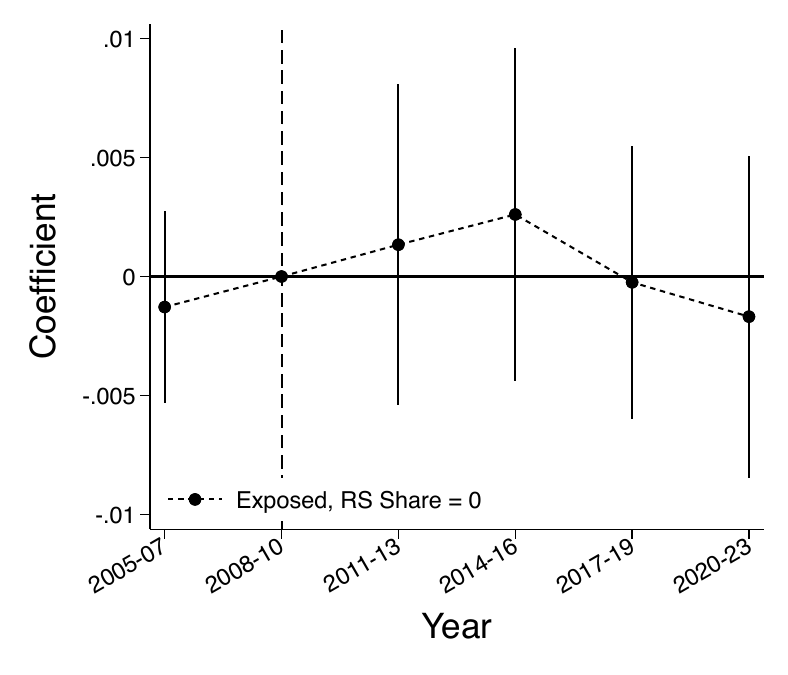}
		\caption{Non-RS Effects, Cont. Measure}
	\end{subfigure}
	\begin{subfigure}{0.49\textwidth}
		\includegraphics[width = \textwidth]{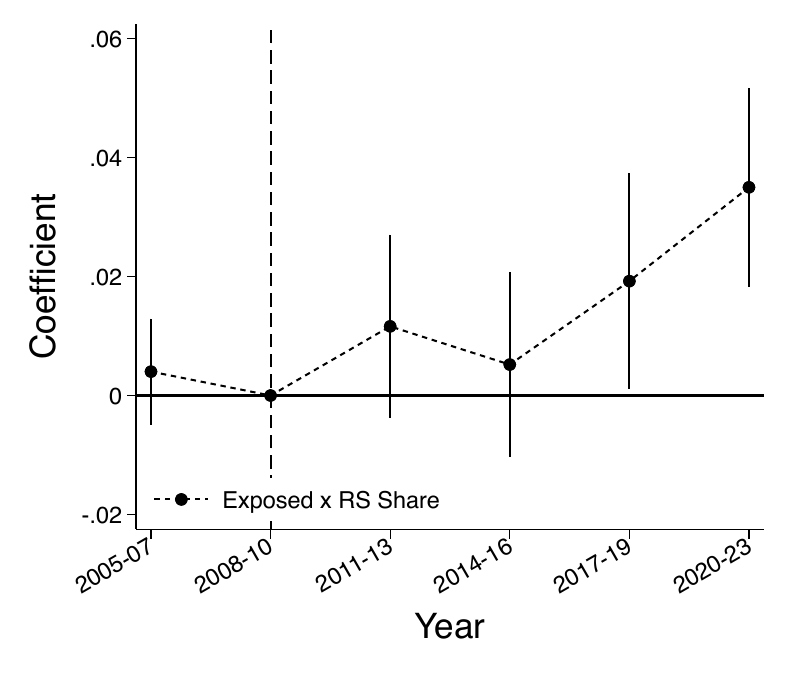}
		\caption{Cross-term Effects, Cont. Measure}
	\end{subfigure}
    \vspace{-1em}
	\caption{Liability Rebalancing --- Ordinary Life Lapsation}
	\label{app:fig:lapse rates regs}
	\floatfoot{Note: This figure reports regression results for equations (\ref{eq:issuance reg}) and (\ref{eq:issuance reg rs}) using ordinary life insurance lapsation rates as the dependent variable. Lapse rates are defined as total amounts lapsed in year $t$ divided by gross amount of insurance in force in year $t-1$. We drop observations in which lapse rates are greater than 1, which corresponds to 0.5\% of the sample. In panels (a) and (c), estimates are presented as relative to non-exposed insurers; black circles represent the effects of all exposed insurers, red diamonds represent the effects of exposed RS insurers, and pink triangles represent the effects of exposed non-RS insurers. In panels (b) and (d), the black circles represent the difference between exposed RS and exposed non-RS insurers. The regressions are weighted by insurers' assets. Vertical lines represent 90\% confidence intervals using standard errors clustered at the insurer level.}
\end{figure}

\begin{figure}[t!]
        \includegraphics[width = \textwidth]{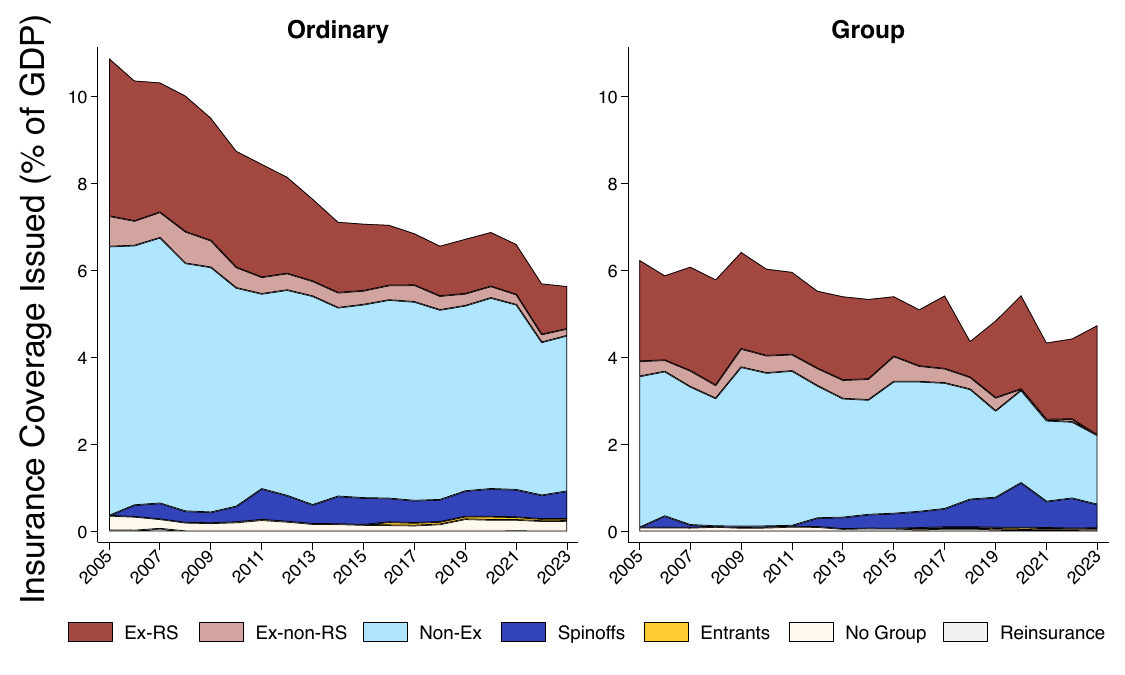}
	\caption{Aggregate Issuance By Product}
	\label{fig:aggregate iss dynamics}
    \vspace{-0.3em}
	\floatfoot{Note: This figure reports real aggregate life insurance issuance as a percentage of real GDP from 2005 to 2023. The first panel reflects ordinary life issuance, and the second panel reflects group life issuance. Red areas represent exposed insurance groups, light red areas represent non-RS-exposed insurance groups, light blue areas represent non-exposed insurance groups, dark blue areas reflect insurance companies that belonged to either the exposed or non-exposed insurance groups in the pre-crisis period but have since spun off, yellow areas represent new entrants relative to the pre-crisis period, white areas represent insurers not in a life insurance group, and gray areas reflect reinsurance companies.}

\end{figure}

\end{appendix}


\begin{thebibliography}{}

\bibitem[Alfaro et~al., 2026]{alfaro2024lash}
Alfaro, L., Bahaj, S.~A., Czech, R., Hazell, J., and Neamtu, I. (2026).
\newblock {LASH} risk and interest rates.
\newblock {\em Working Paper}.

\bibitem[Barbu, 2023]{barbu2023ex}
Barbu, A. (2023).
\newblock Ex-post loss sharing in consumer financial markets.
\newblock {\em Working Paper}.

\bibitem[Barbu et~al., 2024]{barbu2024a}
Barbu, A., Humphrey, D., and Sen, I. (2024).
\newblock The evolution of insurance markets: Insurance provision and capital
  regulation.
\newblock {\em Working Paper}.

\bibitem[Barbu and Sen, 2024]{barbu2024b}
Barbu, A. and Sen, I. (2024).
\newblock Hedging through product design.
\newblock {\em Working Paper}.

\bibitem[Berends et~al., 2013]{berends2013sensitivity}
Berends, K., McMenamin, R., Plestis, T., and Rosen, R.~J. (2013).
\newblock The sensitivity of life insurance firms to interest rate changes.
\newblock {\em Economic Perspectives, Federal Reserve Bank of Chicago}, 37(2).

\bibitem[Briggs et~al., 2026]{briggs2023risky}
Briggs, J., Rogers, C., and Tonetti, C. (2026).
\newblock Risky insurance: Life-cycle insurance portfolio choice with
  incomplete markets.
\newblock {\em Working paper}.

\bibitem[Bundick et~al., 2024]{bundick2024introducing}
Bundick, B., Smith, A.~L., and Van~der Meer, L. (2024).
\newblock Introducing the {Kansas City Fed's} measure of policy rate
  uncertainty ({KC PRU}).
\newblock {\em Economic Review, Federal Reserve Bank of Kansas City}, 109(7).

\bibitem[Carr, 1975]{carr1975}
Carr, W. H.~A. (1975).
\newblock {\em From three cents a week: The story of the Prudential Insurance
  Company of America}.
\newblock Prentice Hall.

\bibitem[Coleman et~al., 2007]{coleman2007robustly}
Coleman, T.~F., Kim, Y., Li, Y., and Patron, M. (2007).
\newblock Robustly hedging variable annuities with guarantees under jump and
  volatility risks.
\newblock {\em Journal of Risk and Insurance}, 74(2):347--376.

\bibitem[Damast et~al., 2025]{damast2025homeowners}
Damast, D., Kubitza, C., and S{\o}rensen, J.~A. (2025).
\newblock Homeowners insurance and the transmission of monetary policy.
\newblock {\em Working Paper}.

\bibitem[Dobbyn, 2015]{ambest2015yielding}
Dobbyn, T. (2015).
\newblock Yielding to the market.
\newblock {\em AM Best's Review}.

\bibitem[Domanski et~al., 2017]{domanski2017hunt}
Domanski, D., Shin, H.~S., and Sushko, V. (2017).
\newblock The hunt for duration: not waving but drowning?
\newblock {\em IMF Economic Review}, 65:113--153.

\bibitem[Drexler, 2024]{Drexler2024InterestRateRisk}
Drexler, A.~H. (2024).
\newblock Measuring interest rate risk sensitivity around {FOMC} announcements.
\newblock {\em Chicago Fed Letter}, (499).

\bibitem[Ellis et~al., 2025]{ellis2025impact}
Ellis, C., Ellul, A., Jotikasthira, C., and Xu, J. (2025).
\newblock The impact of risk management mandates: Evidence from variable
  annuities.
\newblock {\em Working Paper}.

\bibitem[Ellul et~al., 2022]{ellul2022insurers}
Ellul, A., Jotikasthira, C., Kartasheva, A., Lundblad, C.~T., and Wagner, W.
  (2022).
\newblock Insurers as asset managers and systemic risk.
\newblock {\em The Review of Financial Studies}, 35(12):5483--5534.

\bibitem[Froot, 2001]{froot2001market}
Froot, K.~A. (2001).
\newblock The market for catastrophe risk: a clinical examination.
\newblock {\em Journal of Financial Economics}, 60(2-3):529--571.

\bibitem[Ge, 2022]{ge2022financial}
Ge, S. (2022).
\newblock How do financial constraints affect product pricing? evidence from
  weather and life insurance premiums.
\newblock {\em The Journal of Finance}, 77(1):449--503.

\bibitem[Giambona et~al., 2025]{giambona2025hedging}
Giambona, E., Kumar, A., and Phillips, G.~M. (2025).
\newblock Hedging, contract enforceability, and competition.
\newblock {\em The Review of Financial Studies}, 38(7):2034--2087.

\bibitem[Gottlieb and Smetters, 2021]{gottlieb2021lapse}
Gottlieb, D. and Smetters, K. (2021).
\newblock Lapse-based insurance.
\newblock {\em American Economic Review}, 111(8):2377--2416.

\bibitem[Greenwood and Vissing-Jorgensen, 2018]{greenwood2018impact}
Greenwood, R.~M. and Vissing-Jorgensen, A. (2018).
\newblock The impact of pensions and insurance on global yield curves.
\newblock {\em Working Paper}.

\bibitem[Gron, 1994]{gron1994capacity}
Gron, A. (1994).
\newblock Capacity constraints and cycles in property-casualty insurance
  markets.
\newblock {\em The RAND Journal of Economics}, 25(1):110--127.

\bibitem[Gropper and Kuhnen, 2025]{gropper2025wealth}
Gropper, M.~J. and Kuhnen, C.~M. (2025).
\newblock Wealth and insurance choices: Evidence from {US} households.
\newblock {\em The Journal of Finance}, 80(2):1127--1170.

\bibitem[Guardian, 2023]{guardian2023}
Guardian (2023).
\newblock Prepared and protected: How life insurance supports financial
  wellness for those you love.
\newblock Technical report, Guardian Life Insurance Company.

\bibitem[Hartley et~al., 2017]{hartley2017explains}
Hartley, D., Paulson, A., and Powers, K. (2017).
\newblock What explains the decline in life insurance ownership.
\newblock {\em Federal Reserve Bank of Chicago, Economic Perspectives},
  41(8):1--20.

\bibitem[Hartley et~al., 2016]{hartley2016measuring}
Hartley, D., Paulson, A., and Rosen, R.~J. (2016).
\newblock Measuring interest rate risk in the life insurance sector.
\newblock In Hufeld, F., Koijen, R. S.~J., and Thimann, C., editors, {\em The
  Economics, Regulation, and Systemic Risk of Insurance Markets}, chapter~6,
  pages 124--150. Oxford University Press.

\bibitem[Heinrich et~al., 2026]{heinrich2026liability}
Heinrich, N., Verani, S., and Yu, P.~C. (2026).
\newblock Liability structure and monetary transmission: Evidence from life
  insurers.
\newblock {\em Working Paper}.

\bibitem[Huber, 2022]{huber2022}
Huber, M. (2022).
\newblock Regulation-induced interest rate risk exposure.
\newblock {\em Working Paper}.

\bibitem[Inkmann et~al., 2011]{inkmann2011deep}
Inkmann, J., Lopes, P., and Michaelides, A. (2011).
\newblock How deep is the annuity market participation puzzle?
\newblock {\em The Review of Financial Studies}, 24(1):279--319.

\bibitem[Kirti and Singh, 2024]{kirti2024}
Kirti, D. and Singh, A.~V. (2024).
\newblock The insurer channel of monetary policy.
\newblock {\em Working Paper}.

\bibitem[Knight, 1920]{knight1920}
Knight, C.~K. (1920).
\newblock {\em The History Of Life Insurance In The United States To 1870: With
  An Introduction To Its Development Abroad}.
\newblock Kessinger Publishing.

\bibitem[Knox and S{\o}rensen, 2024]{knox2024insurers}
Knox, B. and S{\o}rensen, J.~A. (2024).
\newblock Insurers’ investments and insurance prices.
\newblock {\em Working Paper}.

\bibitem[Koijen et~al., 2024]{koijen2024aggregate}
Koijen, R.~S., Lee, H.~K., and Van~Nieuwerburgh, S. (2024).
\newblock Aggregate lapsation risk.
\newblock {\em Journal of Financial Economics}, 155:103819.

\bibitem[Koijen et~al., 2016]{koijen2016health}
Koijen, R.~S., Van~Nieuwerburgh, S., and Yogo, M. (2016).
\newblock Health and mortality delta: Assessing the welfare cost of household
  insurance choice.
\newblock {\em The Journal of Finance}, 71(2):957--1010.

\bibitem[Koijen and Yogo, 2015]{koijen2015cost}
Koijen, R.~S. and Yogo, M. (2015).
\newblock The cost of financial frictions for life insurers.
\newblock {\em American Economic Review}, 105(1):445--475.

\bibitem[Koijen and Yogo, 2021]{koijen2021evolution}
Koijen, R.~S. and Yogo, M. (2021).
\newblock The evolution from life insurance to financial engineering.
\newblock {\em Geneva Risk and Insurance Review}, 46(2):89--111.

\bibitem[Koijen and Yogo, 2022]{koijen2022fragility}
Koijen, R.~S. and Yogo, M. (2022).
\newblock The fragility of market risk insurance.
\newblock {\em The Journal of Finance}, 77(2):815--862.

\bibitem[Koijen and Yogo, 2023]{koijen2023understanding}
Koijen, R.~S. and Yogo, M. (2023).
\newblock Understanding the ownership structure of corporate bonds.
\newblock {\em American Economic Review: Insights}, 5(1):73--91.

\bibitem[Kubitza et~al., 2025]{kubitza2023life}
Kubitza, C., Grochola, N., and Gr{\"u}ndl, H. (2025).
\newblock Life insurance convexity.
\newblock {\em Journal of Banking \& Finance}, 178:107502.

\bibitem[Li et~al., 2026]{li2026improving}
Li, W., Moenig, T., and Augustyniak, M. (2026).
\newblock Improving fund mapping: An application to variable annuities.
\newblock {\em Journal of Risk and Insurance}.

\bibitem[Li, 2026]{li2024}
Li, Z. (2026).
\newblock Long rates, life insurers, and credit spreads.
\newblock {\em Working Paper}.

\bibitem[{LIMRA}, 2024]{limra2024barometer}
{LIMRA} (2024).
\newblock 2024 insurance barometer study.
\newblock Technical report, LIMRA and Life Happens.

\bibitem[Ozdagli and Wang, 2019]{ozdagli2019interest}
Ozdagli, A.~K. and Wang, Z.~K. (2019).
\newblock Interest rates and insurance company investment behavior.
\newblock {\em Working Paper}.

\bibitem[Panko, 2012]{panko2012}
Panko, R.~J. (2012).
\newblock Two sides of interest-rate risk.
\newblock {\em AM Best's Review}.

\bibitem[Scism, 2023]{wsj2023}
Scism, L. (2023).
\newblock Life insurance is profitable again, but too late for many insurers.
\newblock {\em Wall Street Journal}.

\bibitem[Sen, 2023]{sen2023regulatory}
Sen, I. (2023).
\newblock Regulatory limits to risk management.
\newblock {\em The Review of Financial Studies}, 36(6):2175--2223.

\bibitem[Tang, 2026]{tang2022}
Tang, J. (2026).
\newblock Regulatory competition in the {US} life insurance industry.
\newblock {\em Journal of Political Economy}, Forthcoming.

\bibitem[Verani and Yu, 2024]{verani2024s}
Verani, S. and Yu, P.~C. (2024).
\newblock What’s wrong with annuity markets?
\newblock {\em Journal of the European Economic Association}, 22(4):1981--2024.

\bibitem[Wenning, 2024]{wenning2024}
Wenning, D. (2024).
\newblock National pricing and the geography of {U.S.} life insurers.
\newblock {\em Working Paper}.

\bibitem[Zanjani, 2002]{zanjani2002pricing}
Zanjani, G. (2002).
\newblock Pricing and capital allocation in catastrophe insurance.
\newblock {\em Journal of Financial Economics}, 65(2):283--305.

\end{thebibliography}
\end{document}